\documentclass[twocolumn]{aastex63}

\newcommand\nw{nW m$^{-2}$ sr$^{-1}$}

\received{28 February, 2025}
\revised{29 July, 2025}
\accepted{12 August, 2025}

\shorttitle{The Cosmic Infrared Background Experiment-2 (CIBER-2)}
\shortauthors{Zemcov et al.}

\usepackage{comment}
\usepackage{multirow}
\usepackage{soul}
\begin{document}

\title{The Cosmic Infrared Background Experiment-2: An Intensity Mapping Optimized Sounding-rocket Payload to Understand the Near-IR Extragalactic Background Light}

\correspondingauthor{M. Zemcov}
\email{mbzsps@rit.edu}

\author[0000-0001-8253-1451]{Michael Zemcov}
\affiliation{Rochester Institute of Technology, Rochester, NY 14623, USA}
\affiliation{Jet Propulsion Laboratory (JPL), National Aeronautics and Space Administration (NASA), Pasadena, CA 91109, USA}

\author[0000-0002-5710-5212]{James J. Bock}
\affiliation{California Institute of Technology, Pasadena, CA 91125, USA}
\affiliation{Jet Propulsion Laboratory (JPL), National Aeronautics and Space Administration (NASA), Pasadena, CA 91109, USA}

\author[0000-0002-3892-0190]{Asantha Cooray}
\affiliation{University of California, Irvine, CA 92697, USA}

\author[0000-0002-5698-9634]{Shuji Matsuura}
\affiliation{Kwansei Gakuin University, Sanda, Hyogo 669-1330, Japan}

\author{Dae-Hee Lee}
\affiliation{Korea Astronomy and Space Science Institute (KASI), Daejeon 34055, Republic of Korea}
\affiliation{Korea Advanced Institute of Science and Technology (KAIST), Daejeon 34141, Republic of Korea}

\author[0000-0001-9925-0146]{Candice Fazar}
\affiliation{Rochester Institute of Technology, Rochester, NY 14623, USA}

\author[0000-0002-9330-8738]{Richard M. Feder}
\affiliation{Berkeley Center for Cosmological Physics, University of California, Berkeley, CA 94720,
USA}
\affiliation{Lawrence Berkeley National Laboratory, Berkeley, CA 94720, USA}

\author[0009-0003-5681-2956]{Grigory Heaton}
\affiliation{California Institute of Technology, Pasadena, CA 91125, USA}

\author{Ryo Hashimoto}
\affiliation{Kwansei Gakuin University, Sanda, Hyogo 669-1330, Japan}

\author{Phillip Korngut}
\affiliation{California Institute of Technology, Pasadena, CA 91125, USA}

\author{Toshio Matsumoto}
\affiliation{Institute of Space and Astronautical Science (ISAS), Japan Aerospace Exploration Agency (JAXA),\\Sagamihara, Kanagawa 252-5210, Japan}

\author[0000-0001-9368-3186]{Chi H. Nguyen}
\affiliation{California Institute of Technology, Pasadena, CA 91125, USA}

\author[0000-0002-5756-5964]{Kazuma Noda}
\affiliation{Rochester Institute of Technology, Rochester, NY 14623, USA}

\author[0000-0002-8292-2556]{Won-Kee Park}
\affiliation{Korea Astronomy and Space Science Institute (KASI), Daejeon 34055, Republic of Korea}

\author[0000-0002-6468-8532]{Kei Sano}
\affiliation{Kyushu Institute of Technology, Kita-Kyushu, Fukuoka 804-0015, Japan}

\author[0000-0002-8405-9549]{Kohji Takimoto}
\affiliation{Institute of Space and Astronautical Science (ISAS), Japan Aerospace Exploration Agency (JAXA),\\Sagamihara, Kanagawa 252-5210, Japan}

\author{Toshiaki Arai}
\affiliation{Institute of Space and Astronautical Science (ISAS), Japan Aerospace Exploration Agency (JAXA),\\Sagamihara, Kanagawa 252-5210, Japan}

\author{Seung-Cheol Bang}
\affiliation{Korea Astronomy and Space Science Institute (KASI), Daejeon 34055, Republic of Korea}

\author[0000-0002-3886-3739]{Priyadarshini Bangale}
\affiliation{Rochester Institute of Technology, Rochester, NY 14623, USA}
\affiliation{Department of Physics, Temple University, Philadelphia, PA 19122, USA}

\author{Masaki Furutani}
\affiliation{Kwansei Gakuin University, Sanda, Hyogo 669-1330, Japan}

\author{Viktor Hristov}
\affiliation{California Institute of Technology, Pasadena, CA 91125, USA}

\author{Yuya Kawano}
\affiliation{Kwansei Gakuin University, Sanda, Hyogo 669-1330, Japan}

\author{Arisa Kida}
\affiliation{Kwansei Gakuin University, Sanda, Hyogo 669-1330, Japan}

\author{Tomoya Kojima}
\affiliation{Kwansei Gakuin University, Sanda, Hyogo 669-1330, Japan}

\author[0000-0003-2565-1558]{Alicia Lanz}
\affiliation{The Observatories of the Carnegie Institution for Science, Pasadena, CA 91101, USA}

\author{Chika Matsumi}
\affiliation{Kwansei Gakuin University, Sanda, Hyogo 669-1330, Japan}

\author[0000-0002-7485-7563]{Dale Mercado}
\affiliation{Rochester Institute of Technology, Rochester, NY 14623, USA}

\author{Shunsuke Nakagawa}
\affiliation{Kyushu Institute of Technology, Kita-Kyushu, Fukuoka 804-0015, Japan}

\author{Tomoya Nakagawa}
\affiliation{Kwansei Gakuin University, Sanda, Hyogo 669-1330, Japan}

\author{Shuta Nakahata}
\affiliation{Kwansei Gakuin University, Sanda, Hyogo 669-1330, Japan}

\author{Ryo Ohta}
\affiliation{Kwansei Gakuin University, Sanda, Hyogo 669-1330, Japan}

\author[0009-0008-5671-9061]{Dorin Patru}
\affiliation{Rochester Institute of Technology, Rochester, NY 14623, USA}

\author{Mai Shirahata}
\affiliation{Institute of Space and Astronautical Science (ISAS), Japan Aerospace Exploration Agency (JAXA),\\Sagamihara, Kanagawa 252-5210, Japan}

\author{Hiroko Suzuki}
\affiliation{Kwansei Gakuin University, Sanda, Hyogo 669-1330, Japan}

\author[0000-0003-3881-3202]{Aoi Takahashi}
\affiliation{Institute of Space and Astronautical Science (ISAS), Japan Aerospace Exploration Agency (JAXA),\\Sagamihara, Kanagawa 252-5210, Japan}

\author{Momoko Tamai}
\affiliation{Kwansei Gakuin University, Sanda, Hyogo 669-1330, Japan}

\author[0009-0002-8148-7502]{Serena Tramm}
\affiliation{Rochester Institute of Technology, Rochester, New York 14623, USA}
\affiliation{Jet Propulsion Laboratory (JPL), National Aeronautics and Space Administration (NASA), Pasadena, CA 91109, USA}

\author[0000-0001-7143-6520]{Kohji Tsumura}
\affiliation{Tokyo City University, Tamazutsumi, Setagaya-ku, Tokyo 158-8557, Japan}

\author{Yasuhiro Yamada}
\affiliation{Kwansei Gakuin University, Sanda, Hyogo 669-1330, Japan}

\author[0000-0001-6491-1901]{Shiang-Yu Wang}
\affiliation{Institute of Astronomy and Astrophysics, Academia Sinica, Taipei 10617, Taiwan R.O.C.}

\begin{abstract}
The background light produced by emission from all sources over cosmic history is a powerful diagnostic of structure formation and evolution. At near-infrared wavelengths, this extragalactic background light (EBL) is comprised of emission from galaxies stretching all the way back to the first-light objects present during the Epoch of Reionization.  The Cosmic Infrared Background Experiment 2 (CIBER-2) is a sounding-rocket experiment designed to measure both the absolute photometric brightness of the EBL over $0.5{-}2.0 \, \mu$m and perform an intensity mapping measurement of EBL spatial fluctuations in six broad bands over the same wavelength range.  CIBER-2 comprises a 28.5 cm 80~K telescope that images several square degrees to three separate cameras.  Each camera is equipped with an HAWAII-2RG detector covered by an assembly that combines two broadband filters and a linear-variable filter, which perform the intensity mapping and absolute photometric measurements, respectively. CIBER-2 has flown three times: an engineering flight in 2021; a terminated launch in 2023; and a successful science flight in 2024.  In this paper, we review the science case for the experiment; describe the factors motivating the instrument design; review the optical, mechanical, and electronic implementation of the instrument; present preflight laboratory characterization measurements; and finally assess the instrument's performance in flight.
\end{abstract}


\section{Introduction}

The extragalactic background light (EBL) is the line-of-sight integral of the electromagnetic radiation intensity produced over cosmic history \citep{Cooray2016, Hill2018}. At near-infrared (IR) wavelengths, the EBL is the summed emission from nucleosynthetic and gravitational processes in galaxies all the way back to the first objects present during the Epoch of Reionization (EoR).  While many sources of the near-IR EBL are readily studied using deep surveys \cite[e.g.,][]{Carter2025}, the total EBL also captures emission from fainter and diffuse populations that may remain undetectable in even the deepest surveys with current facilities \citep{Cooray2016}.    

Over the last two decades, observations of spatial fluctuations in the EBL have emerged as an independent way to gain insight into the populations responsible for its
production \citep{Cooray2004, Kashlinsky2004}.  This technique is different from and complementary to mapping large-scale structure with galaxies, as it is sensitive to diffuse structure not accounted through resolved source surveys \citep{Bernal2022}.  This method, sometimes called broadband intensity mapping (BBIM), has already been used extensively at near-IR wavelengths.  Measurements of EBL fluctuations, after masking detected sources in imaging data, have been attempted with Spitzer/IRAC, Akari, CIBER-I, Hubble/WFC3 over the wavelength range of $\lambda = 0.5$ to 5 $\mu$m \citep{Kashlinsky2005, Kashlinsky2007, Thompson2007,
  Matsumoto2011, Kashlinsky2012, Cooray2012, Zemcov2014, Ketron2015,
  Seo2015,Matsumoto2019,Kashlinsky2025,Feder2025}. These measurements all exhibit a deviation from the power spectrum expected for unresolved stars and galaxies at angular scales larger than an arcminute.  
No consensus has been reached on the source of this emission: interpretations range from UV-bright sources during the EoR redshifted to the near-IR today \citep{Kashlinsky2005, Kashlinsky2007, Kashlinsky2012,
  Ketron2015,Kashlinsky2025}, to the diffuse emission from stars tidally stripped
from galaxies during hierarchical structure formation (intra-halo light, or IHL; see e.g., \citealt{Cooray2012, Zemcov2014}), to black holes in faint galaxies
\citep{Matsumoto2019}, to exotic explanations like axion decay in the
local Universe \citep{Gong2015, Kalashev2019}.  
The populations responsible likely have
important implications for both the history of light emission and the
astrophysical populations sourcing the diffuse near-IR background.

The second Cosmic Infrared Background Experiment (CIBER-2) is a sounding-rocket borne instrument designed to measure anisotropy in the EBL in six broad bands covering the optical and near-IR. Using high-sensitivity, wide-angle, multicolor anisotropy measurements, CIBER-2 is designed to elucidate the history of IHL production and carry out a deep search for extragalactic background fluctuations associated with the EoR. The CIBER-2 images also include a small region illuminated by continuously variable filters that can be used to perform an absolute spectrophotometric measurement of the diffuse sky brightness.  CIBER-2 maximizes the sensitivity to near-IR surface brightness observed over a short sounding-rocket flight and offers a fivefold increase in sensitivity to surface brightness compared with its predecessor, CIBER-1 \citep{Bock2013}. Crucially, CIBER-2 extends wavelength coverage down to 0.5 \micron\ (compared to 0.9 \micron\ for CIBER-1), well into the optical band, which constrains fluctuations from the Lyman break of early galaxies to $z \sim 6$.  Aspects of the instrument design and characterization have been described in the literature \citep{Lanz2014, Shirahata2016, Park2018, Nguyen2018, Takimoto2020, Matsuura2024}, but our goal in this paper is to provide a complete description of the design and as-built implementation of the CIBER-2 payload. 

The paper is organized as follows. In Section 2 we describe the mission requirements and mission architecture, in Section 3 we describe the detailed instrument implementation, in Section 4 we describe the laboratory characterization campaign, in Section 5 we describe the flight history, in Section 6 we describe the flight performance of the CIBER-2 instrument, and we conclude with a contextual outlook in Section 7.

\section{Instrument Requirements and Architecture}

CIBER-2 is a wide-field imaging instrument that is optimized for EBL background component separation in the near-IR by maximizing its sensitivity to surface brightness 
and wavelength coverage to match EBL signals of interest.  The mission requirements drive a design that maximizes the wavelength sensitivity, spectral resolving power, and etendue available in a reusable sounding-rocket payload.  
In this section, we motivate the requirements and overall design for the instrument.

\subsection{Requirements}

The design parameters that optimize CIBER-2's ability to perform BBIM observations in a short-duration sounding-rocket flight include the following:
\vspace{-8pt}
\paragraph{Maximized etendue, $A \Omega$}{Maximizing the signal-to-noise ratio (SNR) on a diffuse signal requires maximizing the product of a telescope's collecting area $A$ and field of view (FOV) $\Omega$.  At a fixed sounding-rocket skin diameter, $A$ is maximized by using a large primary optic, and $\Omega$ is maximized by employing wide-field optics to achieve a $2.3^{\circ} \times 2.3^{\circ}$ FOV in each of three cameras.}
\vspace{-8pt}
\paragraph{Pixel scale}{To permit deep source masking, a pixel scale that resolves the point-spread function (PSF) of the instrument minimizes the number of pixels lost at a given masking flux threshold. CIBER-2 uses three off-the-shelf Teledyne HAWAII-2RG (H2RG) detector arrays with 
$2048\times2048$ pixels, $6\times$ the pixel count on CIBER-1, at a $4^{\prime \prime}$ pixel$^{-1}$ plate scale that is a factor of $\sim 2$ larger than the telescope's diffraction limit at the longest wavelength.}
\vspace{-8pt}
\paragraph{Spectral coverage}{Constraining the origin of near-IR EBL fluctuations requires observations throughout the optical and near-IR.  The three cameras together cover $0.5 - 2.0$~\micron~over six wave bands to capture the Lyman break of EoR galaxies expected at $< 1$ \micron.}
\vspace{-8pt}
\paragraph{Cryogenic operation}{The instrument is cooled to temperatures where both the detector dark current and thermally generated photons in the instrument are an insignificant contribution to the measurement's noise budget.  The entire CIBER-2 instrument is cooled to $\lesssim 90 \,$K using liquid nitrogen (LN$_{2}$); the 2.5~\micron~cutoff H2RG detector arrays reach essentially a constant dark current and noise performance below $95 \,$K \citep{Blank2012}.}
\vspace{-8pt}
\paragraph{Airglow mitigation}{In the near-IR, hydroxyl (OH) radicals in the Earth's atmosphere at an altitude of $\sim 100 \,$km cause a diffuse, time-varying signal $\sim 200 {-} 1500\times$ brighter than astrophysical sources on degree scales \citep{Ramsay1992,Adams1996,Allen2000}. 
To observe above airglow, CIBER-2 flies on Black Brant 9 (BBIX) sounding-rockets that have been demonstrated to enable our astrophysical science by the CIBER-1 payload \citep{Zemcov2013}.  At our payload mass we can expect a typical apogee of $315 \,$km with $\sim 300$ s above $200 \,$km for science observations, similar to the CIBER-1 trajectory.}
\vspace{-8pt}
\paragraph{Systematic error control}{Our methodology for systematic error control and photometric calibration in CIBER-2 is based on the flight-demonstrated methods pioneered by CIBER-1 where we used a combination of in-flight and laboratory measurements to determine instrumental characteristics \cite[see][]{Zemcov2014,Takimoto2020}.  The instrument and observations are designed to permit tests to quantify systematic errors using a variety of internal consistency tests.  We observe fields with rich ancillary data (\textit{e.g.,}~from \textit{Spitzer}) to allow cross correlations and other systematic-suppressing combinations that allow us to diagnose the data quality and our scientific interpretation.}

\subsection{Baseline Architecture}

Simply, CIBER-2 comprises a reflecting near-IR optimized telescope coupled to three focal planes through a refractive/reflective optical chain packaged to fit in a $17^{\prime \prime}.26$ diameter rocket skin.
The rocket skin is the exterior of an evacuated section that is closed with a vacuum-tight bulkhead on one end and a sealed shutter door on the other.
The cryogenic insert comprising the telescope, optics, focal plane assemblies (FPAs), and a small liquid nitrogen cryostat is suspended inside of the skin by a thermally insulating suspension system.
The $28.5 \,$cm aluminum (Al) Cassegrain telescope is imaged to three focal planes through a system of lenses, mirrors, and beamsplitters that divide CIBER-2's full spectral range into short, medium, and long wavelength channels. 
The light from each camera is focused on a 
near-IR H2RG detector array \citep{Blank2011} housed in an FPA that is designed to accommodate the mechanical support, optical filtering, thermal regulation, and focusing mechanism for each detector.  
The H2RG detector arrays are operated by a custom-built electronics system located on the warm side of the forward vacuum bulkhead.  A cryogenic cable harness couples the detector cold boards to hermetic connections on the vacuum bulkhead, and these are fed into the warm electronics box. The warm electronics generate the clock and bias signals required to operate each detector, and also amplify, condition, and digitize their analog outputs.  In addition, the warm electronics run a variety of housekeeping and actuated systems, store and transmit testing and flight data, and perform power conditioning and interface to the rocket's electronic systems. 
Design values for key parameters of the CIBER-2 instrument are summarized in Table \ref{table:detector_filter}.  

\noindent\begin{table*}[ht]

	\centering
	\caption{Key CIBER-2 instrument design parameters. Values reflect as-built characteristics measured in the lab (see Table \ref{table:CDS_noise}), except for the ``derived quantities'' that are calculated according to Appendix A of \citet{Bock2013}.  Here, we conservatively assume observations of the COSMOS field at $\epsilon = -9^{\circ}$, which is $\sim 1.7 \times$ brighter than the minimum ZL at the ecliptic poles.  ZL dominates the astrophysical photocurrent at all ecliptic latitudes. 
	\label{table:detector_filter}}
 \hspace{-1.5cm}
	\begin{tabular}{|l|c|c|c|c|c|c|c|}
	    \hline
	    \textbf{Parameter} & \multicolumn{2}{c|}{\textbf{Channel 1}} & \multicolumn{2}{c|}{\textbf{Channel 2}} & \multicolumn{2}{c|}{\textbf{Channel 3}} & Units \\
	    \hline
        
        Aperture &  \multicolumn{6}{c|}{28.5} & cm \\
        Pixel size & \multicolumn{6}{c|}{$4.0 \times 4.0$} & arcsec$^{2}$ \\
        FOV per band & \multicolumn{6}{c|}{$1.1 \times 1.8$ (six WPFs); $1.4 \times 0.20$ (three LVFs)} & degrees$^{2}$ \\
        Frame interval & \multicolumn{6}{c|}{1.35} & s \\
        $T_{\rm obs}$ per field & \multicolumn{6}{c|}{$\sim 50$} & s \\ \hline
		
        Read noise               & \multicolumn{2}{c|}{15}  & \multicolumn{2}{c|}{11}   & \multicolumn{2}{c|}{11}   & e$^{-}$ \\ 

        LVF Spectral Range & \multicolumn{2}{c|}{$1.35{-}2.00$}& \multicolumn{2}{c|}{$0.81{-} 1.39$} & \multicolumn{2}{c|}{$0.50{-}0.87$} & \micron \\ 
        LVF $\lambda/\Delta\lambda$ & \multicolumn{2}{c|}{$21.8{-}22.0$}& \multicolumn{2}{c|}{$21.0{-} 20.9$} & \multicolumn{2}{c|}{$20.5{-}22.0$} & \\ \cline{1-8}

        Band Name & C6 & C5 & C4 & C3 & C2 & C1 & \\
        Window Pane Pivot $\lambda$                   & 1.89     & 1.67         & 1.27     & 1.105             & 0.80     & 0.60            & \micron \\
        Window Pane $\lambda/\lambda\Delta$     & 8.4    & 7.1        & 7.8    & 6.2            & 3.1    & 2.9            & \\ 
        System efficiency & 0.71 & 0.71 & 0.69 & 0.67 & 0.58 & 0.68 & \\ \hline

        Expected sky $\lambda I_{\lambda}$              & 255     & 425         & 545     & 680             & 850     & 1020             & \nw \\ \hline
        \multicolumn{8}{|c|}{Derived Quantities} \\ \hline
        Photo current               & 5.1     & 8.9         & 7.3     & 9.8             & 11.6     & 17.1             & e$^{-}$ s$^{-1}$\\
        Responsivity                & 19.9      & 20.8      & 13.5     & 14.4            & 13.6    & 16.8            & (m$e^{-}$ s$^{-1}$)/(\nw)\\
        $\delta \lambda I_{\lambda}$ & 17.1    & 21.0        & 29.2    & 31.3            & 35.9    & 35.3            & \nw \\
        $\delta F_{\nu}$            & 20.4    & 20.3        & 20.3    & 20.4            & 20.6    & 20.9            & $3\sigma$ AB mag \\
        \hline
	\end{tabular}
\end{table*}

\subsection{Flight Plan}
\begin{figure*}[th]
    \centering
    \includegraphics[width=0.49\textwidth]{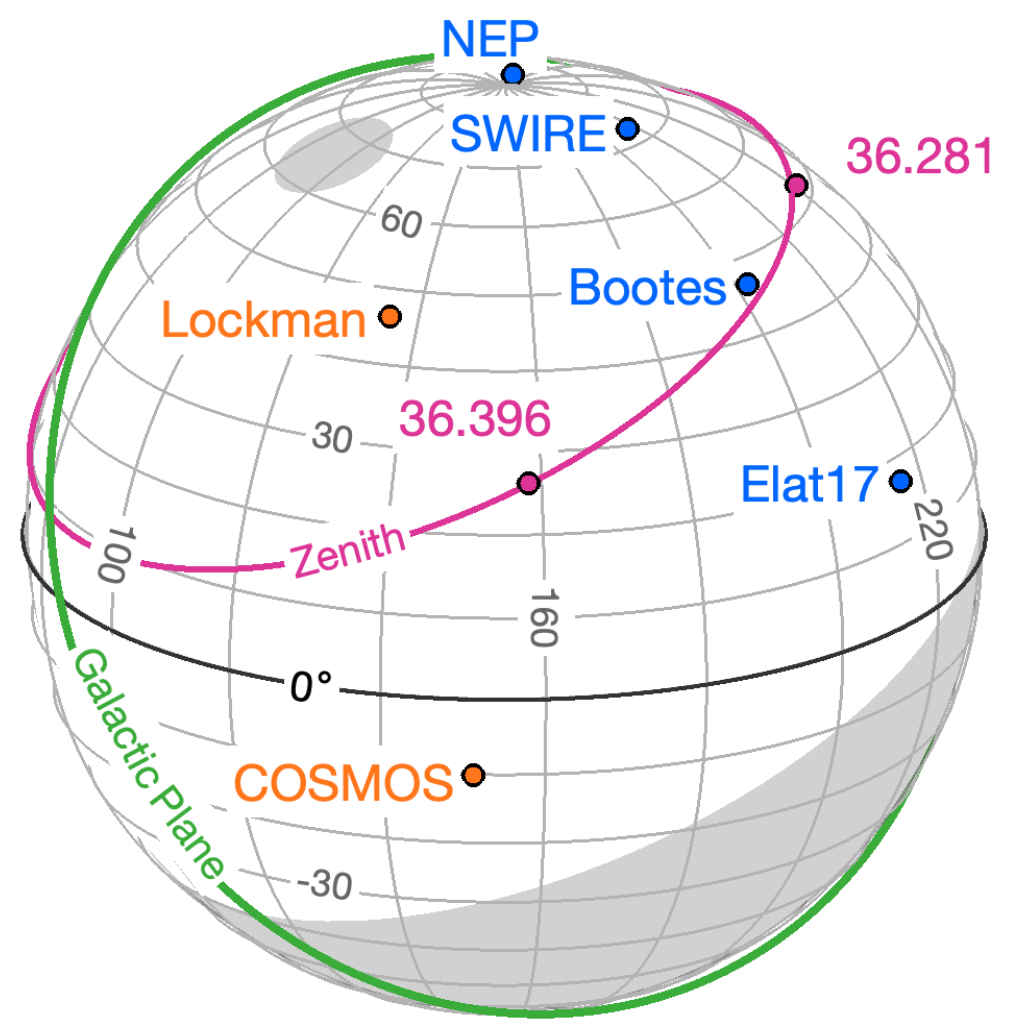} 
    \includegraphics[width=0.49\textwidth]{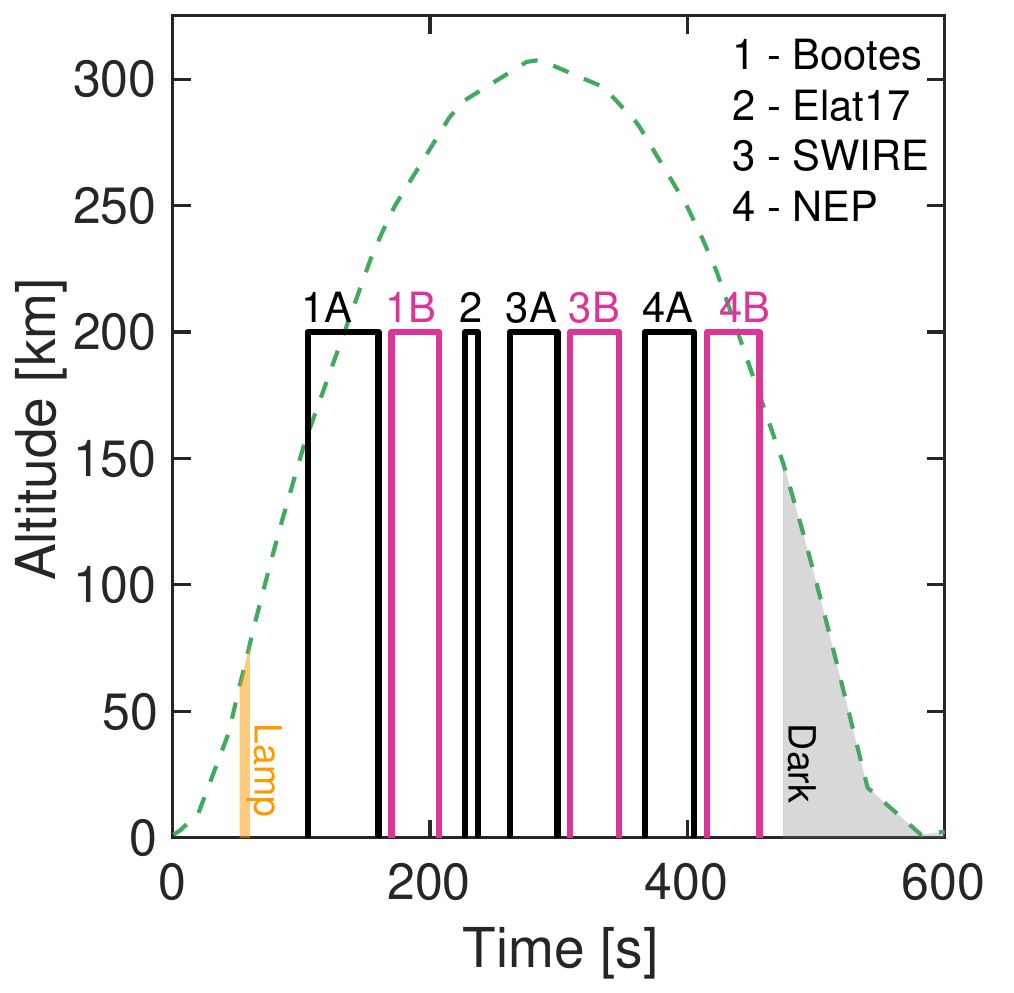} 
    \caption{\label{fig:flightplan}  (Left) Positions of science fields projected on the ecliptic sphere.  Target fields optimized for assessing foregrounds (blue points) and those optimized for intensity mapping (orange points) are visible from WSMR between Nov.~and Jul.~of each year (white area).  The extragalactic fields we consider (ELAIS-N1; north ecliptic pole, NEP; Bootes; Lockman; and COSMOS) have been surveyed extensively in the near-IR and have ample ancillary data for masking and cross correlation.  (Right) A typical flight plan for CIBER-2 modeled on flight 36.281.  The dashed green curve shows the payload's altitude versus time.  Calibration lamp data is taken during ascent (orange zone), and optically dark data is taken during descent (gray zone). }
\end{figure*}

To maximize the science return of each flight, we consider a variety of factors when designing the flight plan.  For a BBIX vehicle, CIBER-2's mass yields $\sim 350 \,$s above $100 \,$km \citep{SRPH2023}. 
We illustrate the observing plan and field choices for a typical flight in Fig.~\ref{fig:flightplan}. 
To provide a reference illumination following a powered flight, we observe three LED lamps installed behind the secondary mirror while the door is still closed.  These can be compared with reference images taken before the flight to check for mechanical shifts,  changes in the optics, and other effects.  The door opens at about $t=100 \,$s, and the payload slews to the first target based on attitude control information from a side-looking star-tracking system \citep{Percival2008}.  

CIBER-2's ``window pane'' filter (WPF) design (see \S \ref{sS:FPAs}) imposes constraints on how we can build up spectral coverage over a flight.  We can either observe multiple fields over a range of ecliptic and Galactic latitudes and forego complex spectral cross correlations, or observe a small number of adjacent and overlapping fields that permit a full set of image cross correlations to be formed.  Both flight plans have been used over CIBER-2's history (see \S \ref{S:flights}). The first option is less efficient since time is lost slewing between widely separated targets, but is more suitable for measuring the variation of ZL and DGL in a wide range of environments.  
In contrast, the second option is optimized for intensity mapping since it permits full spectral correlation.  
In this scheme, the experiment observes each extragalactic field twice, where the second exposure is positioned to overlap with half of the first observation to allow full spectral coverage in the overlap.  The payload then proceeds to the next field and repeats the process.
In either flight plan, targets are observed until the $150 \,$km downleg, at which point the cryogenic optical shutter inside the payload is actuated to provide an in-flight measurement of the read noise and dark current.  The baffle is retracted, and the payload door is closed around the 100 km downleg.  Finally, the payload is spun up, transits the atmosphere, and descends on a parachute to be recovered the next morning.

\section{Detailed Implementation}
\label{S:implementation}

\begin{figure*}[th]
    \centering
    \includegraphics[width=\textwidth]{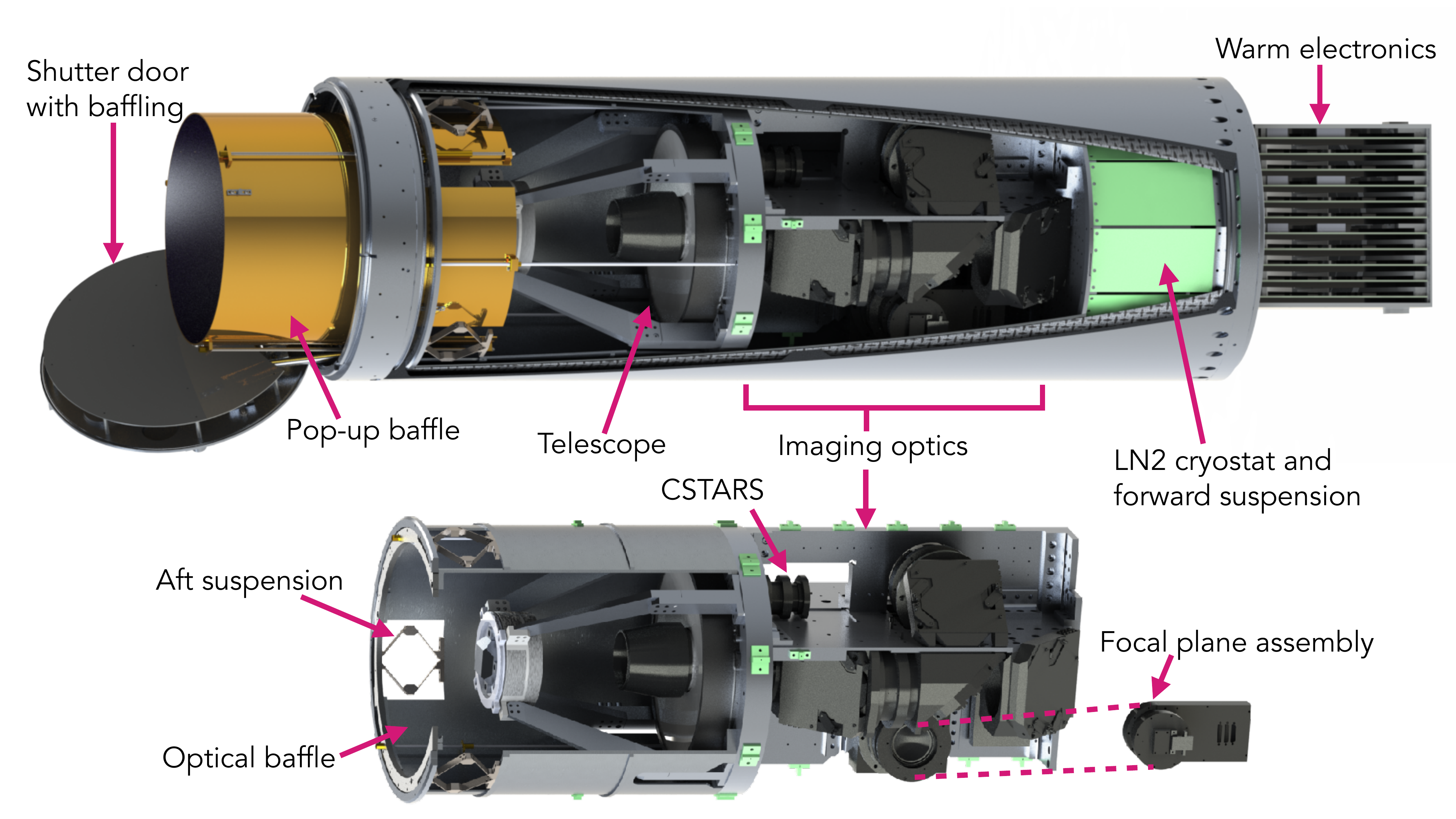} 
    \caption{\label{fig:payload} The overall structure of the CIBER-2 payload.  The upper panel shows a partial section of the instrument section in observing configuration with the door open and pop-up baffle deployed.  The rocket skin sections forward of the experiment have been suppressed.  The lower panel shows the cryogenic insert in partial section, highlighting the telescope and imaging optics described in Section \ref{sS:optics}.  The optics couple to three focal plane assemblies (Section \ref{sS:FPAs}), one of which is shown separated from the optics for clarity.}
\end{figure*}

CIBER-2 can be conceptually broken down into five major subsystems: the optical chain including the telescope and imaging cameras, the FPAs and detectors, the cryostat and cryogenic service, the actuating mechanical systems including the vacuum door and baffle, and the electronic system.  The overall mechanical architecture of the instrument is illustrated in Fig.~\ref{fig:payload}. In this section, we present details of the optical, mechanical, and electronic design and implementation of the instrument.

\subsection{Optical Assembly}
\label{sS:optics}

\begin{figure*}[th]
    \centering
    \includegraphics[width=\textwidth]{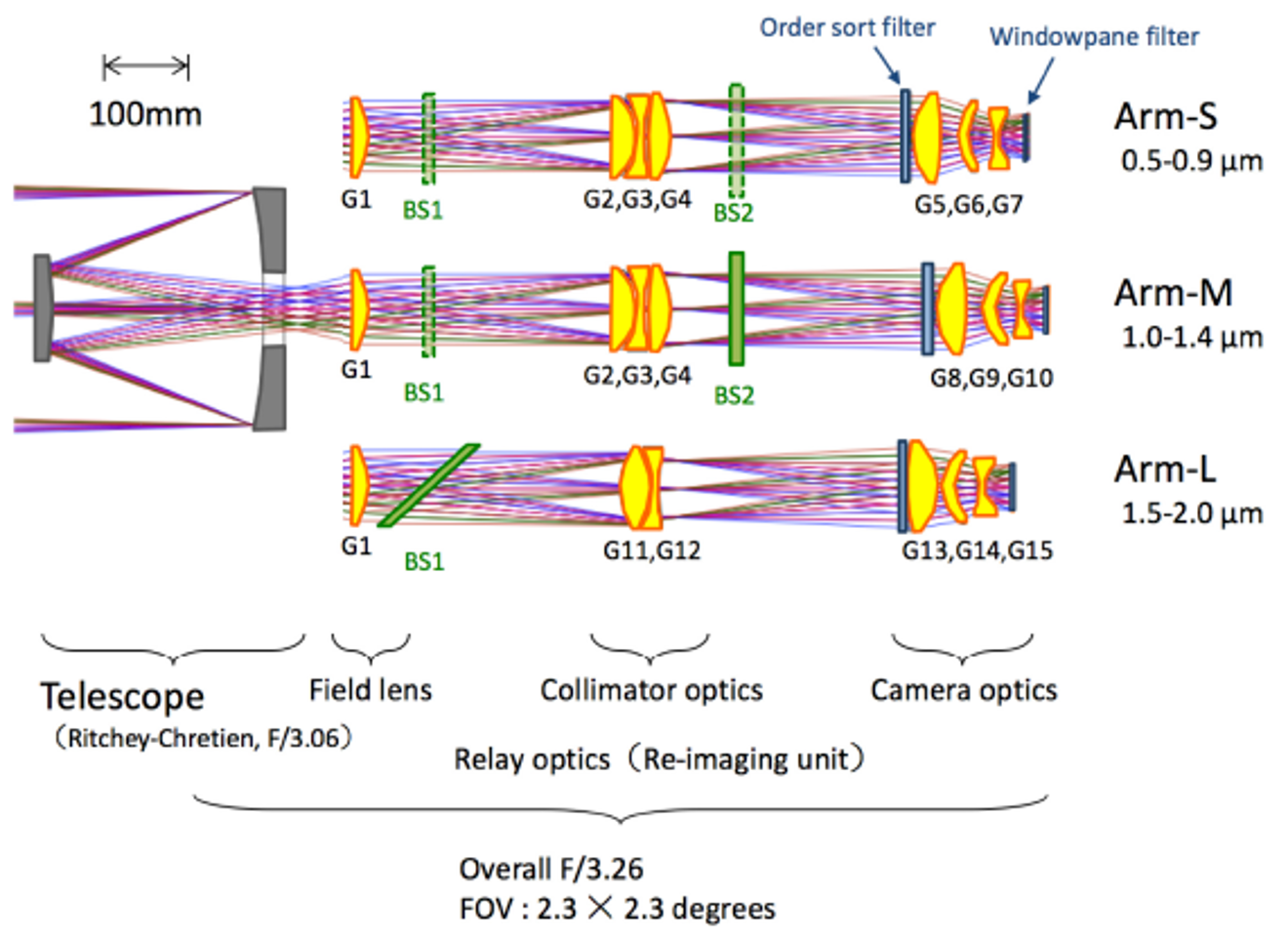} 
    \caption{\label{fig:optics} Design of the end-to-end optical chain consisting of the telescope and the relay lens optics with the light beams (blue, green, and red lines). Each lens optics for the three photometric bands, arm S, arm M and arm L, is shown separately. The G1 lens is common to all photometric bands. The light from G1 is split into two beams for arm S/arm M and arm L by a dichroic beam splitter BS1. The shorter wavelength light is collimated by the lens unit consisting of G2, G3, and G4 and split into two beams for the arm S and arm M focusing by another dichroic beam splitter BS2. The final lens system for each wavelength band focuses the beam onto the detector. The order-sorting filter to block out-of-band light is placed in front of the focusing optics for each band. The window pane filter and the linear-variable filter are placed close to the detector surface.}
\end{figure*}

The CIBER-2 optical assembly focuses sources at infinity onto three detectors sensitive to optical/near-IR light.  Details about the design and testing of the CIBER-2 optical chain are given in \cite{Shirahata2016}; in this subsection, we review the overall design and implementation.  CIBER-2 is designed around a Ritchey-Chr\'{e}tien telescope with primary and secondary mirrors of 285 and 105 mm diameter, respectively, followed by imaging optics consisting of antireflection (AR) coated spherical and aspherical lenses fabricated from S-FPL51, S-FPL53, S-LAL8, S-TIH6, S-BSL7, and fused silica. The combined telescope and the imaging lens optical system has a total $f$-number of $f/3.26$ and a focal length of $930$ mm. The design plate scale is $4^{\prime \prime}$ pix$^{-1}$, which for a $2048 \times 2048$ detector array yields a $2.3 \times 2.3$ deg$^2$ FOV.  The high-level optical design of CIBER-2 is illustrated in Fig.~\ref{fig:optics}.

\subsubsection{Telescope assembly}

CIBER-2's primary and secondary mirrors are constructed from solid Al-alloy RSA6061-T6 by diamond turning blanks to optical-grade surface roughness. The back of the primary mirror is mounted to the base plate through triangular suspension pieces to relieve thermal and mechanical stress. The surface accuracy and roughness of the primary mirror measured with a Fizeau interferometer combined with a computer generated hologram mask are approximately $0.5\lambda$ at 0.633 \micron, which is dominated by a threefold rotational symmetry distortion due to the stress from the triangle-symmetric mirror support. Since CIBER-2 is optimized for studies of large-scale diffuse signals, diffraction-limited imaging is not required, and the wave front error provided by a mirror of this quality is sufficient.

The mirrors are coated with a silver layer to increase the reflectivity over the full passband and then with a TiO$_2$ protective overcoat layer. The coating layers are thin enough not to induce significant stress in the mirror from the difference in the coefficient of thermal expansion between the coating and mirror materials at cryogenic temperatures.

The supporting structure and base plate of the telescope assembly are fabricated from Al 6061-T6 to ensure athermal properties when cooled from room to operating temperature. To reduce the mechanical stress on the primary mirror from the launch vibration, 1 mm thick plates of M2052 damping alloy are placed between the primary mirror suspension pads and the base plate. The secondary mirror is supported by four Al spider arms connected to the base plate.  The primary and secondary are aligned via shimming and cannot be actively adjusted.  All nonoptical parts of the telescope assembly are hard black anodized, and a multiblade light baffle is mounted above the Cassegrain hole to reduce the stray light and cut ghost light paths. 

\subsubsection{Imaging Optics}
\begin{figure}[th]
    \centering
    \includegraphics[width=0.49\textwidth]{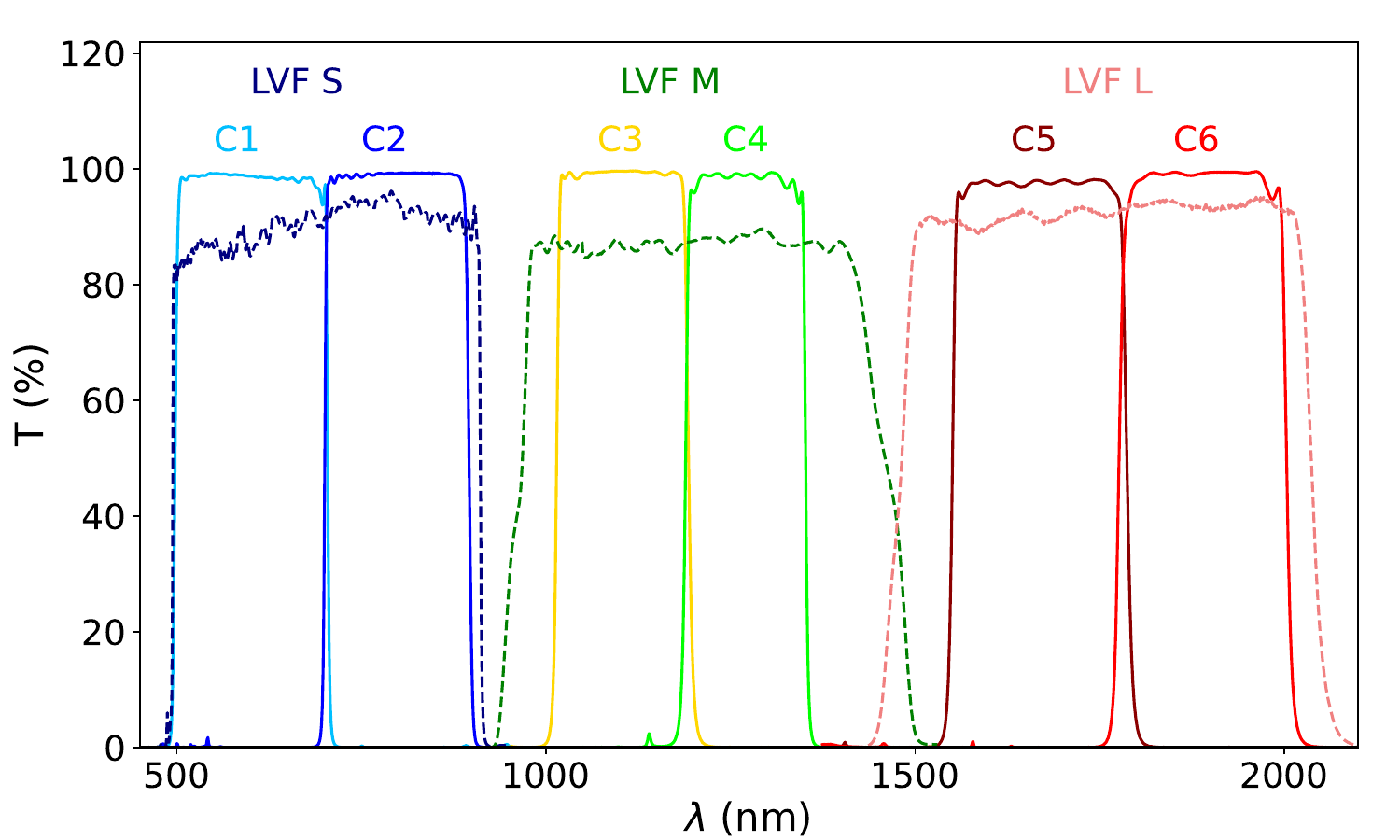} 
    \caption{\label{fig:filters}  Transmittance of the CIBER-2 filters. The solid lines represent the transmittance of the window pane filters, while the dashed lines show the transmittance of the LVFs, multiplied by the transmittance of the beam splitters and order sort filters, which determine the bandpass wavelengths. The LVF transmittance shown corresponds to the wavelength of peak transmittance of each detector column. The transmittance values presented are at room temperature.}
\end{figure}

After passing through the Cassegrain hole, the optical beam is collimated by a common field lens and divided spectrally by a pair of dichroic beam splitters into three different optical paths, which we call arm S, arm M and arm L.  This design permits simultaneous imaging over three primary wavelength ranges to be $0.5{-}0.9 \mu$m, $1.0{-}1.4 \mu$m, and $1.5{-}2.0 \mu$m, respectively.  The first dichroic beam splitter (BS1) separates the incoming light into two paths at a pivot wavelength of 1.45 $\mu$m.  The transmitting path passes to arm S and arm M, and the reflect path passes to arm L.  A second dichroic beam splitter (BS2) separates the light at a pivot wavelength of 0.95 $\mu$m, with arm S as the transmit path and arm M as the reflect path.  The three light paths arrive at their individual focal planes through independent camera optics.  First, order-sorting filters in front of the camera optics block out-of-band leakage and define the band edges.  
To avoid spectral dependencies across the FOV set by the varying angle of incidence on the beamsplitters, the order-sorting filters truncate the band around the pivot wavelengths.
After being imaged by the cameras, a WPF just in front of each detector array divides the FOV into the two wavelengths (see Section \ref{sS:FPAs} for details).  This configuration yields six simultaneous $2.2 \times 1.1$ deg$^{2}$ photometric fields over the three cameras.  The filter set is summarized graphically in Figure \ref{fig:filters}.

A small part of each focal plane is also imaged with a linear-variable filter (LVF) that enables slitless spectroscopy to measure the absolute spectrum of the astronomical sky. The spectral resolution of the LVF is $R\sim20$, and the FOV for the LVF is $1.5\times0.3$ deg$^2$. Slitless spectroscopy with the LVF is orders of magnitude more efficient than the slit spectroscopy our team implemented with the CIBER-1 LRS because of the higher optical throughput for the diffuse light and the higher spatial resolution, which allows us to subtract detected stars accurately.

\subsection{Cryogenic System}

The telescope and optics are mechanically registered through an Al6061 optical bench that is itself attached to a 7 litre LN$_2$ cryostat on the (rocket) forward end.  The cryostat is suspended from the forward vacuum bulkhead via 12 glass fiber reinforced thrust plates based on the CIBER-1 design \citep{Zemcov2013}. On the aft end of the payload where the telescope views the sky, the cryogenic insert is suspended from the rocket skin through three Ti flexures mounted to a static light baffle that is attached to the primary mirror support plate.  The static baffle is hard black anodized on its interior surface and is fabricated from a single Al6061 piece machined to provide structural rigidity during flight operations.

\begin{figure}[th]
    \centering
    \includegraphics[width=0.47\textwidth]{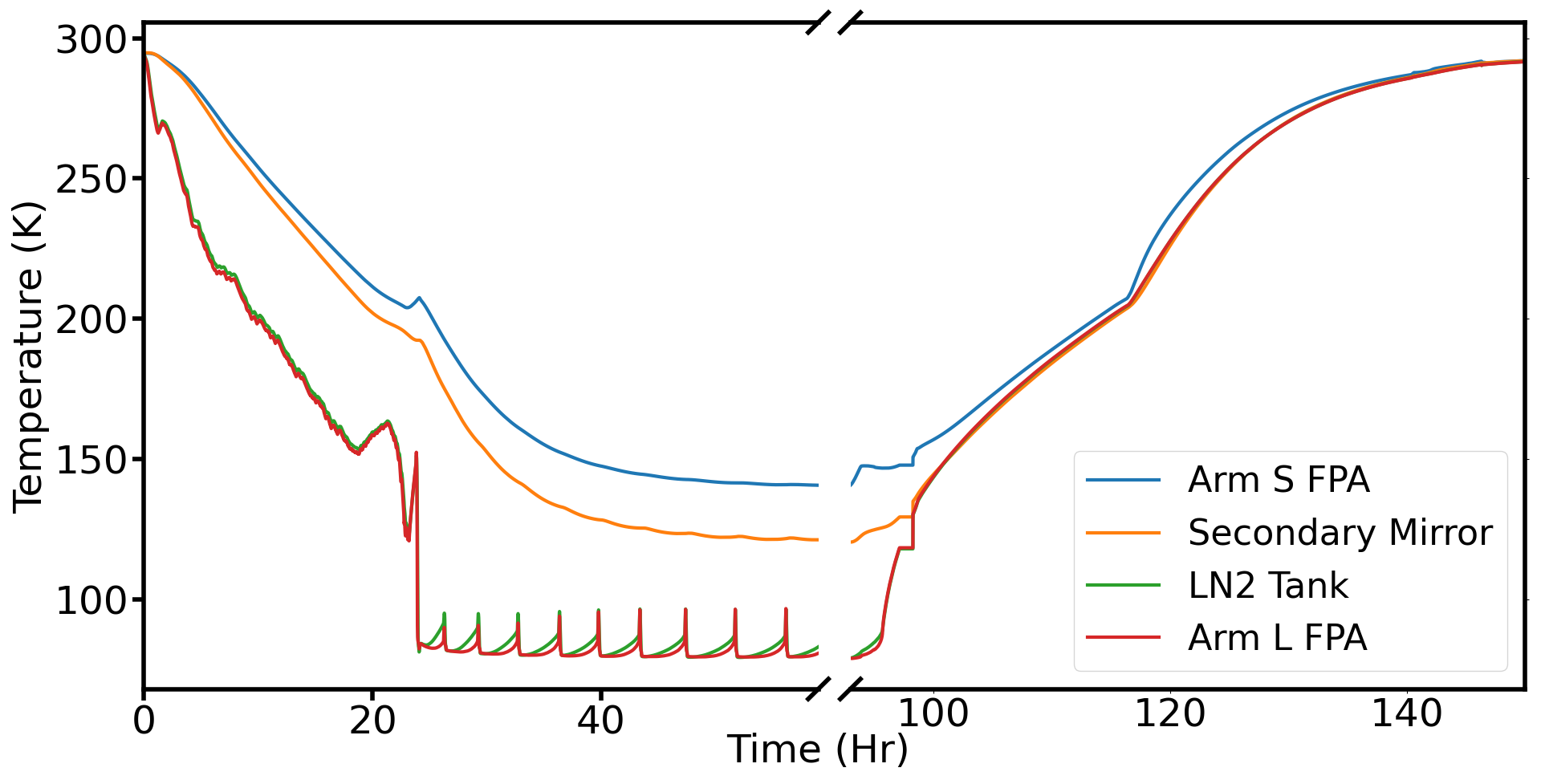} 
    \caption{\label{fig:ciber_temp_curve} Typical CIBER-2 thermal cycle showing four of two dozen thermometers that are monitored by ground equipment during laboratory testing.  The cooling cycle is computer controlled so that $dT / dt < 1 \,$K minute$^{-1}$ to account for the differential thermal contraction of the metal and glass optical assembly.  After 24 hrs the nitrogen tank can be filled, and it must be refilled every 6 hr as the optical assembly cools.  The optics reach a steady-state temperature, and the instrument can be tested 54 hr after the first fill.  Finally, the warming cycle takes about 48 hr from fully cold to ambient temperature.  In this thermal cycle, a small amount of nitrogen gas was injected into the vacuum section to improve conduction to room temperature at 120 hr.}
\end{figure}

Radiation emitted from the vacuum skin needs to be intercepted for the optical chain to reach temperatures $< 150$K.  To block this emission, a 1 mm thick Al-1100 radiation shield running from the bottom of the thrust plates to the top of the static baffle is attached to the cryostat. The shield acts as both a thermal insulation layer and to help seal any potential optical leaks in the radiation shield assembly.  The interior of the shield is hard black anodized to reduce reflections, while the exterior is polished to minimize the emissivity. The shields are wrapped in a five-layer mylar multilayered insulation blanket to further reduce the emissivity.  A typical cooling cycle for testing is shown in Fig.~\ref{fig:ciber_temp_curve}.  

\subsection{Actuating Systems and Baffling}
\label{sS:baffling}

CIBER-2 employs optical baffling to reduce susceptibility to scattered and emitted photons.  These include the static baffle described above, baffling on the shutter door, a deployed pop-up baffle, and a dark shutter.  A calibration lamp system behind the secondary mirror is used to provide a check of in-flight response and optical alignment. 

\begin{figure*}[th]
    \centering
    \includegraphics[width=\textwidth]{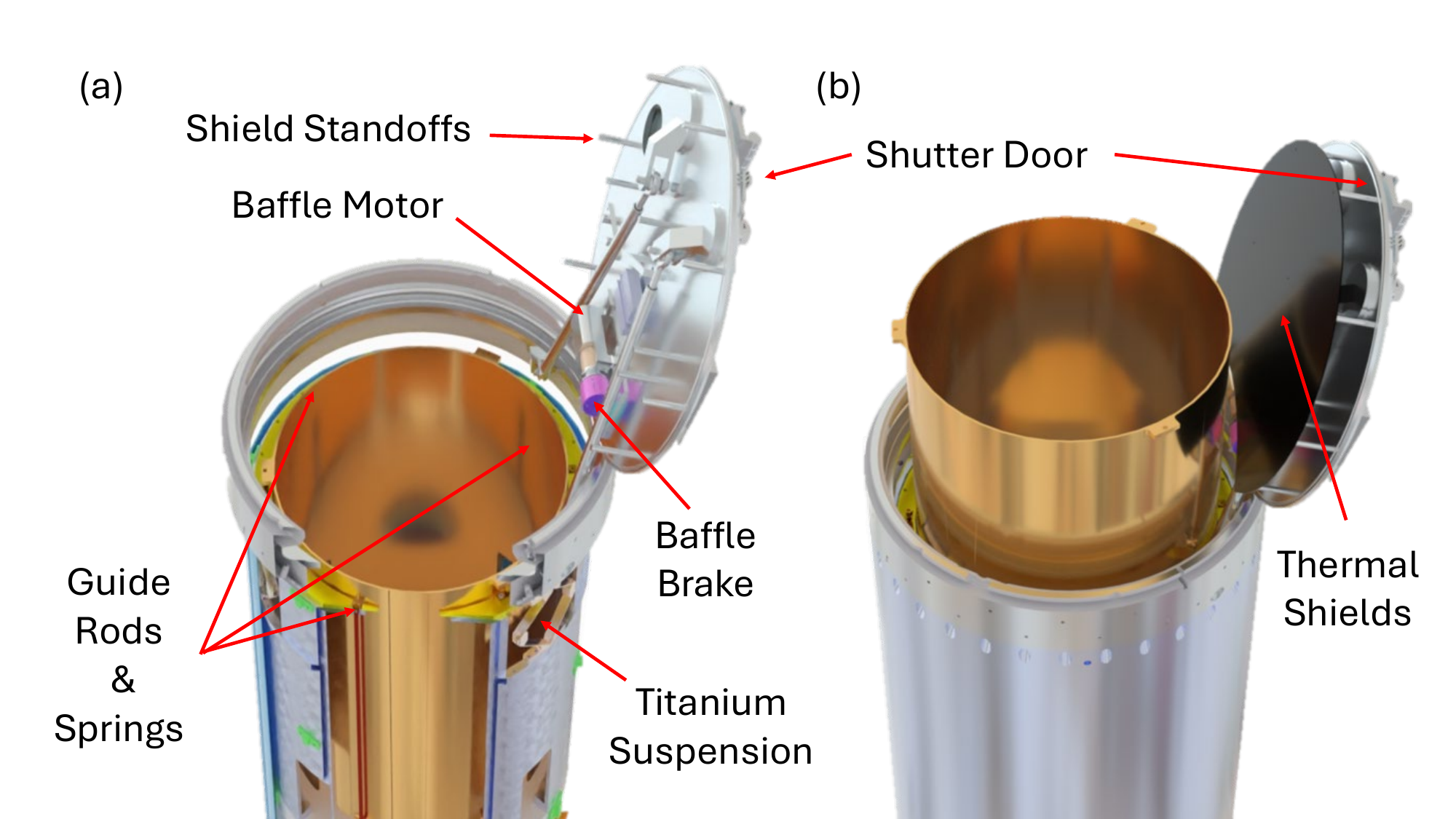} 
    \caption{\label{fig:deployed_baffle} Solid model rendering of the deployed baffle and aft end assembly.  Panel (a) shows the skin in section with the vacuum door open and the pop-up baffle in the stowed configuration to highlight normally hidden parts of the assembly.  Panel (b) shows the deployed pop-up baffle and door with radiation shield covering as it would appear when deployed during astronomical observations.  }
\end{figure*}

During the observing period, the shutter door can couple Earthshine and thermal emission from the rocket into the optical chain. To block this coupling, a cryogenic ``pop-up'' baffle is used, as illustrated in Fig.~\ref{fig:deployed_baffle}. The pop-up baffle has a slightly smaller diameter than the static baffle and when retracted sits within the static baffle assembly. In flight, the baffle is deployed by 13 inches so that the distance between its rim and the rocket interface ring is 10 inches, which blocks scatter paths from the upper part of the rocket skin and door.  The baffle motion is driven by a motor and brake system that is mounted to the shutter door. The motor spools a line attached directly to the fore end of the baffle to its deployed position. During observations, the brake is powered to apply a holding force to prevent the baffle from returning back into the optical assembly. At the end of observations before the rocket door closes, the motor unspools, which allows a set of compressed springs to retract the baffle to its starting position. 

The radiation shields for the cryogenic insert are separated from the optical bench and spines by Vespel SP-1 standoffs but connect directly to the cryostat at the forward end of the insert, which ensures that the radiative load from the 300 K rocket skin flows directly to the cryostat.  A similar radiation shield covers the vacuum door, which is made of 1 mm thick hard black anodized Al and stood off from the door plate using Vespel standoffs.  The shield stabilizes at a temperature of 175 K before launch.

\begin{figure}[th]
    \centering
    \includegraphics[width=0.45\textwidth]{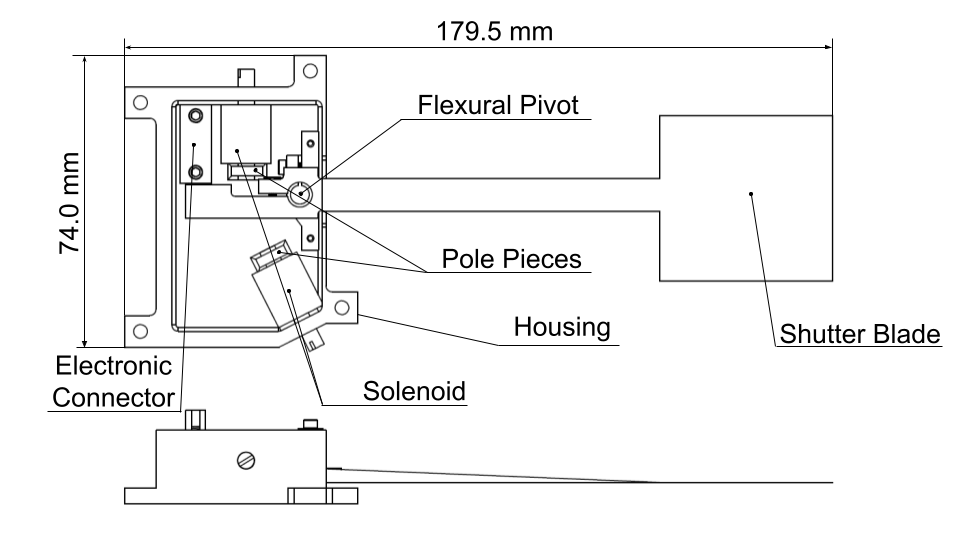} 
    \caption{\label{fig:shutter} Mechanical drawing of the optical shutter assembly. The shutter blade is directly coupled to the main body utilizing a flexural pivot that maintains a constant spring force to swing the shutter to its open or closed positions. Actuation of the shutter is achieved through the use of two solenoid electromagnets, which, when powered, generate a magnetic field that pulls on permanent magnets in the shutter fixture. To reduce stray photon reflections in the closed position, the detector facing side of the blade is blackened with Fineshut KIWAMI, a black polyurethane foam manufactured by Koyo Orient Japan Co., Ltd. }
\end{figure}

An optical shutter assembly is attached to the underside of the telescope support plate and is used in flight to measure the dark current before and during flight. The cold shutter design is based upon a successful shutter in CIBER-1 \citep{Zemcov2013}, but scaled larger in size and weight by approximately a factor of 2 from the CIBER-1 version. The design is illustrated in Fig.~\ref{fig:shutter}.  The shutter consists of an anodized aluminum blade, counterbalanced by a weight and permanent magnet at the back end, that is mounted on a flexural pivot. The shutter housing contains two powered electromagnets, one on either side of the shutter blade, which become polarized when power is applied and attract the weighted end of the shutter blade. The optical shutter can be moved between open and closed positions reliably and repeatedly in a cryogenic environment using the onboard electronics.

\begin{figure*}[th]
    \centering
    \includegraphics[width=1.0\textwidth]{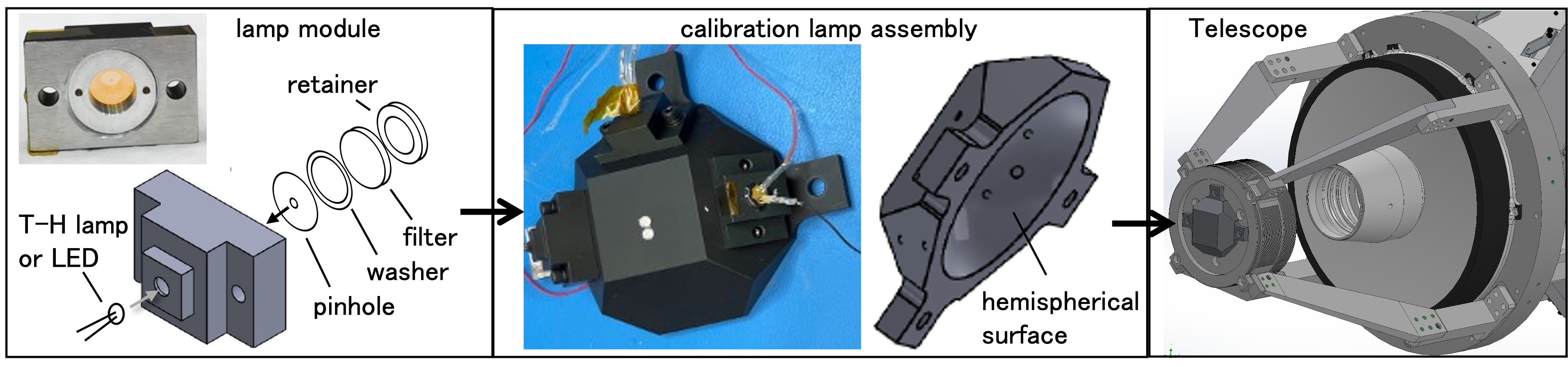} 
    \caption{\label{fig:cal_lamp} CIBER-2 calibration lamp assembly.  (\textit{Left panel}) The lamp modules for arm S, M, and L each consist of an LED or a tungsten-halogen light source coupled to an optical filter and pinhole to attenuate the intensity to the appropriate level. (\textit{Middle panel}) The lamp modules are mounted to the calibration lamp assembly with a hemispherical, reflective, granular inner surface that isotropizes the aperture illumination pattern. (\textit{Right panel}) The calibration lamp assembly is mounted to the back of the telescope's secondary mirror, and when powered illuminates the lens optics through the Cassegrain hole of the telescope.}
\end{figure*}

To check the alignment of the optics during tests and in flight, we employ calibration lamps housed in a small enclosure mounted to the back of the secondary mirror.  The interior of the calibration lamp assembly is  hemispherical to mimic an integrating sphere, which provides uniform illumination of the detector with a uniformity of about 1\%. Light from sources placed at stations around the hemisphere couples to the optical path through four pinholes in the secondary mirror at the image of the Cassegrain hole. The light source is a miniature tungsten-halogen lamp for the arm M and L and a light emitting diode for the arm S. These are housed in small modules with pinholes and optical filters to adjust the light levels and spectra, as illustrated in Fig.~\ref{fig:cal_lamp}.

\subsection{Focal Plane Assemblies}
\label{sS:FPAs}

\begin{figure*}[htb] 
\gridline{\fig{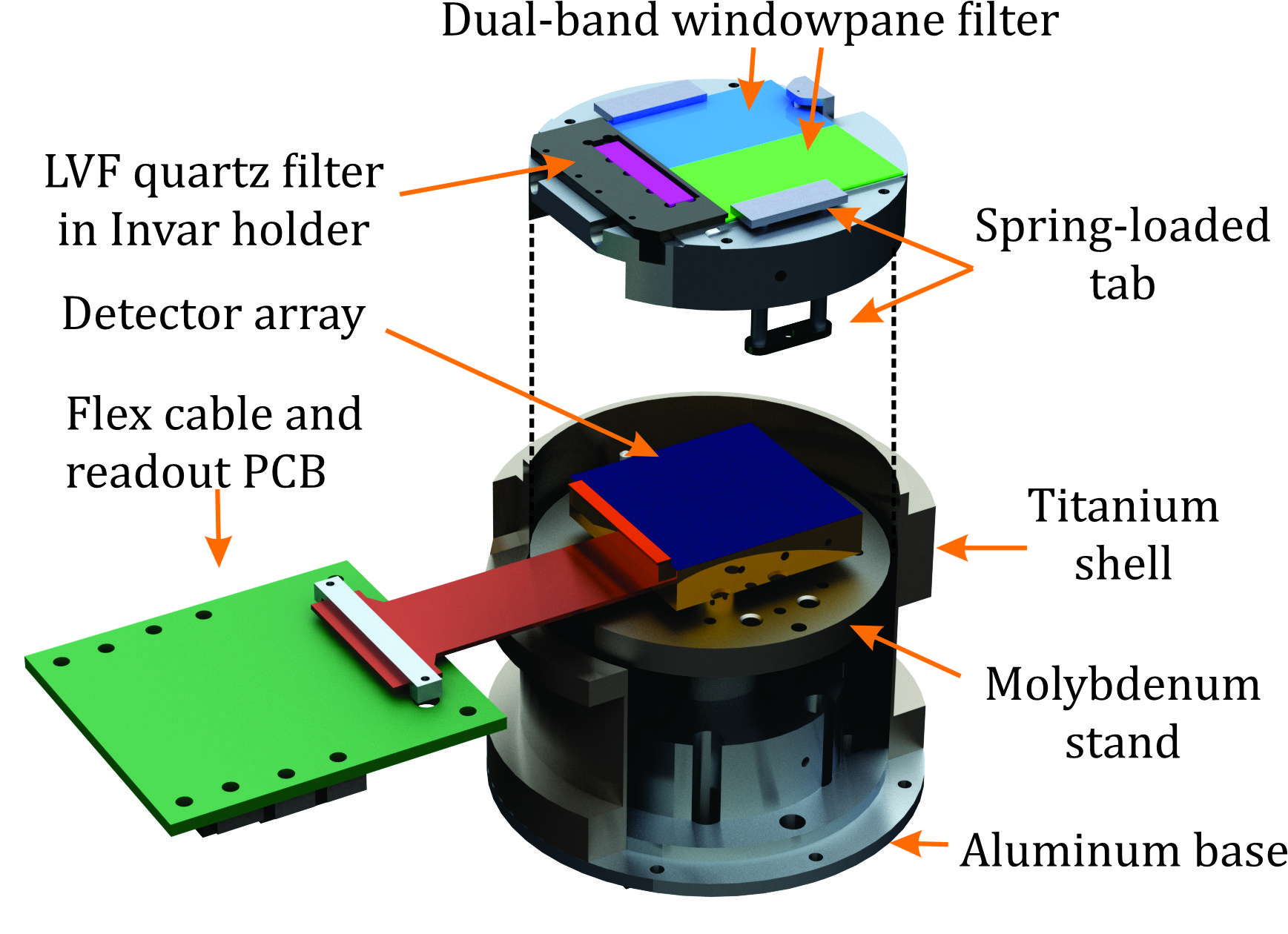}{0.45\textwidth}{(a)}\fig{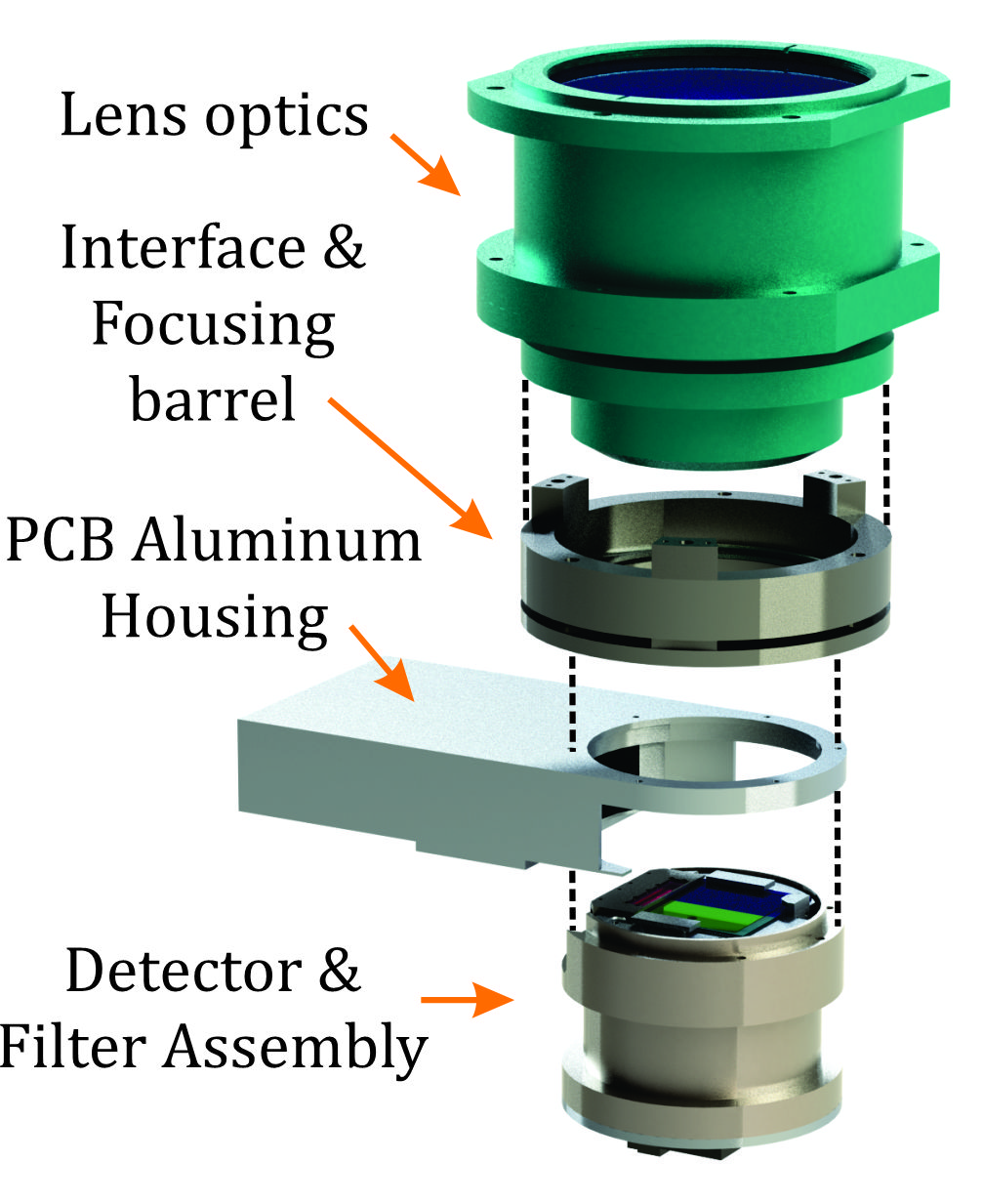}{0.26\textwidth}{(b)}\fig{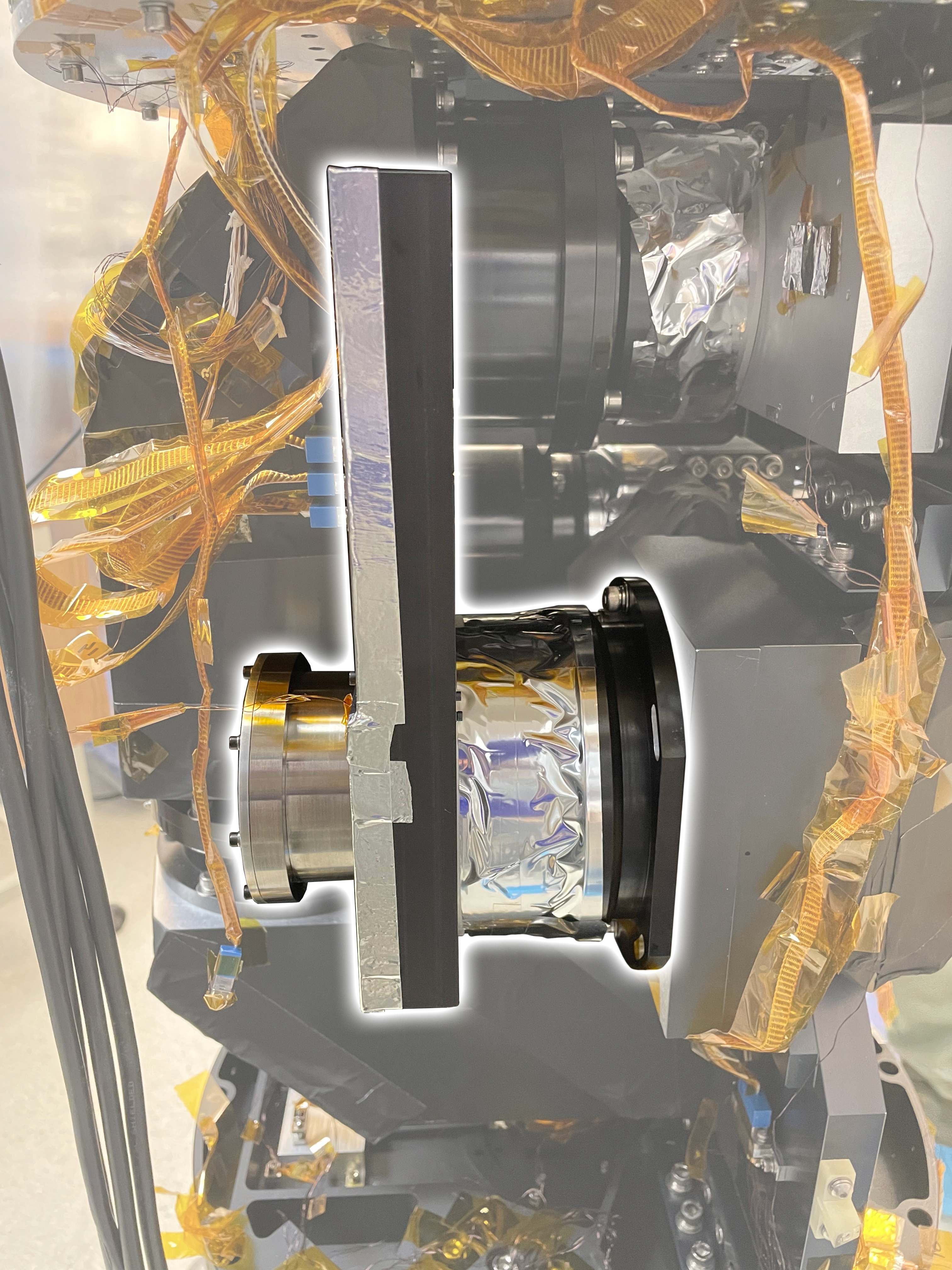}{0.25\textwidth}{(c)}}
\caption{Cutaway renders and in-context image of the focal plane assembly and its interface to the CIBER-2 optics.  The black dashed lines denote surfaces of contact in the exploded views. (a) The dual-band windowpane filter is installed with a pair of spring-loaded tabs going through the aluminum base, while the Invar LVF holder is bolted to this base. Invar was chosen for having similar thermal expansion to the LVF quartz.  (b) The FPA is connected to the lens barrel using an adjustable interfacing barrel. (c) Assembled FPA housing interfaced to lens optics and installed for arm L. \label{fig:FPA}}
\end{figure*}

The light from each optical arm is focused onto a dedicated $2.5 \, \mu$m cutoff Teledyne H2RG array \citep{Beletic2023} housed in an FPA.  Each FPA houses the detector, a filter assembly to provide spectroscopy and multi-wave band imaging capacity, and a readout chain including a printed circuit board (PCB) in a metal enclosure customized to couple to the optics of each arm.  The FPA system is illustrated in Fig.~\ref{fig:FPA}.

The filter assembly includes a dual-band IR-coated quartz windowpane filter manufactured by Materion held above the detector that illuminates $\sim$85\% of the photosensitive area. 
The other 15\% of the area is covered by an LVF manufactured by Academia Sinica glued directly onto a nickel-iron (Invar) holder. 
The windowpane filter and the LVF holder are fastened to an Al housing mounted to the molybdenum stand that also holds the detector.  The filter housing is designed to maintain a temperature-independent 400 $\micron$ distance between the rear surface of the filters and the front surface of the detector to minimize internal reflections that cause optical ghosting.

Each FPA is housed inside a compact metal light-tight enclosure designed to  ensure a safe $<2$ K minute$^{-1}$ cooling rate and mechanical strength to withstand the stress of launch.
The molybdenum detector stand is supported by a thick Al base and mechanically coupled to the optics interface through a thin titanium-alloy shell to control the heat flow.  The detector is electrically coupled to the readout electronics through a flexible cable that connects to the focal plane PCB that provides signal conditioning and interfaces to the main cryogenic wiring harness.  Each focal plane board is housed in a long, thin extension designed with labyrinthine mechanical interfaces to the titanium shell, and its lid to minimize light leaks behind the detector.

The interface between the detector housing and the optics permits the optical focus of each arm to be changed independently. 
To connect an FPA unit to the optics, we use an Al light-tight barrel fastened at three points to the optical housing.
To avoid shifting the FPA-optics alignment during launch, one mounting point is pinned with a stainless steel dowel.
The distance between the last optic and the detector can be adjusted by shimming the interfacing barrel to place each H2RG array at the plane of the best focus.

\subsection{Electronics}
\label{sS:electronics}

\begin{figure*}[th]
    \centering
    \includegraphics[width=\textwidth]{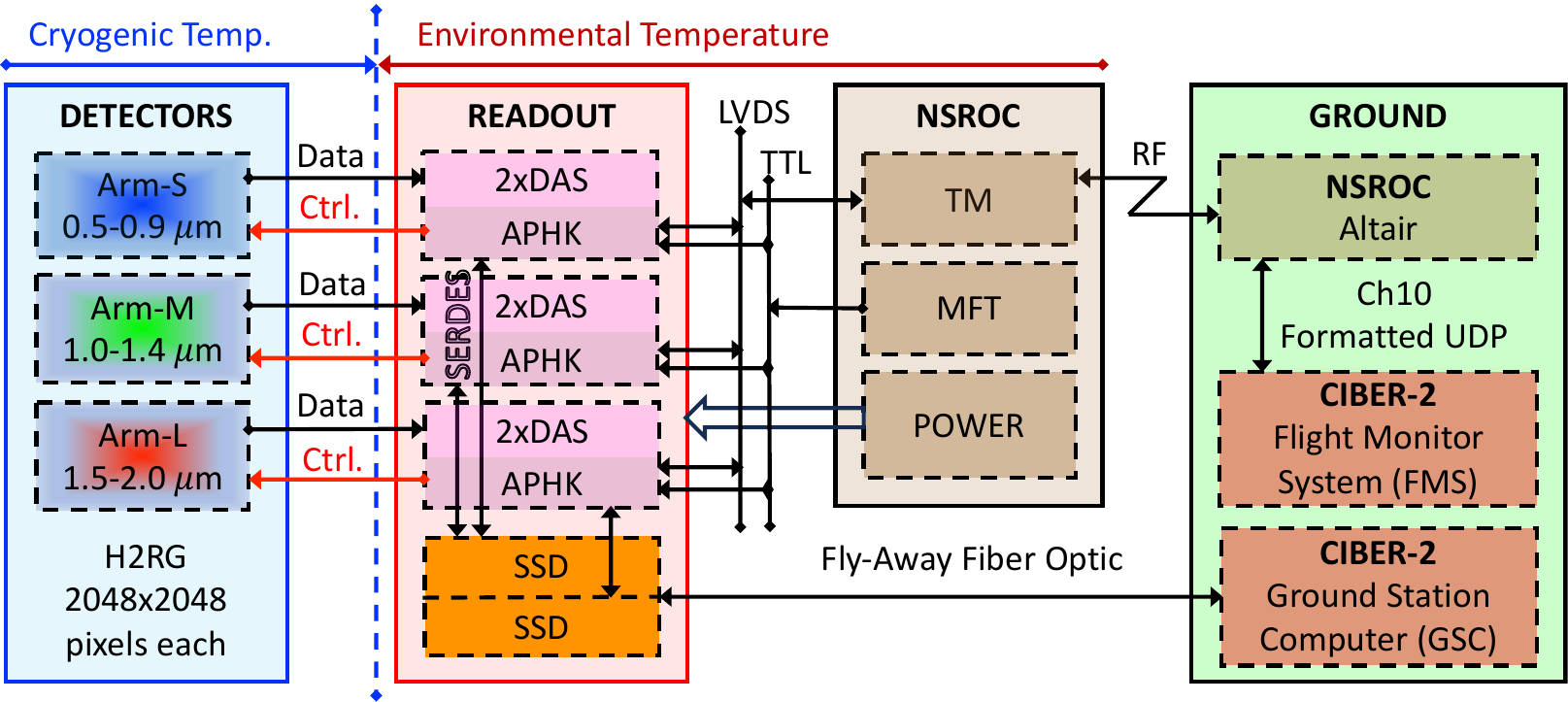} 
    \caption{\label{fig:rocket_elec_block}  High-level block diagram of the CIBER-2 cold and warm electronic systems. 
    The readout is separated into three main units: the cryogenic electronics at left; the ambient-temperature flight electronics in the middle; and the ground station on the right.  The fundamental unit of the CIBER-2 detector electronics are the array processing and housekeeping (APHK) board and dual data acquisition system boards.  This unit is repeated three times, with one servicing a given detector.  These boards address and bias the detector, and then amplify, digitize, and store the data they produce.  The readout units connect to the solid state drive (SSD) onboard data storage boards, which together comprise the ``warm'' (ambient-temperature) electronics enclosed in a red box in the figure.  The warm electronics are coupled to the detectors and other cryogenic electronic systems through ``cold'' (LN2-temperature) harnessing that service the focal plane interface boards (FPBs) and detectors through electrical connectors on the FPAs (see Section \ref{sS:FPAs}).  These components are enclosed in a blue box in the figure.  The rocket systems supply power, timed events and attitude control information, and telemetry, indicated by a brown box in the figure.  Finally, the ground system is used during preflight testing and during flight to monitor the instrument status and store data, and is enclosed in the green box.  The ground station electronics comprise two different systems, one of which handles a single channel of each detector via a telemetered link, and the other parses and stores the full data stream during lab testing and reads from the SSD system following flight.}
\end{figure*}

CIBER-2 uses custom-built electronics that generate the detector clock signals, amplify and digitize the detector outputs, provide a variety of instrument health and status (housekeeping) data, store the data, and provide interfaces to the rocket power, multifunction timer (MFT), and telemetry (TM) systems.  A connectivity diagram showing the relationships between these systems is shown as Fig.~\ref{fig:rocket_elec_block}.  The entire warm assembly is housed in a radio-frequency tight Al box that is mounted to the forward side of the CIBER-2 instrument in a compartment that also houses the CSTARS readout electronics and attitude control system (ACS) star-tracking telescope (to be described in D.~Patru et al.~2025, in preparation). 

\subsubsection{Detector Readout}

For a given H2RG detector array, multiplexer clocking signals are generated by the APHK board and sent to the array through approximately 1.5 m long signal-paired manganin wires.  The output of the detector is connected to the data acquisition system (DAS) boards through a similar harness.  Each array is read out via 32 simultaneous row channels configured in slow readout mode, meaning that each hardware amplification chain must handle a $64 \times 2048$ strip of pixels. By convention, we consider the 32 independent channels to be arrayed along the horizontal direction, so the electronic ``fast'' address direction is horizontal.  The pixels are sampled at approximately 100kHz, leading to a full frame acquisition time of 1.35s.  The signals are amplified and digitized with 16 bit resolution, and then passed to the APHK board that packetizes the output for transmission to both the onboard storage and rocket TM system.

Each H2RG includes a rim of four reference pixels that are not sensitive to light surrounding the $2040 \times 2040$ optically sensitive pixels.  This arrangement results in four reference pixel rows at the start and end of each channel, as well as four additional reference columns in the first and last readout channels. These reference pixels can be used to correct for channel-level offsets produced by the low-frequency components of system $1/f$ noise \citep{Mosely2010,Rauscher2017}. We implement nonsequential row readout, or row-chopping, to reduce spatial noise at the large scales relevant to the instrument's scientific objectives. Instead of the standard method of reading each row in each channel in their physical order, a predefined skip parameter $S$ controls the number of rows the readout system counts over between each read. When the image is rearranged back to its original row positions, readout noise is modulated resulting in some noise power being shifted to smaller scales less relevant to CIB anisotropy measurement. Here, a skip parameter of $S=25$ rows was chosen for all three arrays to optimize noise reduction on $\theta>5'$ scales.  A full description of the method is provided in \citet{Heaton2023}.

\subsubsection{Housekeeping Boards}

The field programmable gate array (FPGA) APHK boards generating the clocking signals described above also perform six housekeeping functions:  (1) monitoring the status of various independently wired hardware subsystems including the rocket door and the pop-up baffle, (2) monitoring and controlling the status of hardware subsystems within the telescope, including the cold shutter and calibration lamp, (3) responding to flight events by driving detector array resets, (4) temperature monitoring of components in the experiment section, (5) temperature monitoring and control of the FPAs, and (6) data stream generation.

The MFT box housed with the rocket systems tracks the timing of the flight and at the appropriate times supplies the signals that instruct the APHK FPGA to change the state of subsystems or institute array resets.  It also controls the activation of the pop-up baffle and rocket door, the status of which the APHK includes in the data stream.  
The cold shutter and calibration lamp driver circuits are carried over from CIBER-1 and are controlled by the APHK FPGA.  These systems are described in detail in \citet{Zemcov2013}.  The pop-up baffle system is housed in a small box separate from the main electronics box, and is powered independently from the readout electronics to prevent noise injection or detector bias variations when the motor is active.  On receiving the command to actuate from the MFT, a small electronic brake is released, the pop-up baffle deployment stepping motor is actuated for 4.0 s to deploy the baffle, and the brake is reset.  On receiving the command to retract the baffle, the process is reversed but uses springs to actively retract the baffle.

The H2RG arrays are charge-integrating detectors, meaning that the voltage across the detector diodes must be reset to flush the accumulated charge and begin a new exposure.  At launch, the arrays are reset, and the clock timing is synchronized across all three arrays.  This synchronizing reset is called a T1 reset and takes multiple seconds to accomplish, corresponding to multiple frame reads of the detector.  During flight, each change in the subsystem status triggers independent resets on each of the three detectors.  These resets are fast and only encompass a single reset frame in sequence with the existing clocking, maximizing exposure time.  The collection of frames during the charge integration between resets corresponds to a single exposure and the slope-fit of these frames can be calibrated to the observed flux.  While each observation in flight corresponds to a single exposure, exposures between observations enable system diagnostics. 
Thus, the APHK board responds to signals such as the rocket door opening and closing, shutter open or close commands, and calibration lamp enable or disable commands, and asserts array resets after these events to start a new exposure.  The array is also reset in response to the ACS ``on target'' signal that indicates the ACS should be settled to within $1^{\prime \prime}$ of the target position.  This starts the observation exposure.  The status of the reset and the bits indicating subsystem status are recorded in the data stream and can be used as a cross-check of our on-sky measurements.

The CIBER-2 housekeeping system supports up to six channels of thermometry for each channel, up to 18 thermometers in total. DT-670 silicon diode sensors situated throughout the vacuum section of the instrument are used as the sensing device. As in CIBER-1, the diodes are biased at a constant current of $10\,\mu$A with the diode voltage logged once per frame in the housekeeping data. These voltage measurements are later converted to temperatures using standard calibration curves.

Lastly, the APHK board firmware receives and assembles array data from the DAS boards, transmitting to the solid state drive (SSD) boards via a serializer--deserializer (SERDES) cable.  With 18 bits of resolution, the theoretical dynamic range of the H2RG voltage output is covered with a 1.6~e$^{-}$ least significant bit, several factors less than the read noise of the detector.  Data are packaged into packets containing a single sample of eight pixels for 4 of 32 outputs along with housekeeping information including data type, array address information, subsystem status bits, and thermometry data.  In order to monitor the array status in real time, one channel per array along with housekeeping information requiring a minimum of 4.9 Mbits s$^{-1}$ (Mbps) is sent from the APHK board to the TM via low-voltage differential signaling to be transmitted to the flight monitoring system (FMS) during flight.

\subsubsection{Onboard Storage}
\label{ssS:obs}

The full-sample data stream generated by the APHK board is passed to the onboard storage system using an SERDES interface.  Each storage (SSD) board houses duplicate data storage systems so that the data for two detector arrays are handled by one SSD board.  The SSD boards are connected to a control computer through multimoded fiber optics that provide data acquisition mode changes via an uplink and data streaming via a downlink to the data viewing computer.  The CIBER-2 SSD system is described in full detail in \cite{Park2018}.  This SSD system was successfully flown three times, and no hardware faults due to launch or exposure to space were experienced.

\subsubsection{Rocket Interfaces}
\label{ssS:Rocket_Interfaces}

The rocket systems provide instrument power in addition to TM, attitude control, and timing of events through the MFT.  The instrument power is provided by a $\pm 14\,$V battery supply on board the rocket, with the capability to switch to an external supply during testing operations.  The TM link serves to inform the system of the flight status and transmit data to the FMS on the ground via a dedicated 10 Mbps radio link housed in the rocket payload \citep{SRPH2023}.  
Because the standard 10 Mbps telemetry link provided by the rocket is significantly less than the full data rate of 150 Mbps, we transmit only one channel per detector to assess the data quality in flight.

\subsubsection{Ground Station Support}

When powered, the onboard electronics are designed to continually clock the detectors and sample their output, sending the digitized data stream either directly to the GSC or to onboard storage (Section \ref{ssS:obs}), according to the data collection mode set by the GSC through custom GSE software, hardware, and fiber optic connections.  The FMS and GSC systems comprise the ground station support of the CIBER-2 electronics systems whose flight configuration is illustrated in Fig.~\ref{fig:rocket_elec_block}.

In lab testing mode, we have direct control over the length of time during which the data stream is collected.  Device resets and cold shutter hardware control are communicated independently, enabling many different types of tests to be performed, including those described in Section \ref{sec:lab_char}.  Immediately following data acquisition, the digitized data stream is received, parsed, decoded into raw image FITS files, and stored by custom software on the GSC, bypassing the onboard storage system.  During the decoding process, frames are displayed in their respective channels, and housekeeping flags are toggled, allowing the user to monitor the decoding in real time.  A log reporting statistics and any errors is saved for later reference.

Prior to flight, the onboard storage system is enabled by the GSC, starting flight mode.  In this mode, all data are stored onboard and can be later retrieved and decoded by the GSC.  To ensure this storage process is properly initiated, the system sends back the first 10 frames of data through the fiber optics to the GSE software and then terminates remote communication with the GSC while it continues to record the data on the SSD boards until the storage is full, the stop command is sent (such as during testing), or the electronics are turned off.  Meanwhile, the FMS receives data showing a single channel from each detector along with a selection of housekeeping data through the NSROC TM link (Section \ref{ssS:Rocket_Interfaces}).  These are used to allow real-time diagnostics of the instrument performance during flight.

\section{Laboratory Characterization}
\label{sec:lab_char}

\subsection{Dark Current and Noise}

To evaluate the instrument's noise performance, we collected dark data both on the ground and in flight. Ground dark exposures are acquired with the cold optical shutter in the closed position.  Before each exposure, all three arrays are reset simultaneously to minimize cross talk in the cold harness.  In-flight dark performance is verified using data from the portions of the arrays covered by the LVF holders, as well as an exposure taken near the end of the flight with the optical shutter closed.

A slight misalignment in the position of the cold shutter results in a small light leak around the edge of the shutter in the closed position.  Thermal emission from the instrument vacuum door is the likely source of the light, so the arm S and M filters effectively block it.  However, it is pronounced in the longest wavelength detector. 
To reduce the impact of the leak, we exclude the affected region when measuring the characteristics of the long wavelength detector, which removes $<$10$\%$ of the total pixel count in arm L.  We also observe leaks in the LVF filters, which are due to their poorer out-of-band rejection performance.  This emission is not present in flight images as the door is out of the optical path.  Even in the presence of the light leakage, all arrays demonstrate a dark current well below the expected photon noise from Zodiacal light and airglow.

\begin{figure*}[htb]
   \gridline{\fig{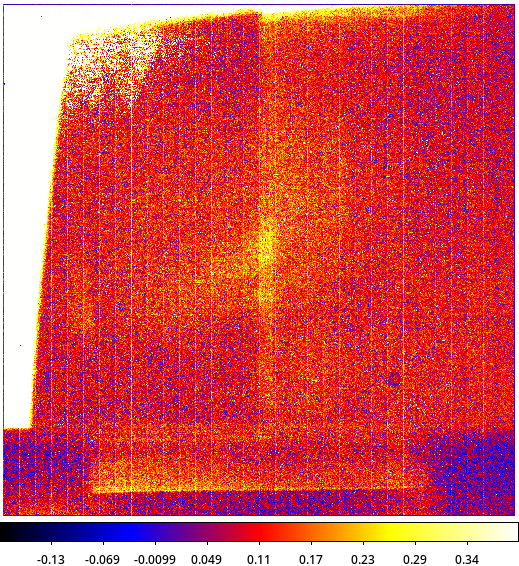}{0.28\textwidth}{(a)}\fig{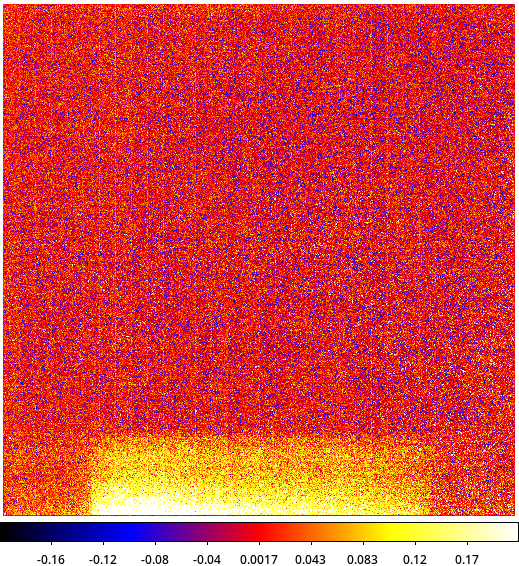}{0.28\textwidth}{(b)}\fig{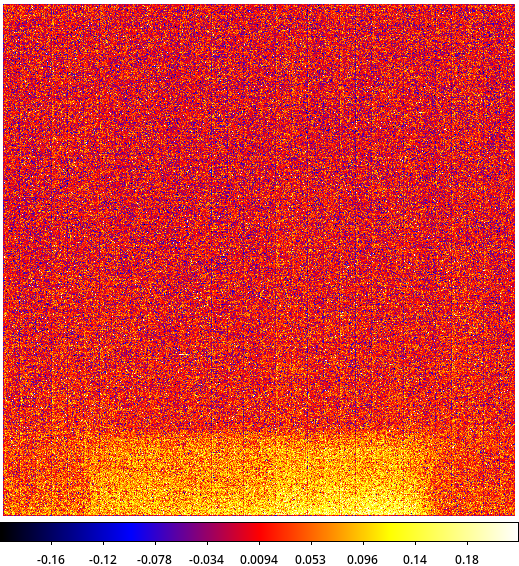}{0.28\textwidth}{(c)}}
    \caption{ \label{fig:darkcurimg}  The median dark current maps of arms (a) L (b) M and (c) S from data taken with the cold optical shutter closed while the vehicle was horizontal on the launcher.  
    The light leak around the closed optical shutter can be seen in all three arm S, producing a noticeable signal in the rectangular LVF region along the bottom of arm S (b) M and (c) S. Oriented differently, the brightest corner of the light leak illuminates the upper left border of (a) arm L with a peak illumination around 1.15~e$^{-}$ s$^{-1}$ and a noticeable shadow under the LVF holder.  The scale of the color gradient goes from -0.2 to +0.4~e$^{-}$ s$^{-1}$ for (a) and from -0.2 to +0.2~e$^{-}$ s$^{-1}$ for (b) and (c).}
\end{figure*}

\begin{figure*}[htb]
\begin{center}
   \gridline{\fig{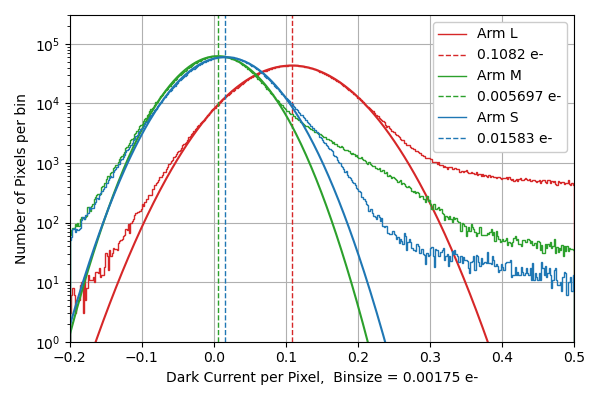}{0.46\textwidth}{(a)}\fig{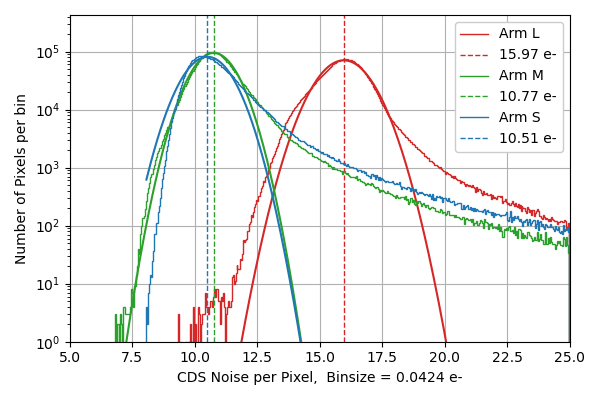}{0.46\textwidth}{(b)}}
   \caption{ \label{fig:CDSnoisehist} (a) Dark current and (b) CDS noise histograms of arm L (red), M (green), and S (blue).  
   A Gaussian fit is overplotted on the data.
   The mean of each fitted Gaussian (dashed vertical line) is reported as the ``peak'' dark current and noise, respectively in Table~\ref{table:CDS_noise}. 
   The spatial distribution of the dark current and noise are shown in Figs.~\ref{fig:darkcurimg} and~\ref{fig:CDSnoiseimg}, respectively.  
 }
\end{center}
\end{figure*}

Ground dark data on the launcher were taken in several different power configurations: with the star tracker on/off, with the payload running on internal/external power, as well as in various TM transmission states. 
Aside from a small number of exposures with excess cross talk noise, the data from all configurations are statistically consistent and can be combined into one large dataset.
The median dark current map shown as Fig.~\ref{fig:darkcurimg} exhibits minor light leaks ($\sim 1$e$^{-}$ s$^{-1}$) along the edge of arm L windowpane filter and the LVFs in arms M and S.
Excluding these specific regions, the dark current of individual pixels in each array is well described by a Gaussian distribution whose mean is reported in Table~\ref{table:CDS_noise}.  
The visible shadow of the LVF holder on arm L along the bottom of panel (a) in Fig.~\ref{fig:darkcurimg} indicates scattered longer wavelength light reaching the detector even with the cold shutter closed.  
Considering only the pixels in the shadowed region of arm L gives a dark current measurement that is consistent with arm M and S.  Histograms of the dark current estimates are shown as Fig.~\ref{fig:CDSnoisehist}.

\begin{figure*}[htb]
   \gridline{\fig{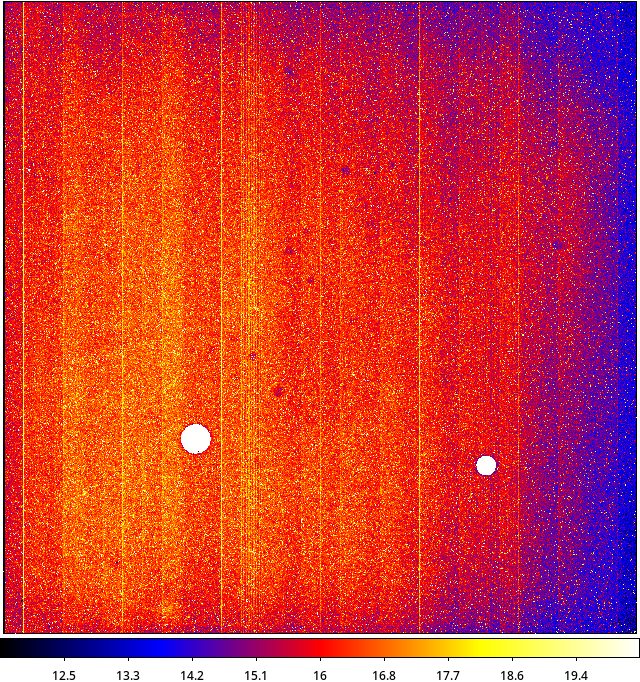}{0.3\textwidth}{(a)}\fig{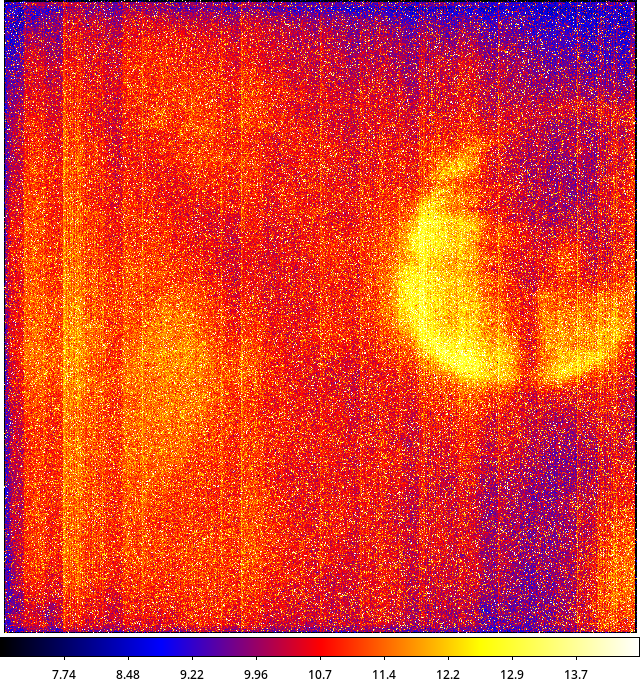}{0.3\textwidth}{(b)}\fig{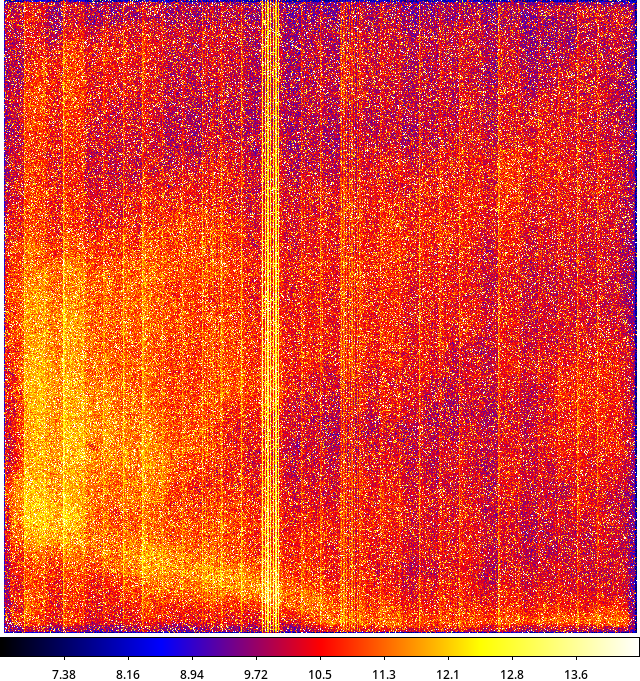}{0.3\textwidth}{(c)}}
    \caption{ \label{fig:CDSnoiseimg} The CDS noise maps of arm (a) L (b) M and (c) S in $e^-$.  
    Multiple spatial features are visible such as the extra noise in the first column of individual amplifier channel (most prevalent in arm L), the ``watermark'' feature in arm M, and the noisy amplifier channel 13 in Arm S.  These features are stable independent of illumination, and detector specific, so can be removed in a fluctuation analysis. 
 }
\end{figure*}

Correlated double-sampling (CDS) noise can be characterized from the same preflight dark data set.
For every pixel, we take the pairwise differences in adjacent readout frames, divide by $\sqrt{2}$, then take the standard deviation of the resulting values to be the pixel CDS.
Excluding the nonresponsive pixels, the CDS noise of remaining pixels follows a Gaussian distribution whose mean is reported as the array CDS.
We find a number of spatial features in the noise maps including excess noise in the first column of each amplifier channel (most prevalent on arm L), a circular ``watermark'' feature in arm M, and a noisy amplifier channel 13 in arm S.  Representative CDS images are shown as Fig.~\ref{fig:CDSnoiseimg} and their corresponding histograms are shown in Fig.~\ref{fig:CDSnoisehist}.
These features are stable and can be removed in a fluctuation analysis.

\begin{table}[ht]
\centering
\caption{Measured dark current and CDS noise.  
To avoid the contamination from the scattered light in arm L, we also report measurements using only the pixels shadowed by the LVF holder, labeled as ``sub'' in parentheses.
\label{table:CDS_noise}} 
\begin{tabular}{l|ccc}
Arm & L (Sub) & M & S \\
 \hline
 Peak Dark Current (e$^{-}$ s$^{-1}$) & 0.11 (0.03) & 0.002 & 0.009 \\
 \hline
 Peak Noise (e$^{-}$) & 16.0 (14.2) & 10.8 & 10.5 \\
\end{tabular}
\end{table}

To avoid contamination from hot pixels and nonresponsive pixels in flight data, we make appropriate masks.  Pixels with a dark current greater than five standard deviations away from the median dark current are masked as hot pixels.  The measured well-depth of pixels in WPF regions of saturated flat field measurements is plotted in a histogram, and pixels with a well-depth lower than three standard deviations below the median pixel response are masked as weakly or nonresponsive pixels.
Due to the large photocurrent gradient in the LVF channels, low-response pixels are difficult to identify at this stage of processing and only the hot pixels are identified and masked in those regions.  Nearest neighbors of masked pixels are also masked to avoid electrical cross talk.

\subsection{Optical Testing Configuration}
\label{sS:optical_testing}
  
\begin{figure}[h]
   \includegraphics[width=0.5\textwidth, trim=2.0cm 0cm 0cm 0cm]{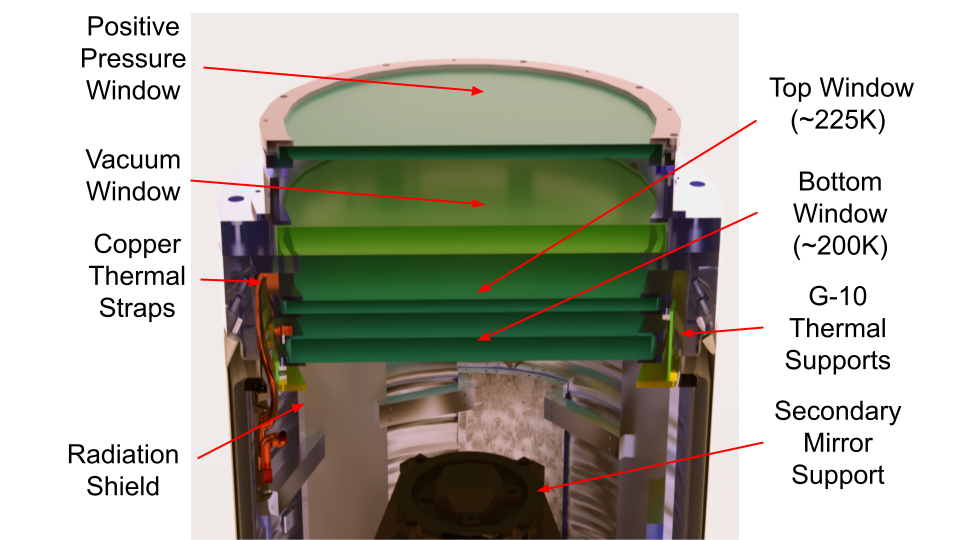}
    \caption{ \label{fig:thermal_windows} The thermal window assembly for optical testing, replacing the shutter door as the vacuum seal.
    During optic testing, GN$_{2}$ is circulated through the space between the vacuum and positive pressure windows to prevent water vapor in the air from condensing on the vacuum window.
    In addition, two cold filter windows (top/bottom) simulate the same thermal load as the shutter door on the optics while also filtering out the 300 K infrared emission background from the lab.
    The filter windows are maintained at $\sim$225 K and $\sim$200 K by copper heat straps mounted to the radiation shield, while G10 standoffs isolate them thermally from the cold optics. 
    All windows are AR coated with measured $>95\%$ transmission in CIBER-2 wavelength range.
 }
\end{figure}

To enable optical testing, we replace the shutter door with a staged assembly of fused-silicate quartz windows to filter the 300~K thermal background from the lab to the level that would occur in the flight configuration (Figure \ref{fig:thermal_windows}). 
A window bulkhead (called the ``vacuum window'') takes the place of the shutter door to seal the payload.
To avoid water condensation on this window bulkhead, a second window is mounted on top of the bulkhead with thermally isolating teflon standoffs creating a compartment into which gaseous nitrogen is circulated (``positive pressure window'').
Inside the payload, two cold windows held at 225~K (``top window'') and 200 K (``bottom window'') serve as filters intercepting the radiation transmitted and emitted by the bulkhead window.
The cold windows are supported mechanically by the optical baffle but thermally isolated from the optics by G10 legs.
To cool the windows independently from the optics, three copper heat straps connect each window to the radiation shield to form a direct thermal path to the LN2 tank.
All windows have the same diameter of 340~mm.
The vacuum and bottom cold windows have the thickness $t=9.5$ mm, while the external window and top cold filter have $t=2$ mm.
To reduce internal reflections that can produce optical ghosting, all windows are AR coated.
The window assembly allows $> 95$\% transmittance over CIBER-2 wavelength range and induces a 9 K temperature gradient between the primary and secondary mirrors, which matches the thermomechanical configuration expected in flight.

\subsection{Detector Gain and Signal Calibrations}
\label{sS:detcalibrations}

To calculate the ADU to electron gain factor in each detector, we use ground exposures with moderate photocurrents far from saturation.
The gain can be extracted by fitting a slope to the measured variance of the photocurrents in predefined bins versus the theoretical photon noise slope variance (\citealt{Garnett1993}, \citealt{Nguyen2021}).  
The measured gains in units of microvolts per electron are measured to be $g_{\rm S} = 3.56$, $g_{\rm M} = 4.53$, and $g_{\rm L} = 4.24$ in the three arms.

As signal accumulates on a detector diode, the diode capacitance changes, which leads to a nonlinear charge accumulation at larger photon flux. 
To measure the nonlinearity of the detector response, we couple a broadband light source to an integrating sphere to produce a uniform diffuse illumination pattern that is then imaged by CIBER-2 \citep{Nguyen2021}.   
We find the average pixel's nonlinear behavior causes the measured signal to deviate from the true signal by $\sim 1\%$ around 7.5\% of saturation, causing pixels with $i \gtrsim 150$ e$^-$ s$^{-1}$ to be subject to significant nonlinear effects.  This is approximately an order of magnitude larger than the photocurrent from diffuse astrophysical sources expected in flight.  

\subsection{Focus}
\label{sS:focus}

To position the detectors at the best focus of the CIBER-2 telescope, we use a 2 \micron\ pinhole illuminating an external 12'' diameter collimator to simulate a point source at infinity following the procedure established for CIBER-1 \citep{Bock2013, Takimoto2020, Nguyen2021}.
The focus of the collimator at infinity is defined as the zero position of the pinhole.
To measure any shift from the best focus, we scan the pinhole along the optical axis of the collimator using a commercial motorized mount.
To determine the distance between the CIBER-2 detectors and the best focus of the imaging optics, $\Delta F _{\rm C2}$, we determine the shift $\Delta F_{\rm col}$ from the zero-point along the optical axis that minimizes the pinhole image.
The two quantities are related by
\begin{equation}
\frac{\Delta F_\textrm{C2}}{\Delta F_\textrm{col}} \approx \left(\frac{f_\textrm{C2}}{f_\textrm{col}}\right)^2
\end{equation}
where
$f_\textrm{C2}$ = 930 mm is the focal length of CIBER-2 telescope,
and
$f_\textrm{col}$ = 865.9 mm is focal length of the collimating telescope.  

Before and after each focus scan, we use a flat mirror and autocollimating microscope to calibrate the encoder coordinate of the focus at infinity of the collimator.
After the zero-point is calibrated, we steer the collimator to produce a point source in the FOV of CIBER-2.
The pinhole is then scanned over a range of at least $\pm$1000~$\mu$m of the best focus position.
For each detector, we sample the focus shift on each side of the dual-band windowpane filter, on the LVF, and on the corners of the detector.

\begin{figure*}[th]
    \centering
    \includegraphics[width=0.9\textwidth]{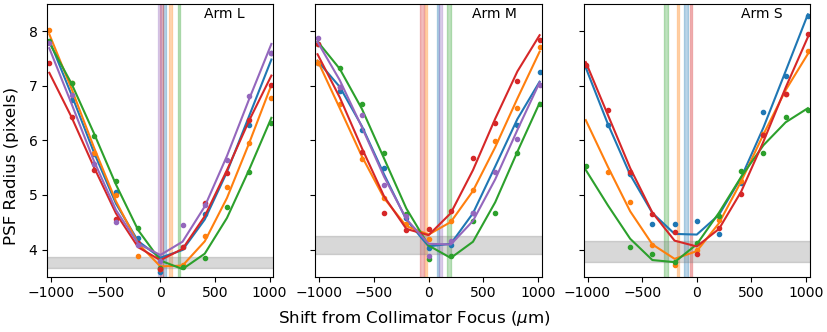} 
    \caption{\label{fig:focus_LMS} The best focus measurement of each arm for multiple positions across the array. The data show the radius of the aperture $R$ enclosing 95\% of the signal detected by our focusing algorithm plotted as a function of shift from the collimator focus, with the best focus at the minimum of the fitted inverted Gaussian curve. (For these data, the radius enclosing $95\%$ of the encircled energy is tracked with the focusing algorithm instead of the typical $85\%$ due to the changing shape of the PSF.)
    The focus position at launch was at the collimator focus, illustrated by zero on the plot.  The vertical shaded bars illustrate the best focus of that measurement, with the width corresponding to the standard deviation of that parameter of the fit.  The horizontal gray shaded region illustrates the mean PSF radius of the best focus measurements with the width corresponding to the standard deviation of those values.  At launch, the detectors were at a focus position that was best to minimize the PSF across all array measurement positions.
    }
\end{figure*}

To reduce the focus shift from the data, we have developed an algorithm \citep{Nguyen2021} that uses a measurement of the encircled energy to isolate and measure the effective radius of the pinhole image at each scan position.  This process is similar to that detailed later in Section~\ref{sec:optical_performance}.  
A model is then fit to the radius of the spot size as a function of the encoder position to identify the position of the best focus.
Although CIBER-1's focus shift is well modeled by a parabola \citep{Bock2013}, we find that an inverted Gaussian curve is a better fit to the shape of CIBER-2 focus shift (Figure \ref{fig:focus_LMS}).  
 
To measure the tilt of the FPAs relative to the optics interface, we measure the focus shift at different positions across the surface of the detector.
The tilt measurements also enable us to identify any optical aberrations from the variations in the shapes of the pinhole images.
Using the results of the cold focus tests, we shim the secondary mirror for large adjustments ($> 500~\mu$m) common to all three detectors, and the optics-FPA interface for finer adjustments and detector-independent positioning.  
The cold focus is then remeasured, and the shimming process is repeated if necessary.  
In Figure \ref{fig:focus_LMS}, we present the final measurements for all three arms prior to the third launch.  
Optical aberrations contribute to a varying PSF shape at different focus positions and locations across the array, particularly in arms S and M, resulting in a noticeable scatter in the measured minimum PSF widths as seen in Figure \ref{fig:focus_LMS}.  
As a result, we evaluate the effectiveness of the shimming based on how well we can minimize the PSF size across all measurement locations across the detector array.

Depending on the vacuum conditions of the cryostat, the temperature of the total optical system can vary by 10 K from one optical test to another conducted in different locations at RIT, WFF, and White Sands Missile Range (WSMR). However, the telescope is designed with athermal optics, so defocus only occurs with a change in temperature gradient. The defocus coefficient against the temperature gradient change is measured to be $\sim$10 $\mu$m K$^{-1}$. Since the temperature gradient is stable within 5 K among the tests, and the temperature affects focus by up to 50 $\mu$m, which is smaller than the focus adjustment accuracy of $\sim$200 $\mu$m, considering collimator stability and the uncertainty in focus determination due to aberration, the temperature gradient has little impact on focus. The temperatures in the final preflight focus test and during the flight observation agree with each other within 3 K, and the thermal stability is sufficient to retain focus.

\subsection{Spectral Response and Photometric Calibration}
\label{sec:abscal_lab}

To measure the spectral response of each pixel in the detector array, we use a monochromatic source coupled to an integrating sphere that illuminates the aperture of the CIBER-2 telescope.  The calibration apparatus is based on that described in \citet{Nguyen2021} and references therein; here, we briefly summarize.  The light source is a standard laboratory-grade tungsten-halogen lamp that illuminates an optical fiber.  The fiber then illuminates the entrance slit of a Newport MS257 1/4 m monochromator.  
The light from the monochromator's exit slit is then fiber coupled to a 30 cm output aperture integrating sphere whose output port fully illuminates the telescope's primary mirror.  To collect the spectral response data, we set the monochromator grating angle to pass a known wavelength range and take an exposure, and repeat for the entire CIBER-2 spectral range.  The step sizes between the scanned wavelengths vary between 10 and 40 nm depending on the spectral resolution of each arm.
The same data set is used to calibrate the windowpane and the LVF regions.
In the case of the LVF, in addition to the spectral response, this calibration also yields the LVF wavelength to detector pixel mapping. From these measurements, we obtain the spectral response of the LVF channel shown in Figure~\ref{fig:filters} and confirmed the design $R=21{-}22$ in all channels, which corresponds to a width of $80{-}200$ pixels in the detector array, depending on wavelength.

To measure the EBL absolute brightness and accurately subtract the ZL foreground, we require precise knowledge  of the absolute photometric calibration of the LVF channels.  A calibration accuracy better than 3\% is required to constrain the near-IR EBL excess to better than $\delta {\lambda}I(\lambda) \lesssim 20 $ \nw\ reported by \citet{Matsuura2017}.  
The photometric brightness calibration requirement for the WPF regions is less stringent, with 10\% calibration accuracy typical in background fluctuation measurements.

Although the final absolute photometric calibration is measured from photometry of stellar standards in the flight data (see Section \ref{sec:inflight_calibration}), we perform preflight calibrations as a consistency check.
To collect absolute calibration data, we couple a tungsten-halogen white light source to the beam-filling integrating sphere. 
We then measure the absolute irradiance at the aperture of the integrating sphere with different calibrated multichannel fiber spectrometers (manufactured by Frame and NIRQuest+, Ocean Insight Inc.) to provide absolute photometric references. 
In theory, we can simply compare the response between CIBER-2 FPA and the calibrated spectrometers to reduce the absolute response of CIBER-2 to a known flux.
In practice, this measurement is complex and has a variety of sources of uncertainty related to the spectral performance of the calibration systems, measurement geometry, and transfer efficiency that need to be quantified.  The preliminary analysis is consistent with the in-flight calibration from stars, but 
additional details on the calibration apparatus, procedure, and data analysis will be presented in a future publication.

\section{Flights}
\label{S:flights}

CIBER-2 has flown three times: a successful engineering flight in 2021; a launch followed by a vehicle failure in 2023; and a successful science mission in 2024. 

\subsection{36.281} For CIBER-2's first flight, we opted to observe several fields in order to obtain sufficient information to assess the foregrounds and cross-check the instrument's performance against existing CIBER-1 data in the same fields.  CIBER-2 was successfully launched for the first time as mission 36.281 from WSMR at 00:25 MDT on 2021 June 7 and reached an apogee of approximately 311 km. All seven celestial targets were successfully acquired, and the pointing control was well within requirements. All events were received correctly, and all deployable systems actuated as intended. The payload was successfully recovered with its data storage devices intact and functional. 

\subsection{36.383} CIBER-2’s second launch took place from WSMR at 23:03 MDT on 2023 April 16. Due to a failure in the range radar system, the flight termination system was actuated 32 s into the flight, and the vehicle was lost. The payload landed $\sim $5 miles downrange of the launcher with a full load of cryogens and the shutter door open, which was significant thermal shock to the system. Fortunately, we were able to refurbish and repair the system in time to make another launch attempt a year later.

\subsection{36.396} For CIBER-2's third flight, we optimized the field selection for intensity mapping measurements, specifically concentrating on the COSMOS and Lockman Hole regions.  CIBER-2's third launch from WSMR occurred at 21:32 MDT on 2024 May 5. The payload reached an apogee of approximately 316 km following a nominal motor burn.  We received all telemetered data on the ground, all flight events were received as planned, and the instrument responded correctly to all inputs before reentry.  The payload was recovered the next morning, and 100\% of the data was recovered successfully from the onboard data recorders.  Due to the upcoming launch of SPHEREx \citep{Crill2020}, no further flights of the CIBER-2 payload are currently planned.

\section{Preliminary Flight Performance}

In this section, we present a characterization of the in-flight performance of the CIBER-2 instrument, focusing on validating the instrument parameters that appear in Table \ref{fig:payload} using the 36.396 flight data.

\subsection{Detector Performance}

The simplest quantifiable characteristics of the electrical performance of the detector arrays are their dark current, noise performance, and image persistence.
To verify dark current performance in flight, we compare the histogram of each pixel's median dark current computed from nine flight-like ground exposures to a histogram of currents from the dark exposure taken at the end of the flight.  Since the in-flight dark exposure was only eight array reads long, the ground exposure data sets are truncated to match in this comparison.  

A visual inspection of the flight dark images with the truncated dark data sets indicates an increased shutter leak in flight due to the shutter not completely engaging in its baffled slot.  Laboratory measurements showing variations of this shutter leak over time make us suspect a temperature-dependent
curling of the optical shutter paddle is present. 
This is likely caused by mechanical stresses due to the different thermal contraction of the metal blade, plastic KIWAMI tape, and acrylic glue. 
Persistence current from bright stars in the previous exposure is also clearly visible in the dark image.  
The dark current $I_{\rm d}$ and CDS noise histograms of the dark data excluding the high-current tails from these effects were fit to a Gaussian distribution,
resulting in the best-fitting parameters given in 
Table~\ref{tab:flight_CDSnoise}.  The noise in these measurements is larger than that shown in Table~\ref{table:CDS_noise} due to the shorter dark exposure in flight, but is statistically identical to ground measurements of the same duration.  With the exception of arm M, the mean flight dark current is statistically higher than ground measurements.  This is likely due to thermal radiation from the hot rocket skin leaking around the light baffling on the affected arm S, and mitigation strategies in data analysis are currently being developed.

\begin{table*}[ht]
\centering
\caption{ Measured dark current and CDS noise from truncated eight frame ground data compared to flight.  Variation in the ground data captured by $\pm$ refers to the standard deviation over repeated measurements. \label{tab:flight_CDSnoise}}
\begin{tabular}{| l | c | c | c | c | c | c |}
\cline{2-7} 
\multicolumn{1}{l|}{}  & \multicolumn{2}{c|}{Arm L} & \multicolumn{2}{c|}{Arm M} & \multicolumn{2}{c|}{Arm S} \\
\cline{2-7}
\multicolumn{1}{l|}{}  & Ground & Flight & Ground & Flight & Ground & Flight \\
\hline
$I_{\rm d}$ (e$^{-}$ s$^{-1}$) & 0.17 $\pm$0.09 & 0.50 & -0.01 $\pm$0.10 & 0.00 & -0.07 $\pm$0.12 & 0.34 \\ 
CDS Noise (e$^{-}$) & 20.6 $\pm$0.1 & 20.9 & 15.3 $\pm$0.7 & 13.9 & 14.3 $\pm$0.1 & 13.9 \\ 
\hline
\end{tabular}
\end{table*}

The dark data from flight 36.396 exhibit a variable cross talk pattern that is not present in ground datasets. This cross talk was caused by an anomalous additional T1-style reset $\sim77$ s into flight on arm M only, desynchronizing arm M's clock and resulting in a stripe pattern on the long and medium wavelength arrays that is normally prevented by a synchronized T1 reset at launch.  Flight data for the short wavelength detector are consistent with ground datasets. The cross talk likely arises from radio-frequency pickup in the cryogenic harness.  Work is in progress to mitigate the impact of this stripe pattern on our science data.

Saturating sources in each detector clearly result in image persistence above the noise current for at least two subsequent observations after illumination, most persisting even into the dark measurement taken after the cold shutter is closed. As described in \citet{Smith2008}, the persistence current in Teledyne HgCdTe detector arrays cannot be eliminated by resetting the detector diode and is therefore expected to contaminate subsequent exposures.  As such, this systematic behavior must be managed in data analysis.  The long-lived persistence current resulting from the pixel saturation in CIBER-2 data is predicted by and follows the model in \cite{Fazar2025} with CIBER-2 detector-specific parameters.  All other nonsaturating sources manifest detectable persistence for at least one subsequent observation after illumination.  Given the initial magnitude of the persistence current and the observed decay, the model for H2RG image persistence suggests that for CIBER-2 the effect should be negligible after two exposures.  Considering this information, between $0.1$\% and $0.4 \%$ of the pixels are affected in any given exposure following the first.

\subsection{Star Catalog Generation}
\label{sec:catalog}

\begin{figure}
    \centering
    \includegraphics[width=0.47\textwidth]{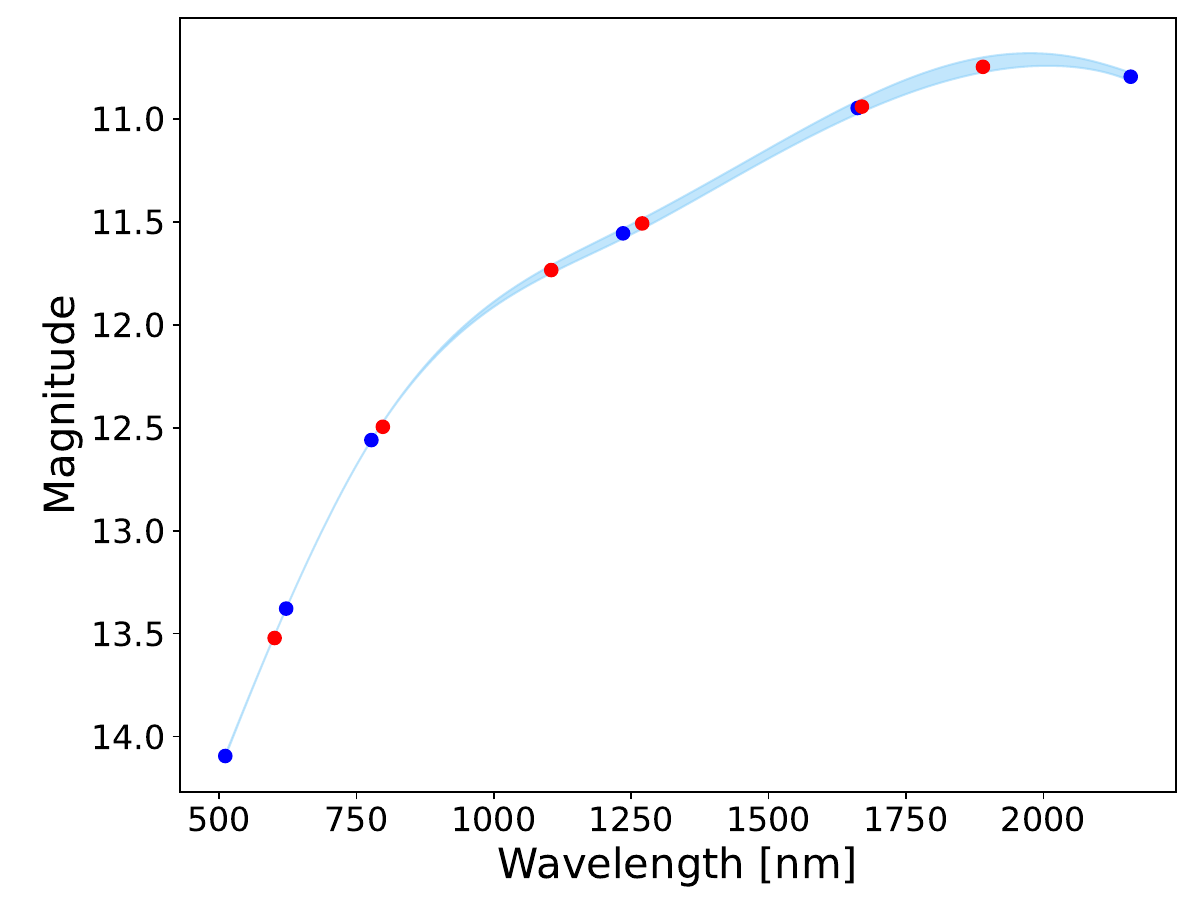}
    \caption{Catalog magnitudes of a star in the Gaia DR3 and 2MASS bands (blue) and the derived magnitudes in the CIBER-2 bands (red). The uncertainties in the CIBER-2 magnitudes were estimated using a Monte Carlo simulation, where catalog photometric errors were propagated through the interpolation. The width of the light blue band indicates the 68\% population of interpolations derived from the Monte Carlo simulation.}
    \label{fig:star_catalog}
\end{figure}

Before proceeding with the flight data, we need to develop a reliable catalog of source positions and magnitudes to enable source masking, PSF stacking, and absolute calibration.  We utilize the Gaia DR3 photometric catalogs covering $G_{\rm BP}$, $G$, and $G_{\rm RP}$ bands at visible wavelengths \citep{Gaia2023} and the Two Micron All Sky Survey (2MASS) catalogs covering $J$, $H$, and $K$ bands at near-IR wavelengths \citep{Skrutskie2006}.   We estimate the magnitudes of stars in the CIBER-2 bands, $m_{\rm C}$, by interpolating the fiducial catalog magnitude using cubic interpolation into a $d \lambda = 1 \,$nm wide grid, $m_{\lambda}$, and then compute 
\begin{equation}
m_{\rm C} = \frac{\int m_{\lambda} T_{\lambda} d \lambda}{\int T_{\lambda} d \lambda},
\end{equation}
where $T_{\lambda}$ is the transmittance profile of each CIBER-2 band.  Figure \ref{fig:star_catalog} shows the catalog magnitudes for one of the stars in the observed fields, along with the derived magnitudes in the CIBER-2 band. 

To assess the uncertainties in these magnitude estimates, we performed a Monte Carlo simulation. For each star, we generate 100 realizations of the fiducial catalog magnitudes in the six bands by sampling from a Gaussian distribution, where the mean is the catalog magnitude, and the standard deviation is the associated photometric error. For each realization, magnitudes in the CIBER-2 bands are computed using the same interpolation and weighting process. The final magnitudes are taken as the mean of these realizations, and the corresponding uncertainties are the standard deviations over the realizations.  These catalogs are called the CIBER-2 Star Catalogs (C2SC) below.

\begin{figure*}[htb]
    \centering
    \includegraphics[width=1\textwidth]{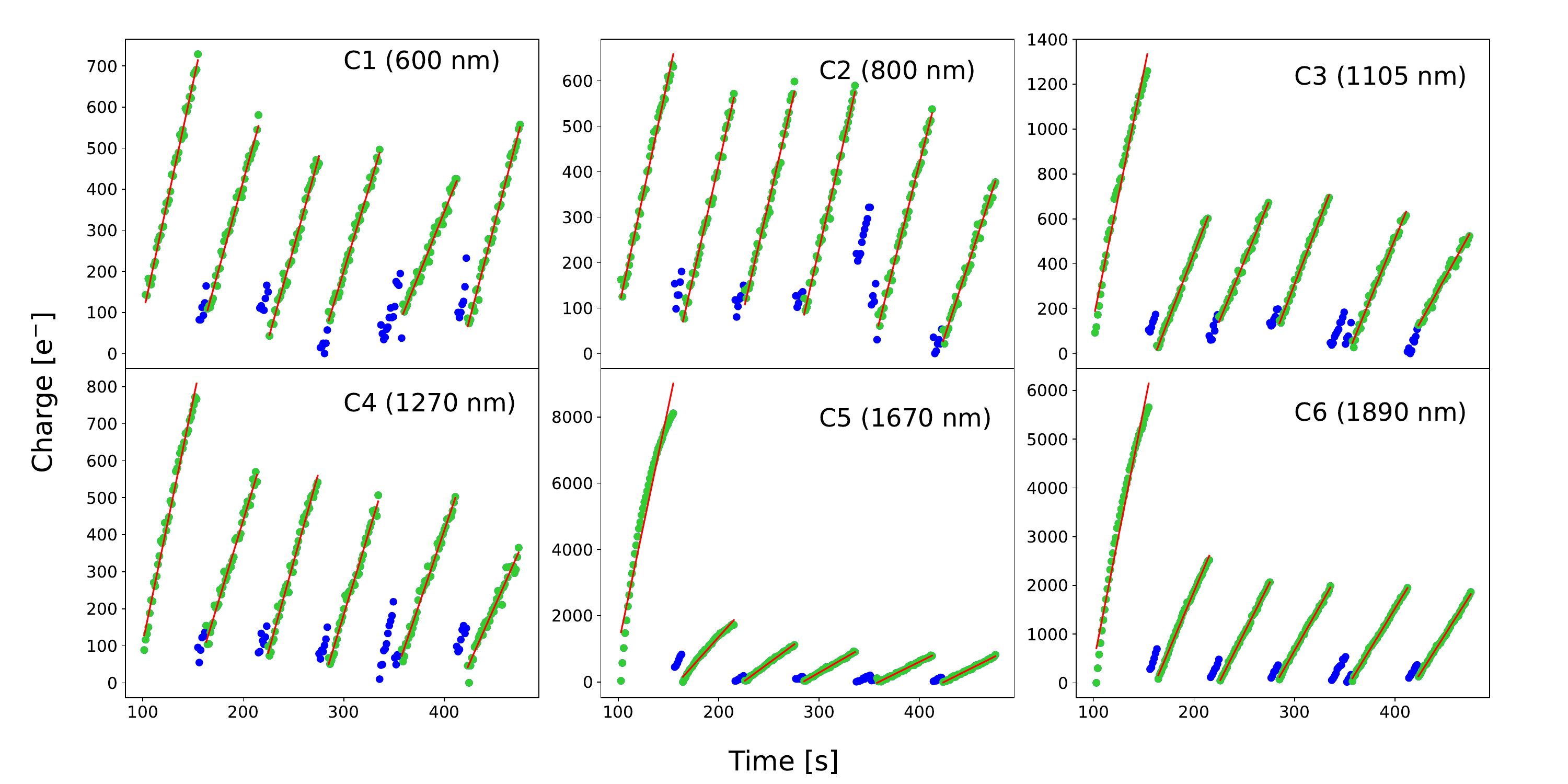} 
    \caption{\label{fig:flightphotocurrent}  Charge accumulation in a single pixel for each band. CIBER-2 observes each target for $\sim 50$ s (green points), then maneuvers to the next target (blue points). The measured photocurrent for each field is typically reported as the slope of the linear fit to the data (red line), although it is clear that at earlier times the photo flux is changing, likely because of airglow contamination. The detectors are reset before and after each maneuver.}
\end{figure*}

\subsection{Photocurrent during Flight}

Detector photocurrent is proportional to incident photon flux and, as such, is the fundamental observable in flight. Figure \ref{fig:flightphotocurrent} illustrates the charge accumulation on a normal pixel during our observations. Under constant illumination, the slope of the line should be constant, so the data for each exposure are typically fit to a linear model to provide an optimal estimate of the photocurrent measured by the pixel \citep{Garnett1993}. Integrating charge with nonlinear behavior is indicative of a time-dependent flux, often from a variety of forms of contamination and airglow or geocoronal emission \citep{Matsuura1994,Zemcov2014,Matsuura2017}.

\begin{figure*}[htb]
    \centering
    \includegraphics[width=1\textwidth]{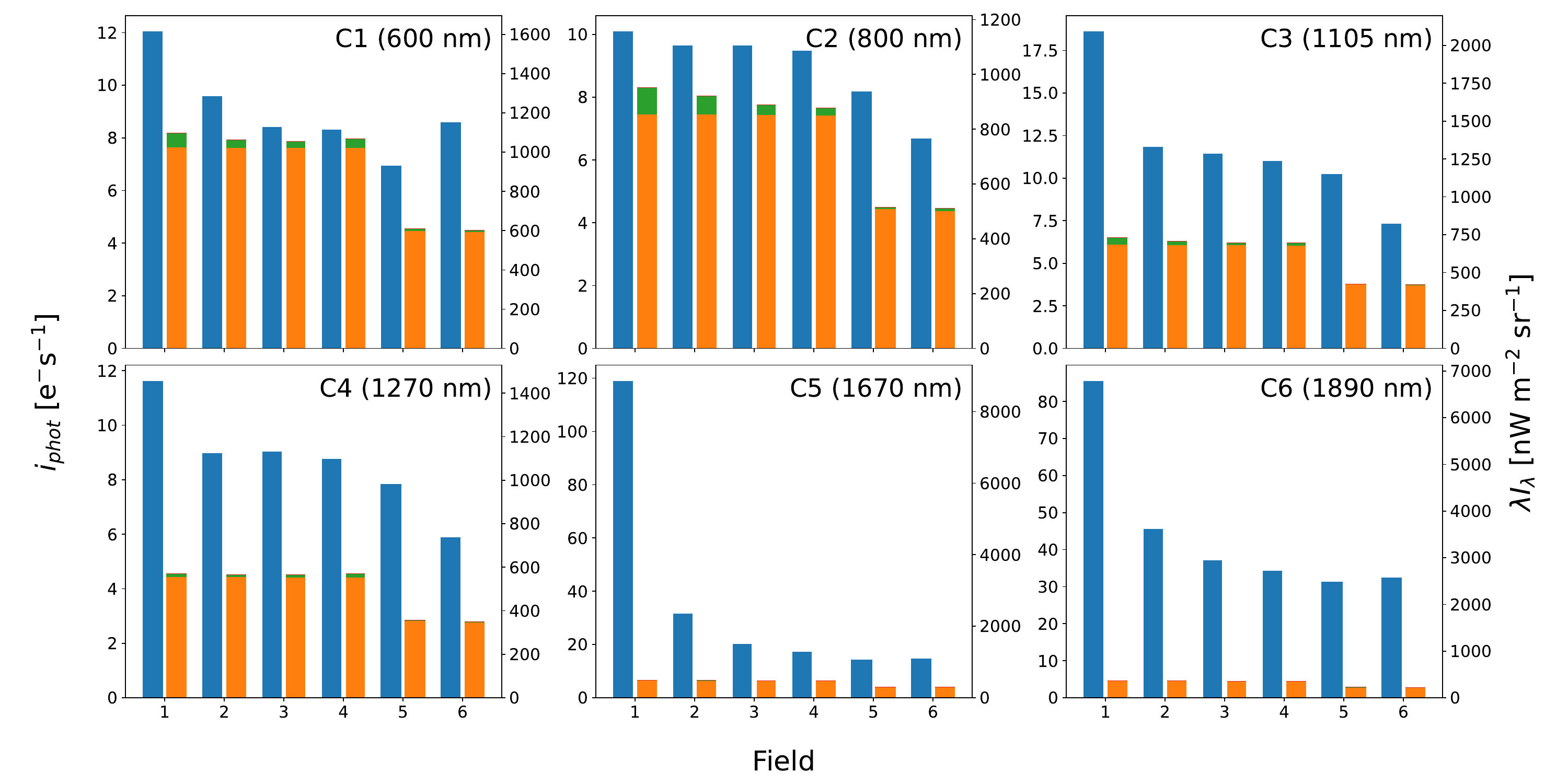} 
    \caption{\label{fig:photocurrent_barchart} Measured photocurrent in each field for each band. The measured photocurrent (blue) is compared with the expected contributions from ZL (orange), DGL (green), and ISL (red).}
\end{figure*}

To estimate the mean photocurrent in each exposure image, we first mask bad pixels and bright stars. Stars brighter than $\{20, 20, 18, 18, 17, 17\}$ Vega mag in the C2SC bands C1--C6 are masked, with the mask size scaling linearly with the star's magnitude. Following masking, a 3$\sigma$ clip is applied to exclude outlier pixels, and the mean photocurrent over the entire WPF image is determined by averaging the remaining values. We present the average photocurrent in each band and field in Fig.~\ref{fig:photocurrent_barchart}.

To compare against the expected sky brightness, we model the intensity from the combination of ZL, DGL, and the residual integrated starlight (ISL) after masking as follows. Both the ZL and DGL sky brightness estimates are calculated using the SPHEREx sky model, which is described in \citet{Crill2020}.  To summarize, the SPHEREx ZL model incorporates new reflectance and thermal emission measurements \citep{Tsumura2010, Planck2013, Matsumoto2015, Kawara2017} into the existing Kelsall ZL model \citep{Kelsall1998}, and is similar to the \textit{JWST} ZL model \citep{Rigby2023}. The DGL model uses the spectral model of \citet{Zubko2004} and the spatial model derived from \textit{Planck} data \citep{Planck2016b} to estimate the near-IR brightness from the scattered interstellar radiation field. Residual ISL is estimated by simulating CIBER-2 images using the C2SC and applying the same astronomical source mask we apply to the real data, and then averaging the remaining pixels. We find that the ISL contributes $1{-}3$ nW m$^{-2}$ sr$^{-1}$ to the total sky brightness. Figure \ref{fig:photocurrent_barchart} compares the photocurrent of the sky brightness in flight with the expected astrophysical contributions. The conversion factor obtained from the in-flight absolute calibration (see Section \ref{sec:inflight_calibration}) is used to determine the relationship between photocurrent and intensity for each band.

\begin{figure}[th]
    \centering
    \includegraphics[width=0.47\textwidth]{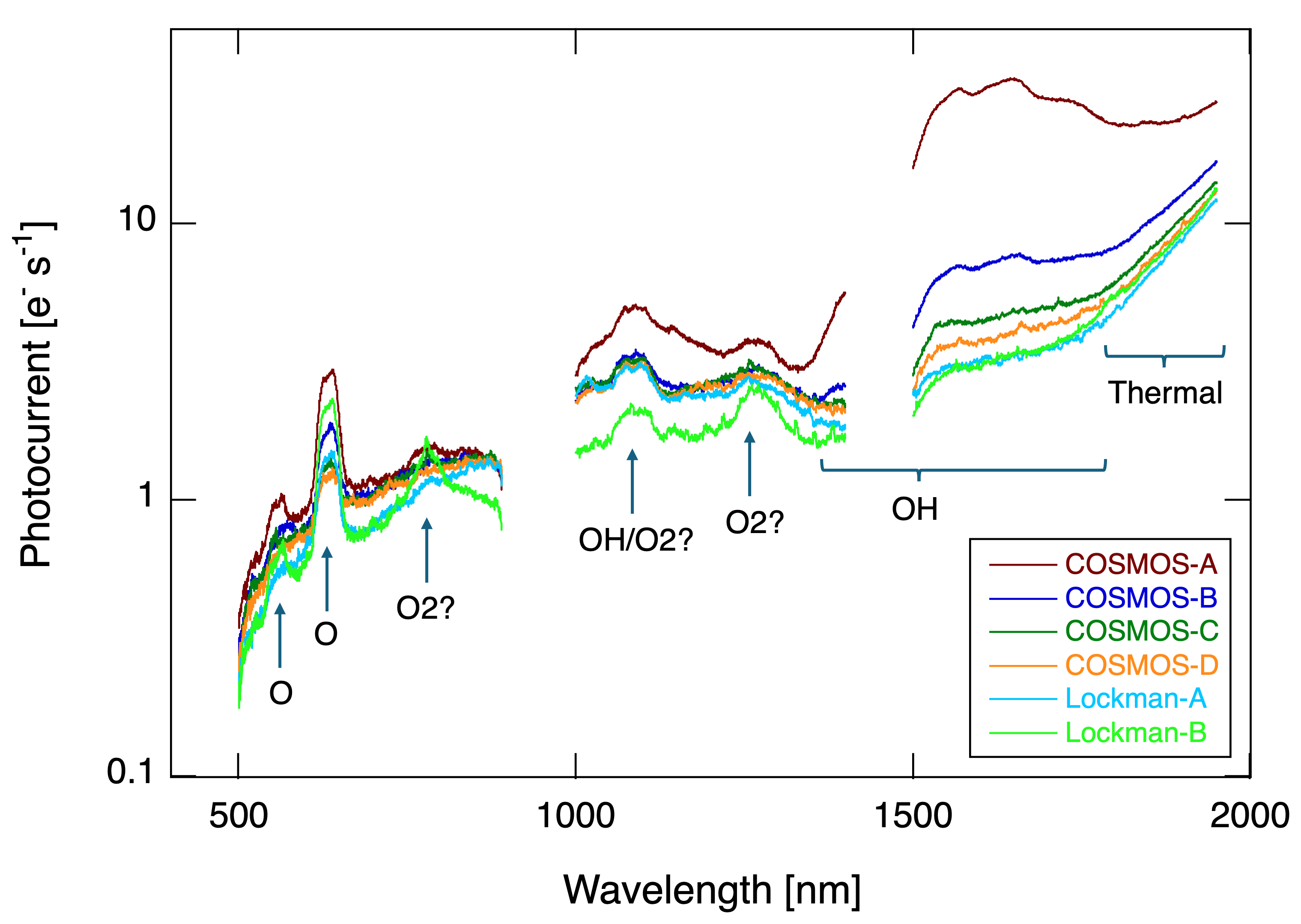} 
    \caption{\label{fig:lvf_photocurrent} The measured photocurrent spectra in the LVF for all six fields. The prominent lines in the visible are due to atomic oxygen in the upper atmosphere at altitudes higher than the payload, and possibly molecular oxygen at lower altitudes. At near-infrared wavelengths, there are broad features due to hydroxyl lines with strong time dependence, as seen in the CIBER-1 flights, and stable features possibly due to molecular oxygen. At the long wavelength end, thermal emission probably from the warm part of the rocket skin contributes significantly.}
\end{figure}

In addition to the broadband channels, we compute the measured photocurrent spectra in the LVF for the six fields, which are displayed in Fig.~\ref{fig:lvf_photocurrent}. The spectra are obtained by computing the median of 200 column pixels in the spatial direction within the LVF aperture and after applying a median filter with width 40 row pixels in the wavelength gradient direction. The average dark current of each detector during flight is estimated by computing the median of the masked pixels located on both sides of the illuminated LVF area in the array and subtracted as a constant value from the raw photocurrent in each field. 

A comparison of Figs.~\ref{fig:photocurrent_barchart} and \ref{fig:lvf_photocurrent} leads to insights into the diffuse sky brightness measured by CIBER-2. Most notable are the differences between the measured and expected brightness early in the flight. These are likely to be caused by residual airglow due to material either brought up with the payload or in the atmospheric exosphere, and at long wavelengths scattered emission from the drag-heated skin \citep{Zemcov2014}. In particular, Fig.~\ref{fig:lvf_photocurrent} shows a large thermal component longward of $1.7 \, \mu$m that decreases over the flight. The entire spectrum is brighter than our preflight expectations, with prominent spectral features from various simple molecules observed early and late in the flight, which is likely due to emission from airglow and outgassed material from the payload.  We also observe an increase in the photocurrent compared with expectations in the C1 band, which we assign to geocoronal emission from oxygen (O) that is known to have a scale height of $\sim300$ km, which is clearly visible in Fig.~\ref{fig:lvf_photocurrent}.  These various increased backgrounds will complicate our data analysis, but we have developed various methods to account for them in the past \citep{Zemcov2014, Matsuura2017}.  Residual uncertainties can be accounted as systematic errors in future analyses. 

\begin{figure*}[th]
    \centering
    \includegraphics[width=0.59\textwidth]{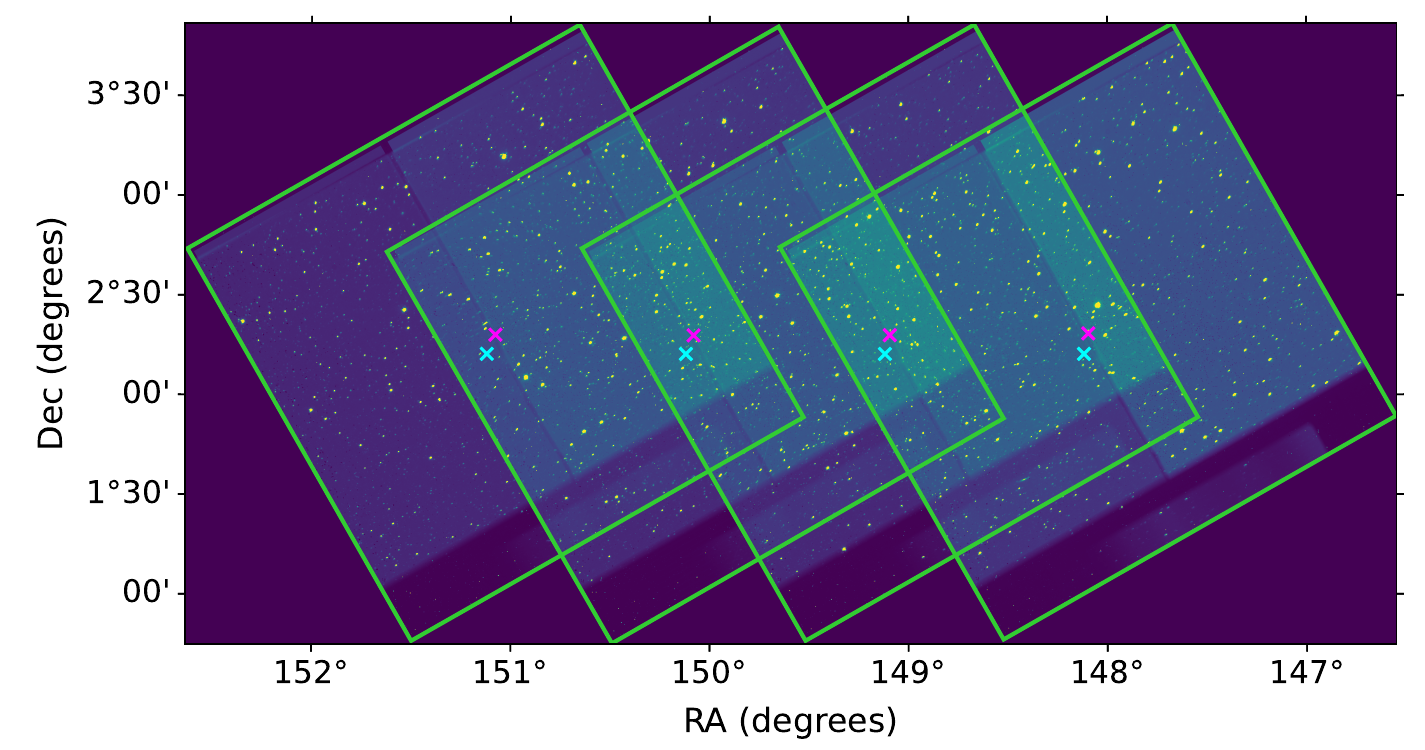} 
    \includegraphics[width=0.4\textwidth]{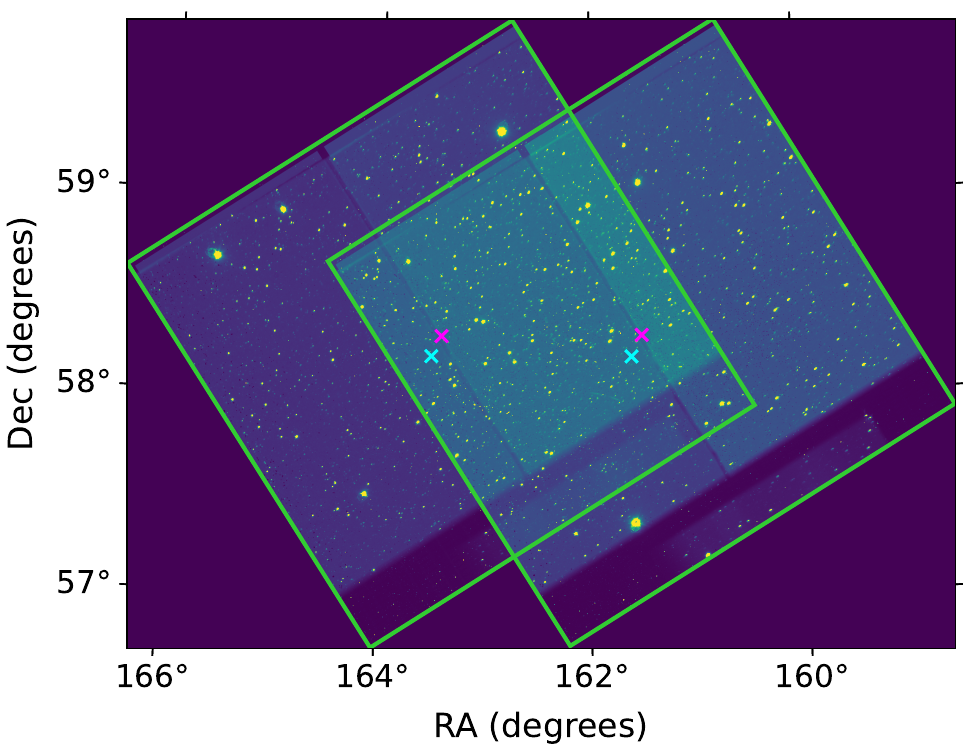} 
    \caption{\label{fig:absolute_pointing} Arm M images of the COSMOS field (left panel) comprising fields 1 to 4 (from right to left), and the Lockman field (right panel) fields 5 and 6 (from right to left). Magenta crosses indicate the central pixel coordinates, and cyan crosses represent the requested coordinates. The difference between the requested and actual pointing is within $8^{\prime}$ for all fields.}
\end{figure*}

\subsection{Astrometric Performance}
\label{sec:astrometric_performance}

We perform astrometric registration for each arm and field using \textsc{astrometry.net} \citep{Lang2010}. The central pixel coordinates of each field were compared to the requested coordinates to assess the absolute pointing accuracy of the experiment. Additionally, the central coordinates were compared across different arms to account for alignment offsets between arms.

In Figure \ref{fig:absolute_pointing} we present registered arm M flight images for the six fields. The difference between the field center coordinates and the requested coordinates increases progressively from Field 1 to Field 6 in all arms, which is likely due to the angle between the side-looking star tracker’s boresight and the CIBER-2 experiment’s boresight changing slowly as the rocket skin cools following the ascent leg \citep{Zemcov2013}.  The absolute astrometric registration difference remains within $8^{\prime}$ over the flight.  

\begin{figure}[th]
    \centering
    \includegraphics[width=0.15\textwidth]{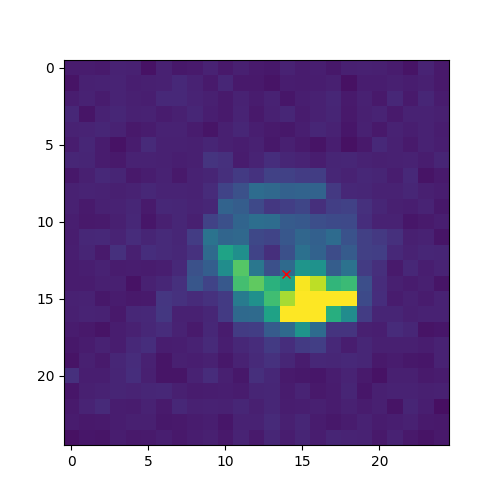} 
    \includegraphics[width=0.15\textwidth]{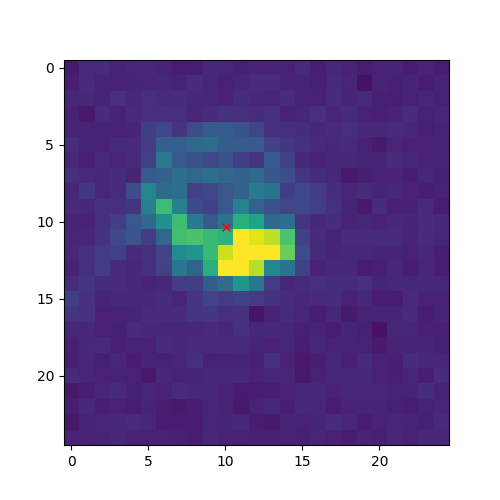}
    \includegraphics[width=0.15\textwidth]{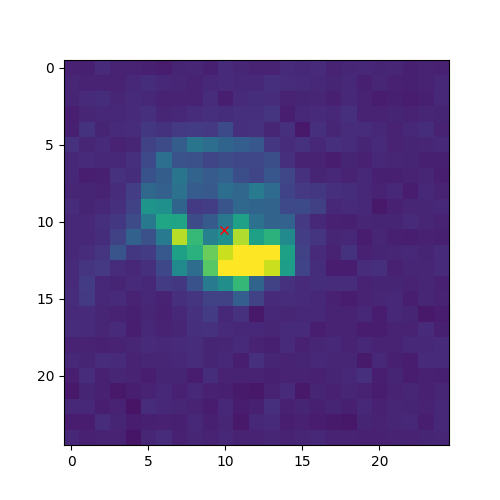}
    \caption{\label{fig:centroid_track} Bright star centroid images from the C1 band, field 5 single-frame differences. The centroid positions (red crosses) are measured from frame-difference images from the beginning (frame 1 $-$ frame 0; left panel), middle (frame 15 $-$ frame 14; middle panel), and end (frame 21 $-$ frame 20; right panel) of the exposure.  The point source wanders by about a beamwidth over the course of an integration due to jitter in the rocket attitude control.}
\end{figure}

The pointing stability of the experiment over an exposure affects the observed size of point sources.  
The pointing drift of the images is quantified by measuring the centroid position of a star in frame-difference images. To determine the centroid, we first subtract the background of the frame-difference image by masking sources using the C2SC and calculating the median of the unmasked pixels. We then extract a star cutout based on the astrometric solution and compute its centroid using the center of mass method. By tracking the position of a bright source in subsequent array reads, we can measure the pointing of the instrument on the $1.35$ s frame-read time scale.  To achieve sufficient SNR without saturating pixels by the end of exposure, bright stars with $m_{\rm vega} \sim 9$ were selected.  Figure \ref{fig:centroid_track} illustrates a star image in single-frame difference images for three frame pairs over a single exposure.  

\begin{figure*}[htb]
    \centering
    \includegraphics[width=0.99\textwidth]{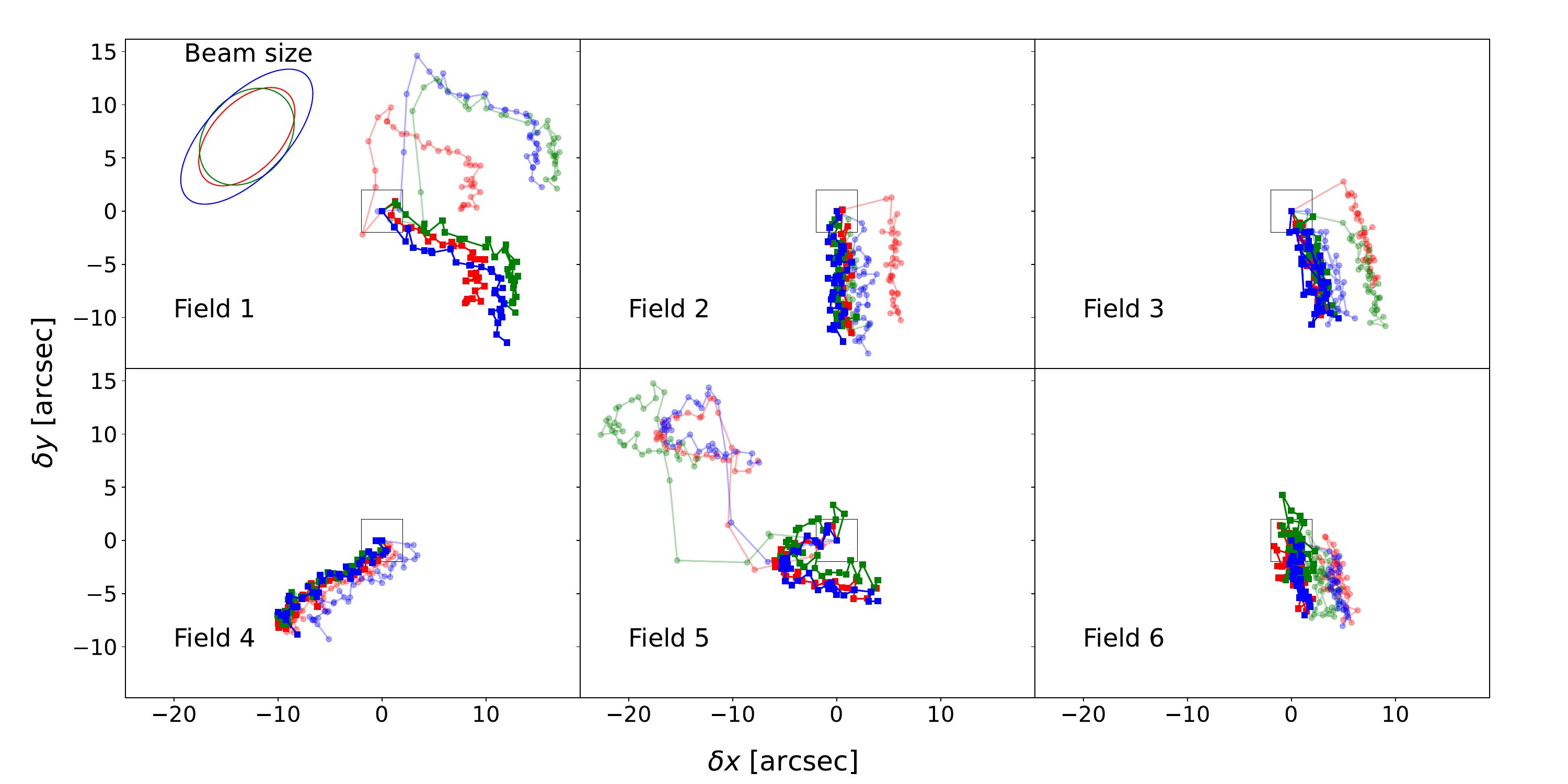} 
    \caption{\label{fig:pointing_drift} Pointing drift in each arm and field over the flight. This figure shows the drift using all frames in an exposure (circles), and then the drift after excluding the first few frames where the pointing instability is large (squares). Drifts for arm L are shown in red, arm M in green, and arm S in blue. The black box indicates the size of one $4^{\prime \prime}.0$ pixel in the CIBER-2 image for comparison. The equivalent optical PSF spot width of each arm are shown as ellipses, and illustrate that the drift in the cut exposures is less than or comparable to the optical beam size.
}
\end{figure*}

A summary of the observed pointing drift for all three arms is shown in Fig.~\ref{fig:pointing_drift}. Significant drift is evident in the first few frames of fields 1 and 5, which is consistent with the drift observed by the rocket star tracker data. To address this, we exclude the first five frames in field 1, the first six frames in field 5, and the first frame in all other fields from subsequent analysis.  This has the effect of reducing the pointing drift to $\lesssim 15^{\prime \prime}$, and thereby reducing the size of the effective PSF, measured by averaging the full width at half-maximum (FWHM) in the $x$- and $y$-directions from an elliptical Gaussian fit, by as much as 15\%.

\subsection{Optical Performance}
\label{sec:optical_performance}

The PSF for each band and field is measured inflight using stars as point sources.  Following the procedure described in \citet{Bock2013}, our estimate requires combining information obtained from two different analyses, one that obtains the ``core'' PSF from faint stars whose signal remains within the saturation limit, and the other that obtains the ``extended'' PSF from bright stars that saturate the PSF's core.  These are then hybridized to generate a full model for the PSF in each field and arm.

For CIBER-2, the core PSF is constructed by stacking stars with Vega magnitudes between 11.5 and 12.5, while the extended PSF is generated by stacking stars with Vega magnitudes between 6 and 9. Stellar magnitudes in the CIBER-2 bands are derived from existing astronomical catalogs, as explained in Section \ref{sec:catalog}. Stars within these magnitude ranges are stacked based on their positions using the astrometric solution that locates stars with subpixel accuracy for both the core and extended PSF. Following \citet{Symons2021}, we use a substacking process to increase the spatial resolution of our PSF model by dividing each pixel into a 9 $\times$ 9 subsampled grid.

\begin{figure*}[htb]
    \centering
    \includegraphics[width=\textwidth]{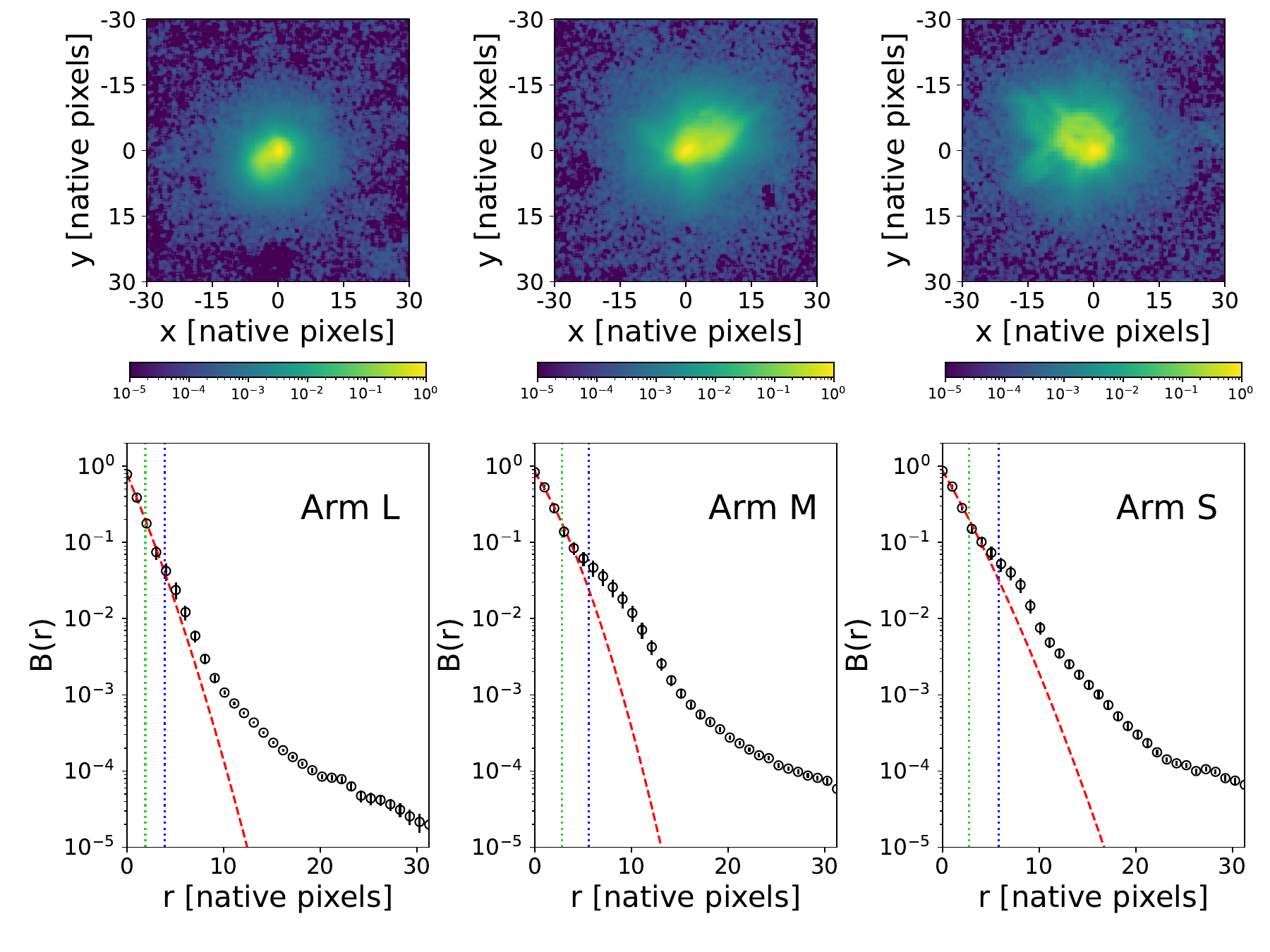}
    \caption{\label{fig:psf_profile} Radial profiles of the stacked PSF for the three arms (L, M, and S, from left to right). The best-fit Gaussian to the core PSF is represented by a red dashed curve. The blue dotted line indicates the radius encircling 85\% of the energy, corresponding to the radius of the red circles in Figure \ref{fig:focus_psf_ee}. The green dotted line marks the half width at half-maximum of the PSF. The PSF width shown in Figure \ref{fig:psfwidth_drift} is defined as the full width at half-maximum.}
\end{figure*}

Before combining the core and extended PSFs, the background is subtracted from the stacked PSF. The background level is estimated by averaging the pixel values where $r>58^{\prime \prime}$ for the core PSF and $r>169^{\prime \prime}$ for the extended PSF. After background subtraction, both PSFs are normalized by fitting a Gaussian and dividing the PSF by the model's integral. Since the extended PSF is saturated, the Gaussian fit is restricted to the region where $5.3<r<11^{\prime \prime}.1$ for the extended PSF. The full PSF is constructed by averaging the core and extended PSFs. For the core PSF, pixels with $r>17^{\prime \prime}$ are excluded from the average, while, for the extended PSF, pixels with values exceeding a defined threshold are excluded. This threshold is calculated as the annular average of the core PSF at $r=8^{\prime \prime}$, where it approximately matches the annular average of the extended PSF. In Figure \ref{fig:psf_profile} we show the radial profiles of the full PSF for the three arms. The core PSF closely follows a Gaussian profile, as shown by the red dashed curve, whereas the extended PSF deviates from the Gaussian profile.

\begin{figure*}[htb]
    \centering
    \includegraphics[width=\textwidth]{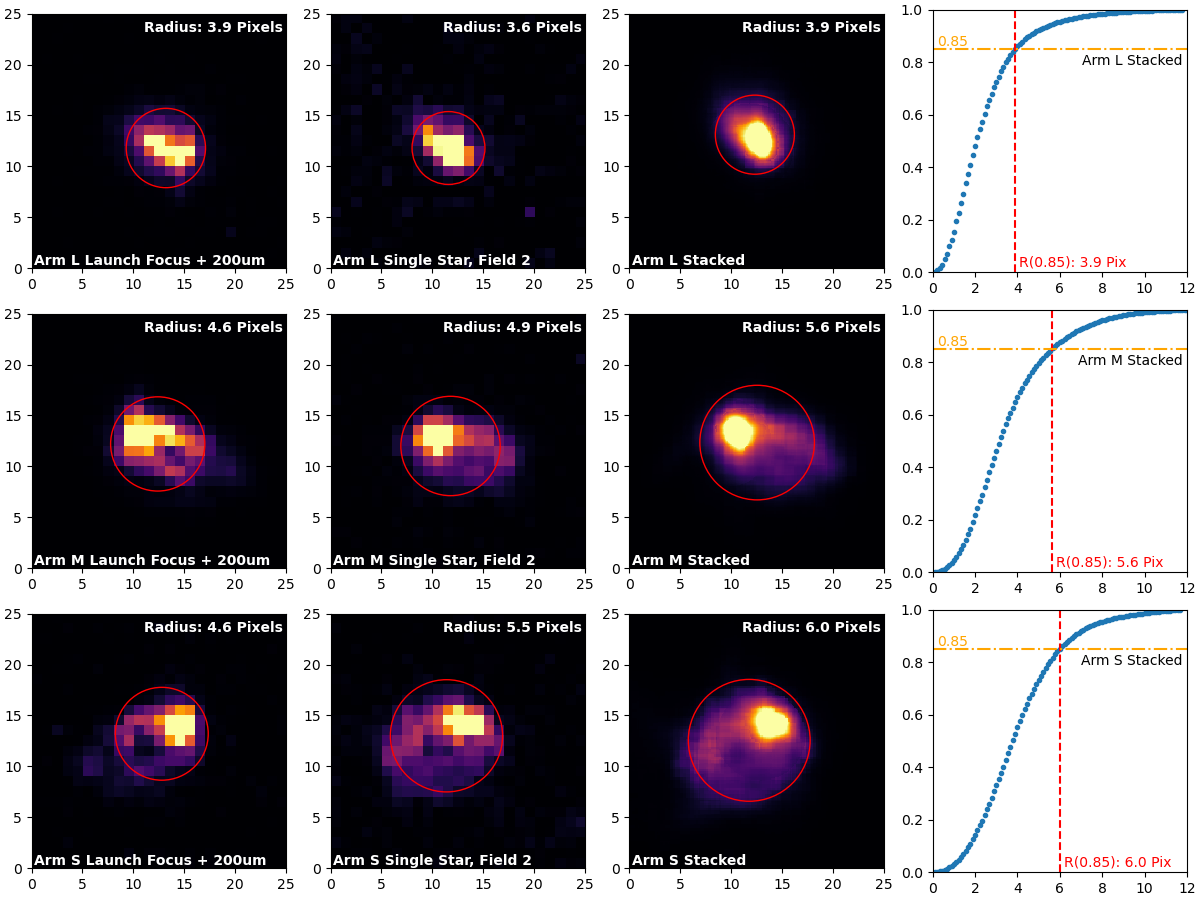} 
    \caption{\label{fig:focus_psf_ee} Ground and flight PSF comparison with stamp size 25 $\times$ 25 pixels.  Panels show (left to right) closest launch focus PSF, a single isolated star PSF from field 2 in flight, 9$\times$ resolution stacked PSF, and encircled energy graph for (top to bottom) arm L, M, and S, respectively.  Each overlaid red circle corresponds to the radius that encloses 85\% of the energy in the upsampled PSF centered on the weighted average signal center. Images are locally background subtracted 
    and scaled from zero to the 99th percentile of the signal in the stamp.}
\end{figure*}

The flight PSF can be compared to lab focus measurements to quantify the instrument's optical performance in flight.  We are particularly interested whether the launch, which is mechanically violent, changed the properties of the optical system. We estimate the PSF shape and size from $25 \times 25$ pixel stamps centered on an isolated star in field 2, which is similar in array location to that sampled by the laboratory measurement. The star used in this comparison is magnitude 11.6, 11.6 and 13.3 for arms L, M, and S, respectively.
The local background of the star is determined by taking the median of all the pixels bordering 
the stamp, and this value is subtracted from the stamp image.  With hot pixels masked, bright pixels due to illumination are isolated by determining the population with SNR $> 3$ in the stamp image.
The stamp image is then oversampled by a factor of 5 and the signal center is determined with a weighted average of positions by considering only these bright pixels.  Lastly, the radius that encircles 85\% of the energy as measured from the signal center is obtained from the oversampled image, providing a measure of the size of the PSF.
The PSF of a single star, stacked PSF, and closest PSF measurement during ground focus testing prior to launch are shown in Fig.~\ref{fig:focus_psf_ee}.  When compared to images of a point source as the system is moved through the best focus during focus measurements, the shape and diameter of these images suggest a coherent optical shift of $\sim 200\ \mu$m following testing, indicating a change in some shared feature of the optical chain between ground and flight conditions.  This is a significantly larger shift than we expect from the change in temperature of the telescope, which we observe to be $< 0.5$ K over a flight (see Section \ref{sS:focus}).

\begin{figure*}[htb]
    \centering
    \includegraphics[width=0.32\textwidth]{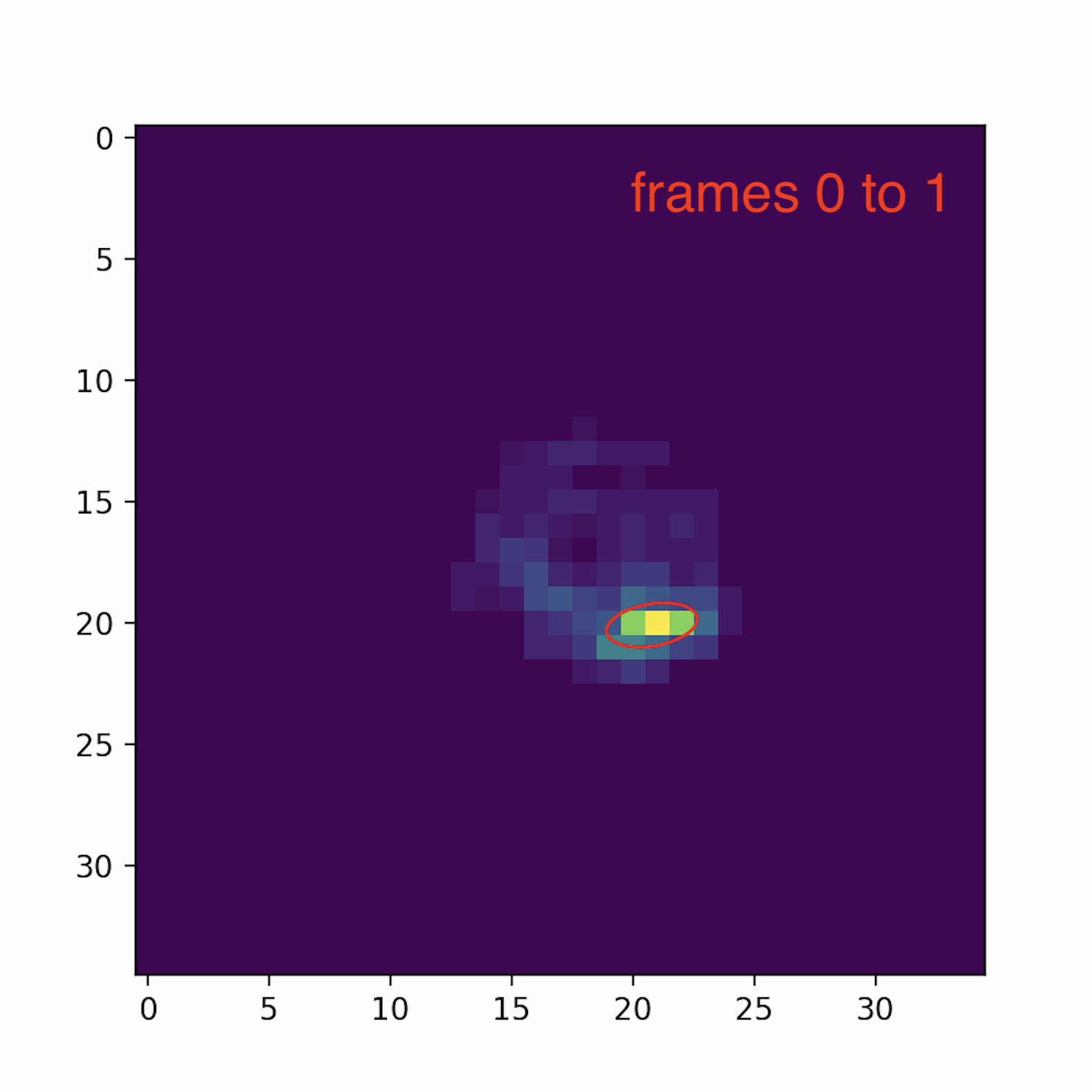} 
    \includegraphics[width=0.32\textwidth]{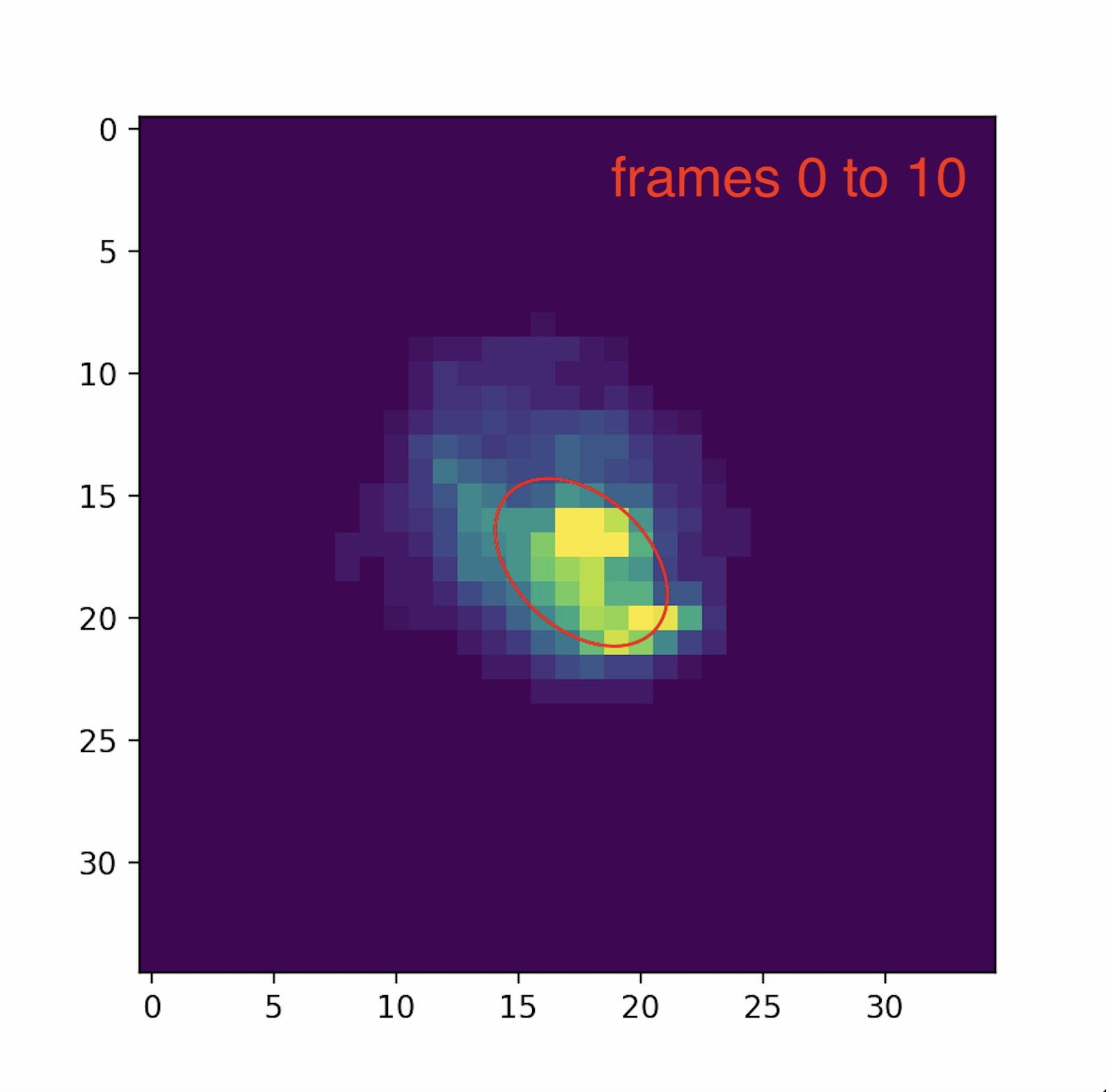}
    \includegraphics[width=0.32\textwidth]{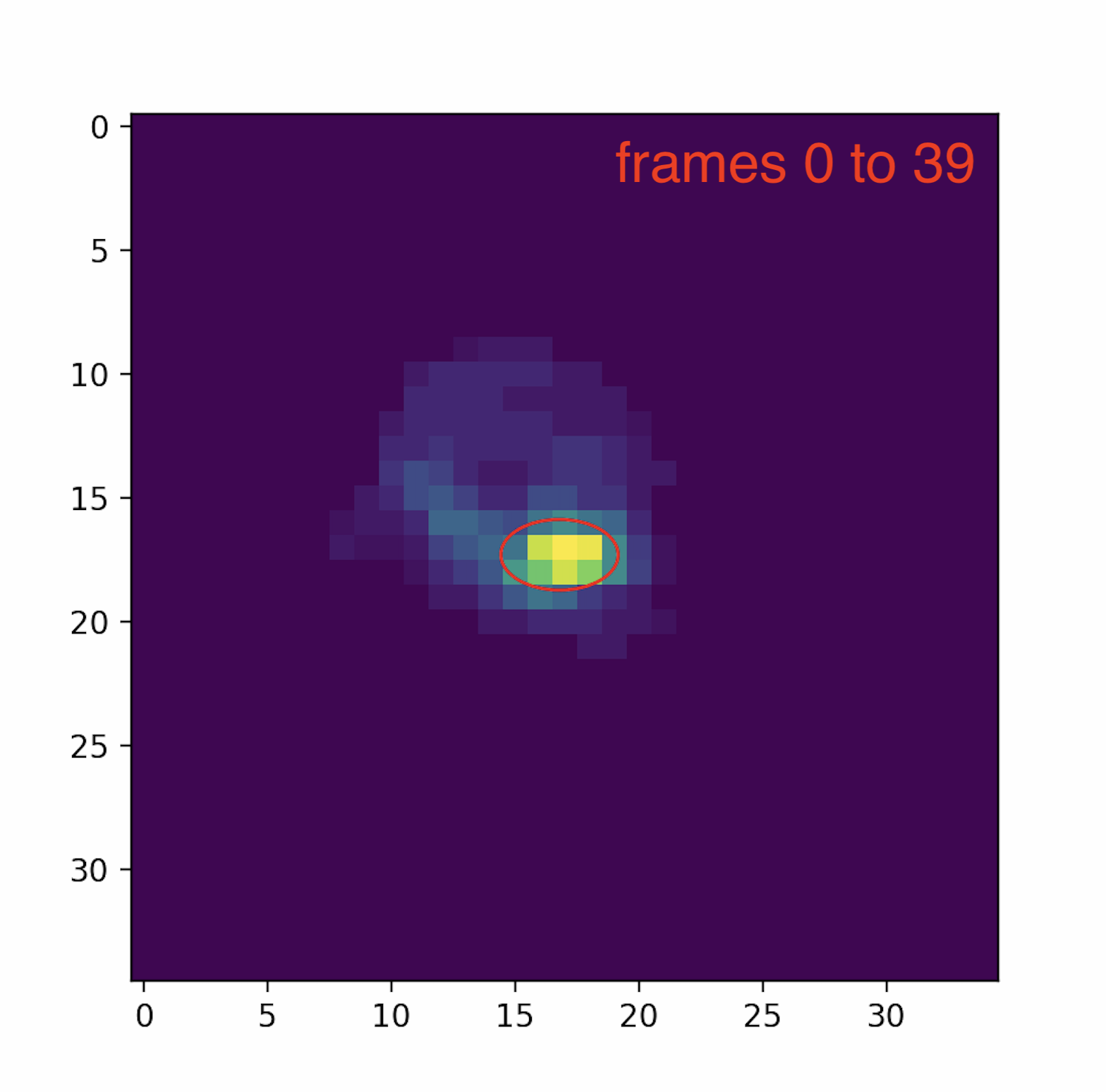}
    \caption{\label{fig:psf_width} Change in PSF width as traced by C1, field 5 line-fit images. The left panel shows a line-fit image using frames $0{-}1$, with a PSF width of 2.76 pixels represented by the red ellipse. The middle panel shows a line-fit image using frames $0{-}10$. The pointing of the experiment shifted, resulting in two bright spots and an increased PSF width of 6.81 pixels. The right panel shows a line-fit image using frames $0{-}39$. At this point, the pointing had been stable for tens of frames, so the PSF width decreases to 3.8 pixels.}
\end{figure*}

To evaluate the impact of pointing drift on the PSF width, we analyzed subexposure images formed by photocurrent fits from frame 0 to frame $i$ ($i=1,2,\dots,n_{\text{frame}}$ where $n_{\text{frame}}$ is the number of frames integrated over up to the total available for an exposure).  These images are equivalent to integrating an exposure for each integration time along the SUTR curve.  These images capture the progressive blurring of the PSF over the exposure as the instrument pointing changes, so the PSF width should be correlated with the observed pointing drift.  For all such images, the PSF width was determined by fitting an elliptical Gaussian to a bright star ($m_{\rm Vega} \sim 9$). The FWHM was calculated in both the $x$- and $y$-directions, and the average of these values was taken as the PSF width. Figure \ref{fig:psf_width} illustrates examples of the subexposure images and the corresponding PSF width measurements.  

\begin{figure*}[htb]
    \centering
    \includegraphics[width=0.49\textwidth]{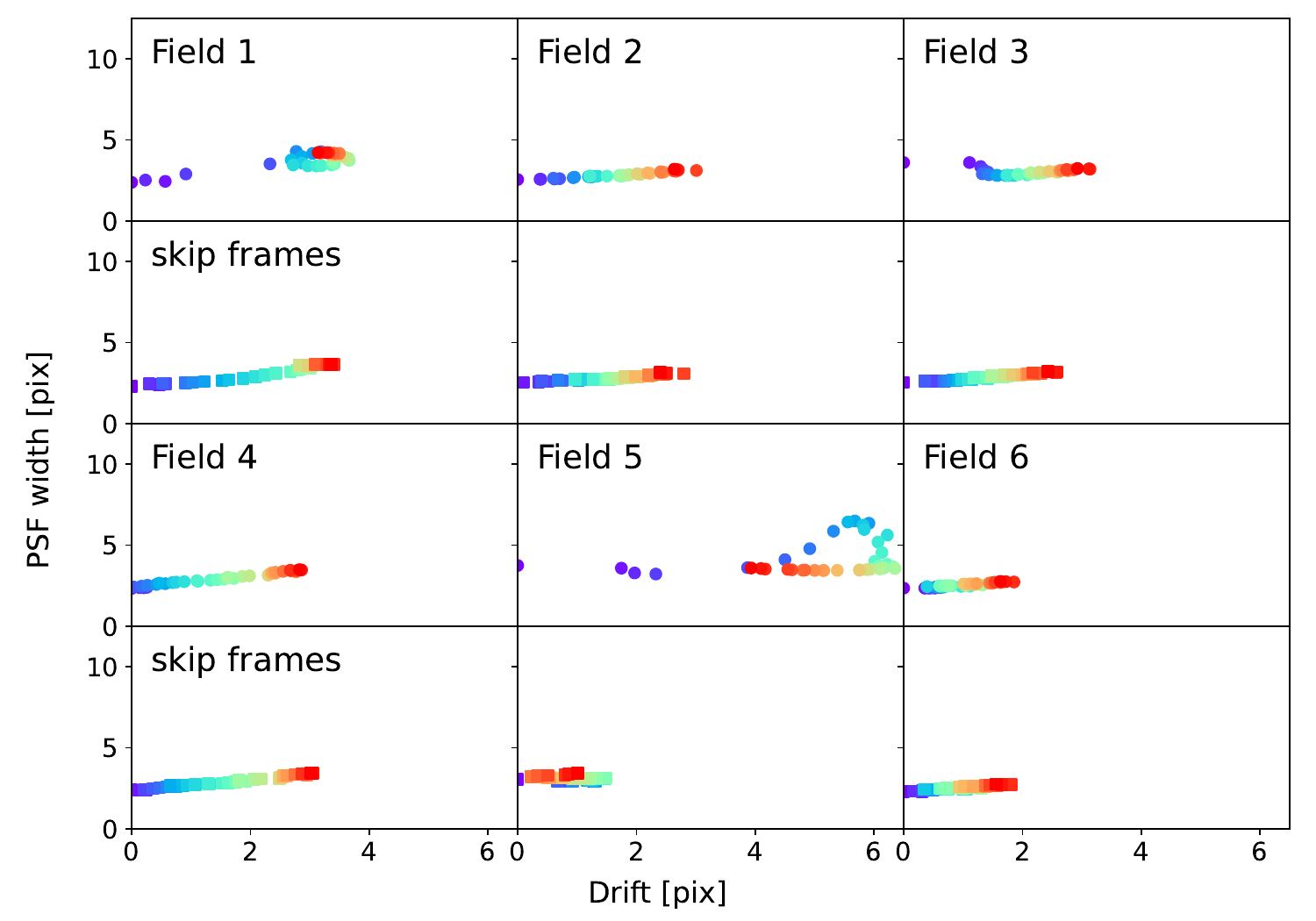} 
    \includegraphics[width=0.49\textwidth]{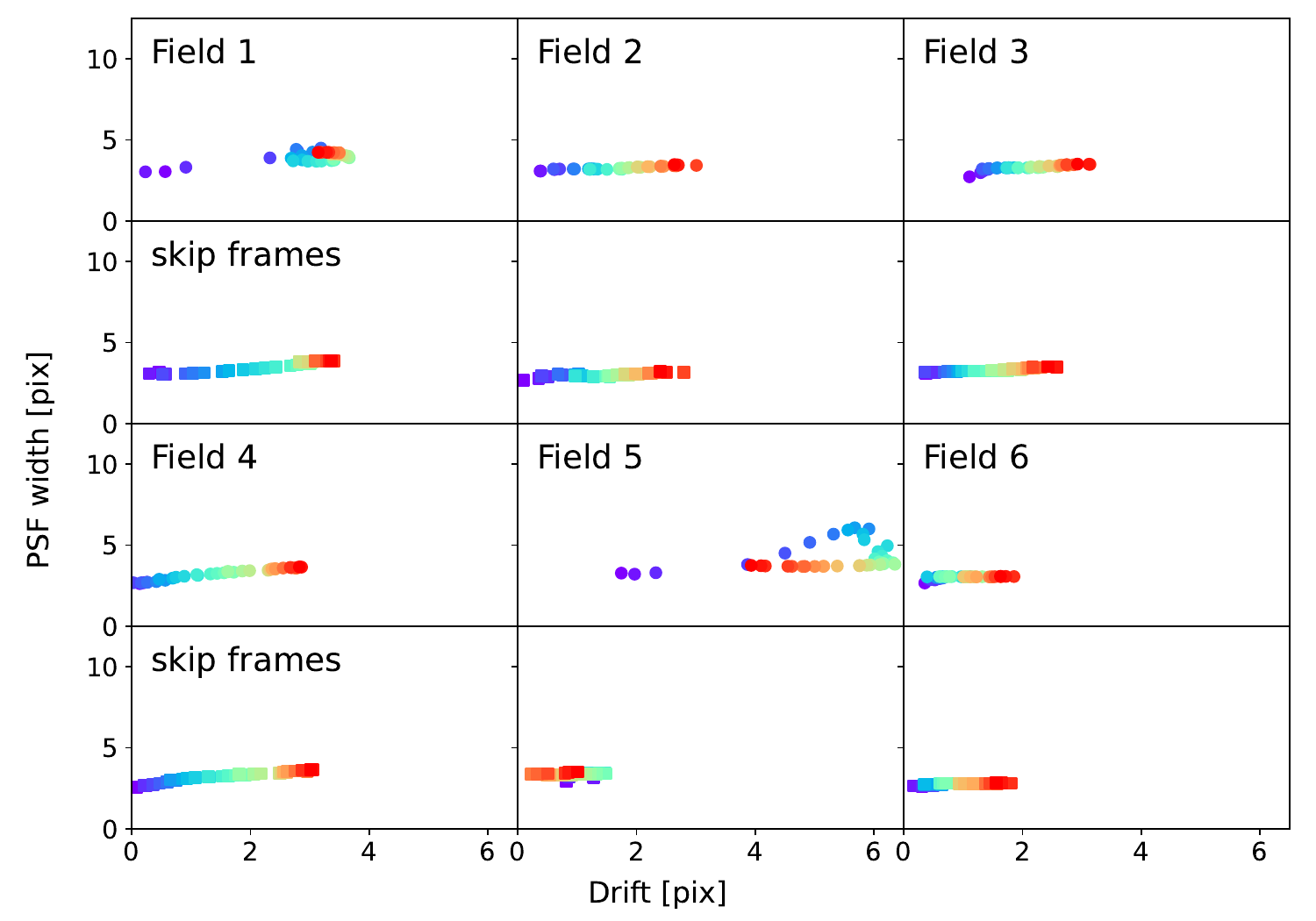} 
    \caption{\label{fig:psfwidth_drift} PSF width as a function of drift for the C3 band, shown for flight data (left) and simulated data (right). Each field presents the results using all frames (circles) and skipping the first few frames (squares). The points represent PSF widths at successive intervals throughout the integration, with colors indicating the progression from the start (violet) to the end (red) of the integration. Both data and simulations show a significant increase in PSF width during the initial frames of fields 1 and 5, attributed to unexpected motions observed in the top panel of Figure \ref{fig:pointing_drift}. In field 1, the pointing shifts by approximately 3 pixels, leading to a PSF width increase of about 2 pixels. Similarly, in field 5, the pointing shifts by 5 pixels, resulting in a PSF width increase of approximately 4 pixels. As the pointing stabilizes and drift diminishes, the PSF width decreases after the first few frames. The other fields exhibit only minor changes in PSF width, as the drift between frames is minimal.} 
\end{figure*}

The relationship between the PSF width and drift in all fields in C3 is illustrated in Figure \ref{fig:psfwidth_drift}.
To validate these results, we simulate frames using the drift data derived in Section \ref{sec:astrometric_performance} and a PSF model. A frame-difference image of a bright star was used as the PSF model to reduce the effect of pointing drifts. The full exposure image is generated by adding a sequence of PSF model images shifted by the measured drift in each frame. The right panel of Figure \ref{fig:psfwidth_drift} shows the PSF width in the simulated images as a function of drift for the same band (circles).  The data and simulation demonstrate good agreement in the PSF width and its variation across frames. 

To check the effect of the drift-based exposure cuts (see Section \ref{sec:astrometric_performance}), Figure \ref{fig:psfwidth_drift} also shows the PSF width in data and simulation with the first few frames excluded from the line-fitting process.  The large increase in PSF width observed in fields 1 and 5 in Figure \ref{fig:psfwidth_drift} is no longer present, as excluding the initial frames removes the significant PSF width increase caused by the large drift.

The as-flown optical performance of the instrument is worse than our expectations.  The PSF is $\{ 5, 22, 30\}$\% broader at arms \{S, M, L\} than our preflight expectations.  More problematically, the PSF has a broad comatic halo that increases the encircled energy considerably above its design values.  This is an issue for studies concerning point sources, which for CIBER-2 include calibration, where the extended PSF must be properly included in calculations.  However, background estimation is tolerant to large PSFs provided they are characterized, so we expect this will not preclude using CIBER-2 data for measurements of the diffuse EBL.

\subsection{Absolute Calibration}
\label{sec:inflight_calibration}

The primary CIBER-2 photometric calibration is performed in flight using high-significance stars of a known magnitude in the sky images. By comparing the measured photocurrent of these stars with their catalog fluxes, we can derive a flux conversion factor, $C_{\textrm{flux}}$. This factor, combined with the pixel solid angle, $\Omega_{\textrm{pix}}$, yields the conversion factor $C=C_{\textrm{flux}}\Omega_{\textrm{pix}}$, which allows conversion from photocurrent to surface brightness.

As the calibration catalogs, we use the C2SC and use all stars that fall into unmasked pixels in the magnitude range listed in Table \ref{table:conversion_factor}. To minimize the effects of detector nonlinearity, we exclude stars whose images accumulated charges exceeding 10\% of the saturation level. 

\begin{table*}[htb]
\caption{A per-band summary of the CIBER-2 photometric calibration.  Magnitudes are given in the Vega system, aperture and annulus radii are given in pixels, and the calibration factors $C$ carry units (nW m$^{-2}$ sr$^{-1}$)/(e$^{-}$ s$^{-1}$).  The uncertainty quoted on the measured $C$ is the formal statistical error from a bootstrap analysis and does not incorporate the absolute uncertainty in the source catalogs. }
\label{table:conversion_factor}
\vspace{-4mm}
\begin{center}
\begin{tabular}{ l c c c c c c }
 \hline
  & C1 & C2 & C3 & C4 & C5 & C6 \\
 \hline

Calibrator Mag. & $10.5 {-} 15$ &
$9.8 {-} 15$ & $10.1 {-} 14$ & $9.7 {-} 14$ & $10.0 {-} 13$ & $9.1 {-} 13$ \\ \hline


Aperture $r$ & 15.3 & 15.3 & 10.8 & 10.8 & 7.9 & 7.9 \\ 
Annulus $r_{\textrm{inner}}$ & 25.5 & 25.5 & 18.0 & 18.0 & 13.2 & 13.2 \\
Annulus $r_{\textrm{outer}}$ & 40.9 & 40.9 & 28.8 & 28.8 & 21.2 & 21.2 \\ \hline

 Design $C$  & 68.5 & 71.2 & 81.2 & 90.0 & 61.3 & 63.9 \\
 
 Measured $C$ & $134.2 \pm 0.5$ & $114.5 \pm 0.5$ & $112.3 \pm 0.3$ & $125.3 \pm 0.2$ & $74.4 \pm 0.2$ & $79.3 \pm 0.2$ \\

Efficiency Loss & 0.51 & 0.62 & 0.72 & 0.72 & 0.82 & 0.81 \\
 
 \hline
\end{tabular}
\end{center}
\end{table*}

\begin{figure}[th]
    \centering
    \includegraphics[width=0.47\textwidth]{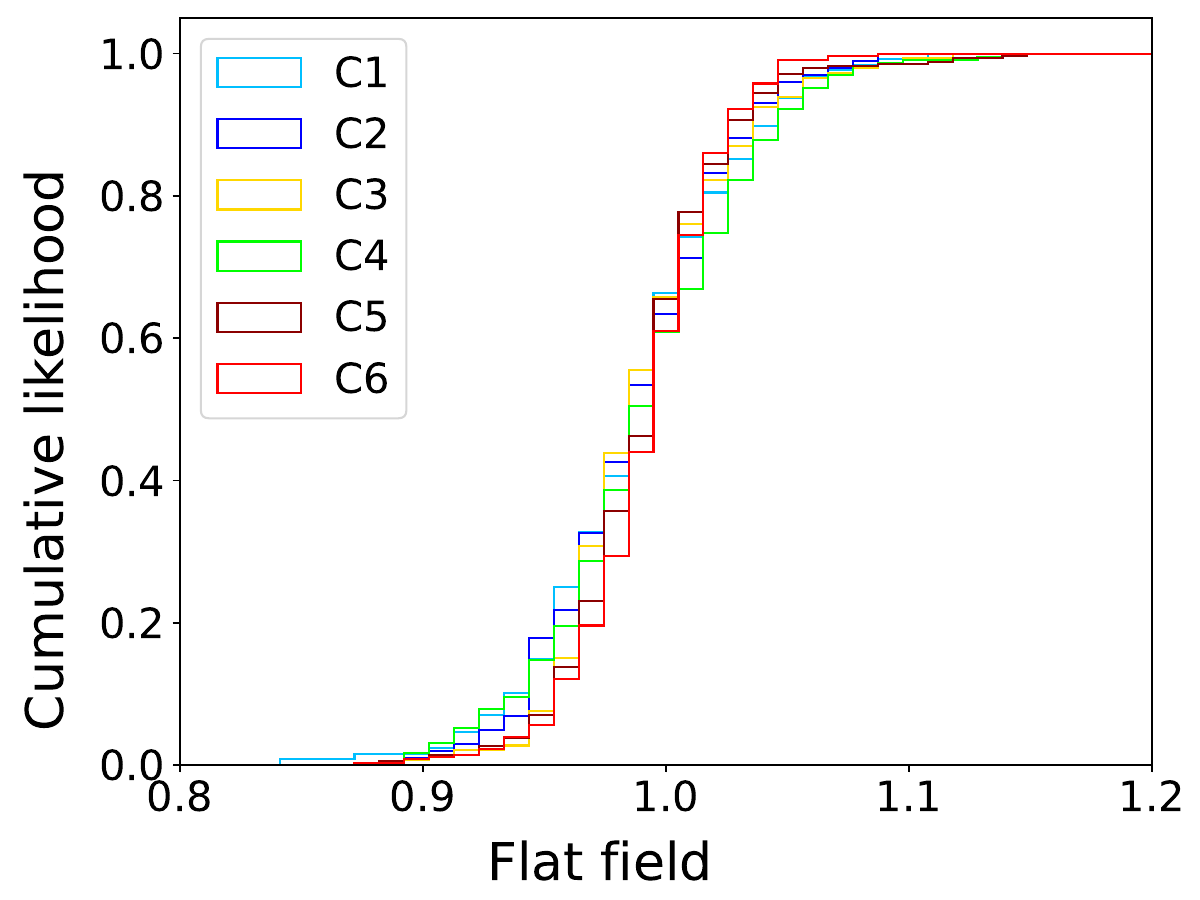} 
    \caption{\label{fig:flat_field_histogram} Cumulative histograms of flat field values at the positions of stars used for calibration in the C1${-}$6 bands. Since the mean values are consistent with 1 across all bands, the effect of flat field variations on the calibration is not biasing in our calibration.}
\end{figure}

The instrumental flat field, which is the relative response of each detector pixel to a uniform illumination at the telescope aperture, must be accounted for during absolute calibration. The flat field for each band is derived from the flight data by masking sources, normalizing each image by its mean value, and averaging the six observed fields. To assess whether the in-flight calibration is impacted by flat field variations, we generate a cumulative histogram of the flat field values at the positions of stars used for calibration for each band, shown as Figure \ref{fig:flat_field_histogram}. We find no evidence for a systematic bias in the calibration due to the positions of sources in the images in any band.  

The photocurrent from stars is measured using aperture photometry following well-established methods \citep{DaCosta1992}.  We perform the aperture photometry on stamps where every source in the C2SC except for the source of interest is masked, which removes their photocurrent contributions from the integral.  Since the PSF varies in each band, we vary the photometric and background apertures depending on the properties of point sources in the image.  
We find that a photometric aperture and background annulus radii, summarized in Table \ref{table:conversion_factor}, capture the flux from sources in the C1${-}$6 bands for the selected stars. The background annulus mean is then calculated after removing outliers using a 3$\sigma$ clip. This mean is multiplied by the number of pixels within the aperture and subtracted from the total signal inside the photometric aperture to isolate the star’s photocurrent.  This estimate of the photocurrent can then be compared with the known flux from the source to estimate the calibration factor.

The flux conversion factor, $C_{\textrm{flux}}$, is defined as
\begin{equation}
    C_{\textrm{flux}} = \frac{\lambda F_{\textrm{ref}}(\lambda_{\rm p})}{\sum i_i},
\end{equation}
where $F_{\textrm{ref}} \propto \lambda^{-1}$ is the reference spectrum, $\lambda_{\rm p}$ is the pivot wavelength of the CIBER-2 band, and $\sum i_i$ is the total signal measured through aperture photometry. Since the spectrum of a star $F_{\lambda}$ is not proportional to $\lambda^{-1}$, we need to apply a color correction factor, $K$. The color correction factor $K$ is given by
\begin{equation}
    K = \frac{\int F_{\lambda} T_{\lambda} \lambda d\lambda}{F_{\lambda}(\lambda_{\rm p})\lambda_{\rm p} \int T_{\lambda} d\lambda}
\end{equation}
where $T_{\lambda}$ is the transmittance profile of CIBER-2 band.

\begin{figure*}[htb]
    \centering
    \includegraphics[width=\textwidth]{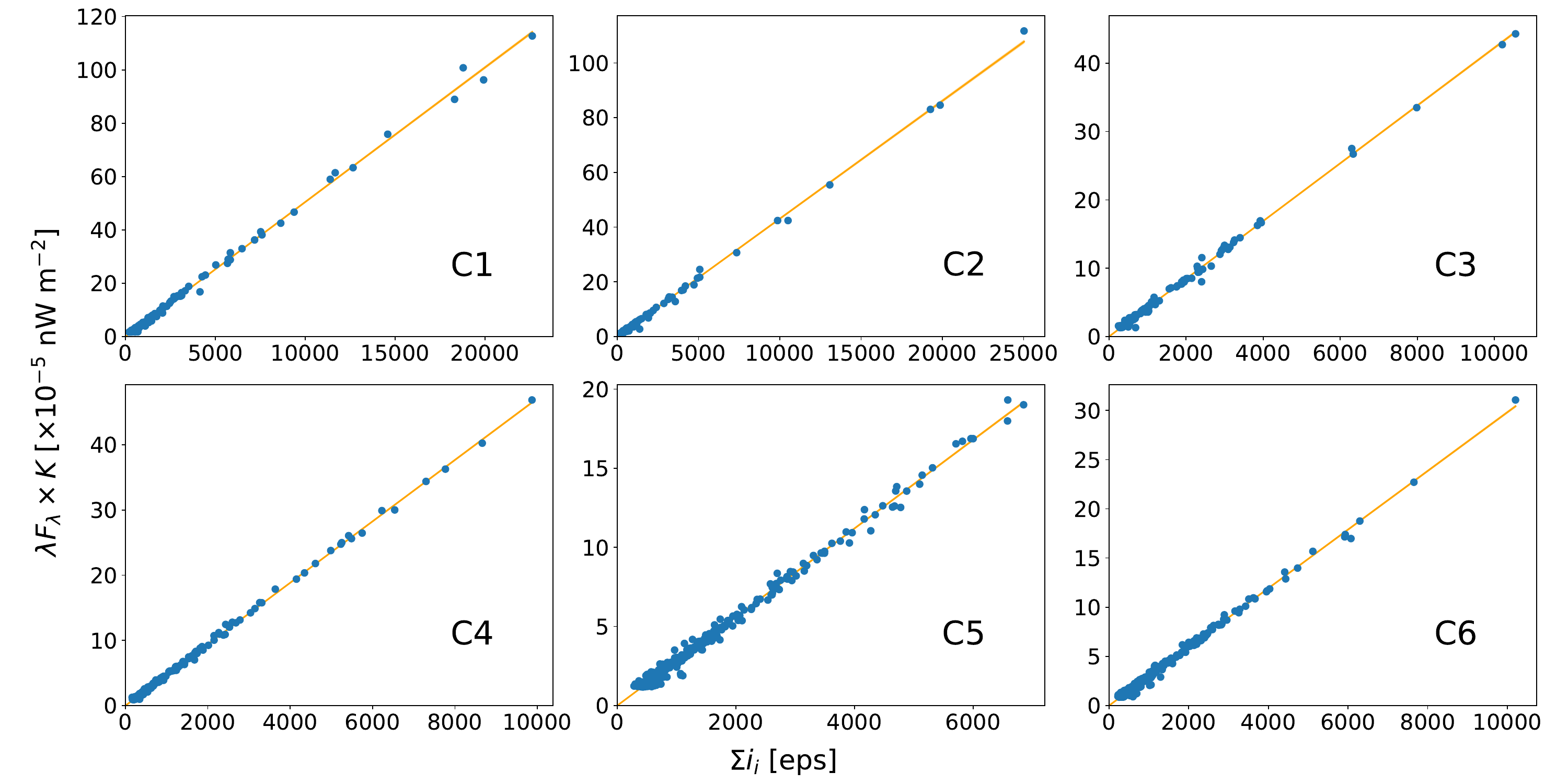} 
    \caption{\label{fig:conversion_factor} The relationship between the catalog flux $\lambda F_{\lambda} \times K$ and the photocurrent $\sum i_{i}$ for each star (blue points). The flux conversion factors are the slope of the fitting (orange lines).}
\end{figure*}

We determine the flux conversion factor for each CIBER-2 band using $\sim100$ stars per field from the sky images. The relationship between catalog fluxes and measured photocurrents that sets the calibration is shown in Figure \ref{fig:conversion_factor}. Flux conversion factors are estimated by fitting these data points to a linear model, where the slope corresponds to the flux conversion factor. To convert the flux conversion factors into surface brightness units, we divide by the pixel solid angle estimated from astrometric registration (see Section \ref{sec:astrometric_performance}). 

The uncertainties in the flux conversion factors are estimated using a Monte Carlo method that requires the generation of 100 mock realizations of the data. A single realization is generated by sampling from the distribution of residuals from the original linear fit and adding these to the best-fit model. For each realization, the slope of the linear fit is calculated, and the standard deviation of these slopes across all realizations is taken as the uncertainty.
Table \ref{table:conversion_factor} summarizes the conversion factors obtained from in-flight calibration and theoretical calculations. 

We observe larger values of $C$ than expected from an instrument model assuming the parameters listed in Table \ref{table:detector_filter}, which we quantify by computing the ratio of the expected conversion factors and those derived from in-flight measurements. A larger value of $C$ means the instrument is less responsive than predicted, and is most likely due to poor optical efficiency somewhere in the optical chain.  The quality and performance of the antireflection coating of the telescope mirrors have visibly degraded between CIBER-2's first flight in 2021 and third flight in 2024.  It is likely that a large fraction of the difference between the flight and expectations is due to the degraded coating, which laboratory tests suggest is due to water and would affect the shortest wavelengths the most.  Improvements to the mirror coating would improve the absolute responsivity of the instrument by as much as a factor of 2 at short wavelengths. 

\section{Conclusions}

CIBER-2 provides timely in-flight end-to-end demonstrations of unique technologies and analysis methods relevant
to the SPHEREx MIDEX mission launched 2025 March 11.  CIBER-2 operates H2RG arrays
to meet the unique and demanding requirements for controlling spatial $1/f$ noise necessary for measuring
clustering anisotropy, a core science program on SPHEREx. Hardware and methods first developed for CIBER-2
have been incorporated and refined for SPHEREx, including stabilized room-temperature readout electronics, reference pixel subtraction, temperature-controlled focal planes, and the data required to develop row-chopping, a new method for skipping rows to modulate $1/f$ noise intrinsic to the array readout out of the low spatial frequencies of scientific interest \citep{Heaton2023,Nguyen2025}. CIBER-2 used LVFs, smaller versions of the filters used for spectroscopy on SPHEREx, for absolute EBL photometry. CIBER-2 is exploring the challenges of implementing these filters in a flight experiment by understanding how to account for spectral leaks and ghost reflections,
and how to measure the spectral response in situ as part of the instrument calibration.

CIBER-2 data will also develop new EBL fluctuation analysis methods that can be incorporated and refined in the SPHEREx data analysis pipeline. Indeed, key analysis modules incorporated into the SPHEREx EBL pipeline
grew from the CIBER-1 data analysis \citep{Zemcov2014,Feder2025}, including PSF estimation through stacking \citep{Symons2021}, source masking and mode-coupling correction, noise modeling, systematic error assessment in power spectra, and component separation methods \citep{Feng2019,Cheng2022}.  This rich interplay will continue as CIBER-2 analysis confronts the new challenge of analyzing data at multiple wavelengths through a large set of cross correlations.

\acknowledgments

We are grateful for the support of the engineers, staff, and technicians at the NASA Wallops Flight Facility and the White Sands Missile Range.

This material is based upon work supported by
the National Aeronautics and Space Administration under APRA research grants NNX10AE12G,
NNX16AJ69G, 80NSSC20K0595, 80NSSC22K0355, and
80NSSC22K1512.
CSTARS was supported in part by USIP NASA grant NNX16AI82A. 
Japanese participation is supported by KAKENHI (15H05744, 18KK0089, 21KK0054, 21111004, 22H00156, and 24KK0071) from Japan Society for the Promotion of Science (JSPS) and the Ministry of Education, Culture, Sports, Science and Technology (MEXT). 
Korean participation is funded by the Pioneer Project from Korea Astronomy and Space Science Institute (KASI). C.H.N.~acknowledges support from NASA Headquarters under  NASA Earth and Space Science Fellowship Program grant 80NSSCK0706.

Our thanks to J.~Battle, K.~Suzuki, T.~Wada, Y.~Onishi, K.~Danbayashi, S.~Hanai, T.~Morford, and S.~Sakai for help in designing, fabricating, and testing various parts of the instrument over its history.  Thanks also to S.~Davis, K.~Gates, M.~Klein, B.~Stewart, and D.~Schake for their contributions to the CSTARS star-tracking instrument. 

\bibliography{ref}{}
\bibliographystyle{aasjournal}

\end{document}